%% file: BTV-12-001_temp.tex
\pdfoutput=1

\documentclass[11pt,twoside,a4paper,cmspaper,final,collab]{cms-tdr}
\def\svnVersion{187805}\def\cmsCernNo{2012-262}\def\cmsCernDate{\today}\def\cmsMessage{Submitted to the Journal of Instrumentation}

\begin{document}\cmsNoteHeader{BTV-12-001}

\hyphenation{had-ron-i-za-tion}
\hyphenation{cal-or-i-me-ter}
\hyphenation{de-vices}

\RCS$Revision: 166817 $
\RCS$HeadURL: svn+ssh://svn.cern.ch/reps/tdr2/papers/BTV-12-001/trunk/BTV-12-001.tex $
\RCS$Id: BTV-12-001.tex 166817 2013-01-24 13:38:02Z adamwo $
\renewcommand{\fixme}[1]{}

\newcommand{\comment}[1]{}
\newcommand{\pb}{\ensuremath{\mathrm{pb}}}%
\newcommand{\rts}{\ensuremath{\sqrt{s}}}%
\newcommand{\ptrel}{\ensuremath{p_\mathrm{T}^\text{rel}}}
\newcommand{\ptmu}{\ensuremath{p_\mathrm{T}^{\mu}}}
\newcommand{\pthat}{\ensuremath{\hat{p}_{\mathrm{T}}}\xspace}
\newcommand{\effb}{\ensuremath{\varepsilon_\cPqb^{\text{tag}}}}
\newcommand{\effc}{\ensuremath{\varepsilon_\cPqc^{\text{tag}}}}
\newcommand{\SFb}{\ensuremath{SF_\cPqb}}
\hyphenation{mis-iden-ti-fi-ca-tion}
\providecommand{\FEWZ}{\textsc{fewz}\xspace}

\providecommand{\bdtag}{\cPqb-jet tagging\xspace}
\providecommand{\bi}{\begin{itemize}}
\providecommand{\ei}{\end{itemize}}
\providecommand{\fixme}[1]{{\large\sffamily{\bfseries{}FIXME:} #1}}
\providecommand{\update}[1]{{\large\sffamily{\bfseries{}UPDATE:} #1}}
\providecommand{\comment}[1]{{\large\sffamily{\bfseries{}COMMENT:} #1}}
\newcommand{\met}{\makebox[2.4ex]{\ensuremath{\not\!\! E_{\mathrm{T}}}}}
\newcommand{\mymet}{\makebox[2.4ex]{\ensuremath{\not\!\! E_{\mathrm{T}}}}}
\newcommand{\ftcm}{flavour tag consistency method\xspace}
\newcommand{\plrm}{profile likelihood ratio method\xspace}
\newcommand{\ftmm}{flavour tag matching method\xspace}
\newcommand{\ftc}{flavour tag consistency\xspace}
\newcommand{\plr}{profile likelihood ratio\xspace}
\newcommand{\ftm}{flavour tag matching\xspace}
\newcommand{\FTCm}{FTC method\xspace}
\newcommand{\PLRm}{PLR method\xspace}
\newcommand{\FTMm}{FTM method\xspace}

\cmsNoteHeader{BTV-12-001} 
\title{Identification of b-quark jets with the CMS experiment}

\date{\today}

\abstract{
At the Large Hadron Collider, the identification of jets originating from
 b quarks is important for searches for new physics and for
 measurements of standard model processes.
A variety of algorithms has been developed by CMS to select b-quark jets
 based on variables such as the impact parameters of charged-particle tracks,
 the properties of reconstructed decay vertices, and the presence or absence
 of a lepton, or combinations thereof.
The performance of these algorithms has been measured using data from
 proton-proton collisions at the LHC and compared with expectations based on
 simulation.
The data used in this study were recorded in 2011 at $\sqrt{s} = 7\TeV$ for a
 total integrated luminosity of 5.0\fbinv.
The efficiency for tagging b-quark jets has been measured in events from
 multijet and t-quark pair production.
CMS has achieved a b-jet tagging efficiency of 85\% for a light-parton misidentification probability of 10\%  in multijet events.
For analyses requiring higher purity, a misidentification probability of only 1.5\% has been achieved, for a 70\% b-jet tagging efficiency.
}

\hypersetup{%
pdfauthor={CMS Collaboration},%
pdftitle={Identification of b-quark jets with the CMS experiment},%
pdfsubject={CMS},%
pdfkeywords={CMS, physics, software, computing}}

\maketitle 

\input{Introduction}
\input{Detector}

\input{Samples}

\input{Algorithms}

\input{EfficienciesMuonJets}

\input{EfficienciesTTbar}

\input{Results}
\input{Mistags}

\input{Conclusions}

\section*{Acknowledgements}
\input{Acknowledgements}

\bibliography{auto_generated}   

\input{Glossary}

\appendix

\cleardoublepage \appendix\section{The CMS Collaboration \label{app:collab}}\begin{sloppypar}\hyphenpenalty=5000\widowpenalty=500\clubpenalty=5000\input{BTV-12-001-authorlist.tex}\end{sloppypar}
\end{document}

%% file: Introduction.tex
\section{Introduction}\label{sec:Introduction}

Jets that arise from bottom-quark hadronization (\cPqb\ jets) are present in many physics processes, such as the decay of top quarks, the Higgs boson, and various new particles predicted by supersymmetric models. 
The ability to accurately identify \cPqb\ jets is crucial in reducing the otherwise overwhelming background to these channels from processes involving jets from gluons (\cPg) and light-flavour quarks (\cPqu, \cPqd, \cPqs), and from \cPqc-quark fragmentation. 

The properties of the bottom and, to a lesser extent, the charm hadrons can be used to identify the hadronic jets into which the \cPqb\ and \cPqc\ quarks fragment.  These hadrons have relatively large masses, long lifetimes and daughter particles with hard momentum spectra.  Their semileptonic decays can be exploited as well.  The Compact Muon Solenoid (CMS) detector, with its precise charged-particle tracking and robust lepton identification systems, is well matched to the task of \cPqb-jet identification (\cPqb-jet tagging). The first physics results using \cPqb-jet tagging have been published~\cite{Chatrchyan:2011yy, Chatrchyan:2011bj, Chatrchyan:2011vp} from the first data samples collected at the Large Hadron Collider (LHC).

This paper describes the \cPqb-jet tagging algorithms used by the CMS experiment and measurements of their performance. 
A description of the apparatus is given in Section~\ref{sec:Detector}. The event samples
and simulation are discussed in Section~\ref{sec:Samples}.
The algorithms for \cPqb-jet tagging are defined in Section~\ref{sec:Algorithms}.
The distributions of the relevant observables are compared between simulation and proton-proton collision data collected in 2011 at a centre-of-mass energy of 7\TeV.  
The robustness of the algorithms with respect to running conditions, such as the alignment of the detector elements and the presence of 
additional collisions in the same bunch crossing (pileup), is also discussed.

Physics analyses using \cPqb-jet identification require the values of the efficiency and misidentification probability of the chosen algorithm, and, in general, these are a function of the transverse momentum (\pt) and pseudorapidity ($\eta$) of a jet. They can also depend on parameters such as the efficiency of the track reconstruction, the resolution of the reconstructed track parameters, or the track density in a jet.  While the CMS simulation reproduces the performance of the detector to a high degree of precision, it is difficult to model all the parameters relevant for \cPqb-jet tagging.  Therefore it is essential to measure the performance of the algorithms directly from data.  These measurements are performed with jet samples that are enriched in \cPqb\ jets, either chosen by applying a discriminating variable on jets in multijet events or by selecting jets from top-quark decays.  
The methods that are used to measure the performance are described in Sections~\ref{sec:EfficienciesMuonJets} and~\ref{sec:EfficienciesTTbar}.
The measurements are complementary: multijet events cover a wider range in \pt, while the results obtained from \ttbar events are best suited for some studies of top-quark physics.
The efficiency measurements are summarized and compared in Section~\ref{sec:Results}.
The measurement of the misidentification probability of light-parton (\cPqu, \cPqd, \cPqs, \cPg) jets as \cPqb\ jets in the data is presented in Section~\ref{sec:Mistags}.

%% file: Detector.tex
\section{The CMS detector}\label{sec:Detector}

The central feature of the CMS apparatus is a superconducting solenoid, of 6\unit{m} internal diameter, which provides a magnetic field of 3.8\unit{T}.  Within the field volume are the silicon tracker, the crystal electromagnetic calorimeter, and the brass/scintillator hadron calorimeter.  Muons are measured in gas-ionization detectors embedded in the steel return yoke.

CMS uses a right-handed coordinate system, with the origin at the nominal interaction point, the $x$ axis pointing to the centre of the LHC ring and the $z$ axis along the counterclockwise-beam direction.
The polar angle, $\theta$, is measured from the positive $z$ axis and the azimuthal angle, $\phi$, is measured in the $x$-$y$ plane.
The pseudorapidity is defined as $\eta \equiv -\ln[\tan(\theta/2)]$.
A more detailed description of the CMS detector can be found elsewhere~\cite{CMS:2008zzk}.

The most relevant detector elements for the identification of \cPqb\ jets and the measurement of algorithm performance are the tracking system and the muon detectors.
The inner tracker consists of 1440 silicon pixel and 15\,148 silicon strip detector modules.
It measures charged particles up to a pseudorapidity of $|\eta| < 2.5$.
The pixel modules are arranged in three cylindrical layers in the central part of CMS and two endcap disks on each side of the interaction point.  The silicon strip detector comprises two cylindrical barrel detectors with a total of 10 layers and two endcap systems with a total of 12 layers at each end of CMS.
The tracking system provides an impact parameter (IP) resolution of about 15 (30)\mum at a \pt of 100 (5)\GeVc.
In comparison typical IP values for tracks from \cPqb-hadron decays are at the level of a few 100\mum.
Muons are measured and identified in detection layers that use three technologies: drift tubes, cathode-strip chambers, and resistive-plate chambers.
The muon system covers the pseudorapidity range $|\eta|< 2.4$.
The combination of the muon and tracking systems yields muon candidates of high purity with a \pt resolution of about 1 to 3\%, for \pt values from 5 to 100\GeVc.

%% file: Samples.tex
\section{Data samples and simulation}\label{sec:Samples}

Samples of inclusive multijet events for the measurement of efficiencies and misidentification probabilities were collected using jet triggers with \pt thresholds of 30 to 300\GeVc.
For efficiency measurements, dedicated triggers were used to enrich the data sample with jets from semimuonic b-hadron decays.
These triggers required the presence of at least two jets with \pt thresholds ranging from 20 to 110\GeVc.
One of these jets was required to include a muon with $\pt > 5 \GeVc$ within a cone of $\DR = 0.4$ around the jet axis, where \DR\ is defined as $\sqrt{(\Delta\phi)^2 + (\Delta\eta)^2}$.
Triggers with low-\pt thresholds were prescaled in order to limit the overall trigger rates.
Depending on the prescale applied to the trigger, the multijet analyses used datasets with integrated luminosities of up to $5.0 \fbinv$.

Data for the analysis of \ttbar events were collected with single- (\Pe\ or $\mu$) and double-lepton (\Pe\Pe\ or \Pe$\mu$ or $\mu\mu$) triggers.
The samples were collected in the first part of the 2011 data taking with an integrated luminosity of $2.3 \fbinv$.
The precision on the \cPqb-jet tagging efficiency from \ttbar\ events is limited by systematic uncertainties.
Using the full dataset collected in 2011 would not significantly reduce the overall uncertainty.

Monte Carlo (MC) simulated samples of multijet events were generated with \PYTHIA 6.424~\cite{Sjostrand:2006za} using the Z2 tune~\cite{Field:2010}.
For \cPqb-jet tagging efficiency studies, dedicated multijet samples have been produced with the explicit requirement of a muon in the final state.

In the simulation, a reconstructed jet is matched with a generated parton if the direction of the parton is within a cone of radius $\DR = 0.3$ around the jet axis.  The jet is then assigned the flavour of the parton.  Should more than one parton be matched to a given jet, the flavour assigned is that of the heaviest parton.
The \cPqb\ flavour is given priority over the \cPqc\ flavour, which in turn is given priority over light partons.
According to this definition jets originating from gluon splitting to $\cPqb\cPaqb$, which constitute an irreducible background for all tagging algorithms, are classified as \cPqb\ jets.

Events involving \ttbar production were simulated using the \textsc{MadGraph}~\cite{Alwall:2011uj} event generator (v.~5.1.1.0), where the top quark pairs were generated with up to four additional partons in the final state.
A top quark mass of $m_\cPqt = 172.5\GeVcc$ was assumed.
The parton configurations generated by \textsc{MadGraph} were processed with \textsc{Pythia} to provide showering of the generated particles.
The soft radiation was matched with the contributions from the matrix element computation using the $k_\mathrm{T}$-MLM prescription~\cite{mlm}.
The tau-lepton decays were handled with \TAUOLA (v.~27.121.5)~\cite{Davidson:2010rw}.

The electroweak production of single top quarks is considered as a background process for analyses using \ttbar\ events, and was simulated using \POWHEG~301~\cite{powheg}.
The production of  $\PW/\Z$  + jets events, where the vector boson decays leptonically, has a signature similar to \ttbar and constitutes the main background.
These events were simulated using \MADGRAPH+\PYTHIA, with up to four additional partons in the final state.
The bottom and charm components are separated from the light-parton components in the analysis by matching reconstructed jets to partons in the simulation.

Signal and background processes used in the analysis of \ttbar events were normalized to next-to-leading-order (NLO) and next-to-next-to-leading-order (NNLO) cross sections, with the exception of the QCD background.

The top-quark pair production NLO cross section was calculated to be $\sigma_{\ttbar}=157^{+23}_{-24}\ \pb$, using \MCFM \cite{Campbell:2010ff}.
The uncertainty in this cross section includes the scale uncertainties, estimated by varying simultaneously the factorization and renormalization scales by factors of 0.5 or 2 with respect to the nominal scale  of $(2m_\cPqt)^2 + (\sum \pt^\text{parton})^2$, where $p_T^\text{parton}$ are the transverse momenta of the partons in the event.
The uncertainties from the  parton distribution functions (PDF)  and the value of the strong coupling constant $\alpha_S$ were estimated following the procedures from the MSTW2008~\cite{mstw08}, CTEQ6.6~\cite{Lai:2010nw}, and NNPDF2.0~\cite{Demartin:2010er} sets.
The uncertainties were then combined according to the PDF4LHC prescriptions~\cite{Botje:2011sn}.

The $t$-channel single top NLO cross section was calculated to be $\sigma_\cPqt = 64.6^{+3.4}_{-3.2}\ \pb$ using \textsc{MCFM}~\cite{Campbell:2010ff,mcfm2,mcfm3,mcfm4}.
The uncertainty was evaluated in the same way as for top-quark pair production.
The single top-quark associated production (\cPqt\PW) cross section was set to $\sigma_{\cPqt\PW}=15.7\pm1.2\ \pb$~\cite{Kidonakis:2010tW}.
The $s$-channel single top-quark next-to-next-to-leading-log (NNLL)  cross section was determined to be $\sigma_{s} = 4.6\pm0.1\ \pb$~\cite{NNLLtop}.

The NNLO cross section of the inclusive production of \PW\ bosons  multiplied by its branching fraction to leptons was determined to be  $\sigma_{\PW\rightarrow \ell\nu} = 31.3 \pm 1.6\unit{nb}$ using \FEWZ~\cite{fewz},
setting the renormalization and factorization scales to $(m_\PW)^2 + (\sum \pt^\text{jet})^2$ with $m_\PW=80.398\GeVcc$.
The uncertainty was determined in the same way as in top-quark pair production.
The normalizations of the \PW+\cPqb\ jets and \PW+\cPqc\ jets components were determined in a measurement of the top pair production cross section in the lepton+jet channel~\cite{CMS-PAS-TOP-11-003}, where a simultaneous fit of the  \ttbar\ cross section and the normalization of the main backgrounds was performed.

The Drell--Yan production cross section at NNLO was calculated using \FEWZ as
$\sigma_{\Z/\gamma^*\rightarrow \ell\ell} (m_{\ell\ell}>20\GeV) = 5.00 \pm 0.27$\unit{nb}, where $m_{\ell\ell}$ is the invariant mass of the two leptons  and the scales were set using  the \Z boson mass $m_\Z=91.1876\GeVcc$~\cite{PDG}.

All generated events were passed through the full simulation of the CMS detector based on \GEANTfour~\cite{GEANT4}.
The samples were generated with a different pileup distribution than that observed in the data.
The simulated events were therefore reweighted to match the observed pileup distribution.

%% file: Algorithms.tex
\section{Algorithms for b-jet identification}\label{sec:Algorithms}

A variety of reconstructed objects -- tracks, vertices and identified leptons -- can be used to build observables that discriminate between \cPqb\ and light-parton jets.  Several simple and robust algorithms
use just a single observable, while others combine several of these objects to achieve a higher discrimination power.  Each of these algorithms yields a single discriminator value for each jet.
The minimum thresholds on these discriminators define loose (``L''), medium (``M''), and tight (``T'') operating points with a misidentification probability for light-parton jets of close to $10 \%$, $1 \%$, and $0.1 \%$, respectively, at an average jet \pt of about 80\GeVc.
Throughout this paper, the tagging criteria will be labelled with the letter characterizing the operating point appended to the acronym of one of the algorithms described in Sections~\ref{sec:ImpactParameter} and \ref{sec:SecondaryVertex}.
The application of such a tagging criterion will be called a ``tagger''.

After a short description of the reconstructed objects used as inputs, details on the tagging algorithms are given in the following subsections, proceeding in order of increasing complexity.  Muon-based \cPqb-jet identification is mainly used as a reference method for performance measurements.
It is described in more detail in Section~\ref{sec:EfficienciesMuonJets}.

\subsection{Reconstructed objects used in \texorpdfstring{\cPqb-jet}{b-jet} identification}\label{sec:RecoObjects}

Jets are clustered from objects reconstructed by the particle-flow algorithm~\cite{PFT-09-001,PFT-10-002}.  This algorithm combines information from all subdetectors to create a consistent set of reconstructed particles for each event.  The particles are then clustered into jets using the anti-$k_T$ clustering algorithm~\cite{antikt} with a distance parameter of 0.5. The raw jet energies are corrected to obtain a uniform response in $\eta$ and an absolute calibration in \pt~\cite{Chatrchyan:2011ds}.  Although particle-flow jets are used as the default, the \cPqb-jet tagging algorithms can be applied to jets clustered from other reconstructed objects.

Each algorithm described in the next section uses the measured kinematic properties of charged particles, including identified leptons, in a jet.  The trajectories of these particles are reconstructed in the CMS tracking system in an iterative procedure using a standard Kalman filter-based method.  Details on the pattern recognition, the track-parameter estimation, and the tracking performance in proton-proton collisions can be found in Refs.~\cite{CMS:2008zzk,TRK-10-001}.

A ``global'' muon reconstruction, using information from multiple detector systems,  is achieved by first reconstructing a muon track in the muon chambers.
This is then matched to a track measured in the silicon tracker~\cite{Chatrchyan:2012xi}.
A refit is then performed using the measurements on both tracks.

Primary vertex candidates are selected by clustering reconstructed tracks based on the $z$ coordinate of their closest approach to the beam line.
An adaptive vertex fit~\cite{AVFitter} is then used to estimate the vertex position using a sample of tracks compatible with originating from the interaction region.
Among the primary vertices found in this way, the one with the highest $\sum (\pt^\mathrm{track})^2$ is selected as a candidate for the origin of the hard interaction, where the $\pt^\mathrm{track}$ are the transverse momenta of the tracks associated to the vertex.

The \cPqb-jet tagging algorithms require a sample of well-reconstructed tracks of high purity.
Specific requirements are imposed in addition to the selection applied in the tracking step.  The fraction of misreconstructed or poorly reconstructed tracks is reduced by requiring a transverse momentum of at least 1\GeVc.
At least eight hits must be associated with the track.
To ensure a good fit, $\chi^2/\mathrm{n.d.o.f.}<5$ is required, where n.d.o.f. stands for the number of degrees of freedom in the fit.
At least two hits are required in the pixel system since track measurements in the innermost layers provide most of the discriminating power.
A loose selection on the track impact parameters is used to further increase the fraction of well-reconstructed tracks and to reduce the contamination by decay products of long-lived particles, e.g. neutral kaons.
The impact parameters $d_{xy}$ and $d_z$ are defined as the transverse and longitudinal distance to the primary vertex at the point of closest approach in the transverse plane.
Their absolute values must be smaller than 0.2\cm and 17\cm, respectively.  Tracks are associated to jets in a cone $\DR < 0.5$ around the jet axis, where the jet axis is defined by the primary vertex and the direction of the jet momentum.
The distance of a track to the jet axis is defined as the distance of closest approach of the track to the axis.
In order to reject tracks from pileup this quantity is required to be less than 700\mum.
The point of closest approach must be within 5\cm of the primary vertex.
This sample of associated tracks is the basis for all algorithms that use impact parameters for discrimination.

Properties of the tracks and the average multiplicity after the selection (except for the variable plotted) are shown in Fig.~\ref{fig:TrackSelection}.
The uncertainties shown in this and all following figures are statistical unless otherwise stated.
The data were recorded with a prescaled jet trigger in the second part of 2011 when the number of pileup events was highest.
The jet \pt threshold was $60 \GeVc$.
The distributions show satisfactory agreement with the expectations from simulation.
The track multiplicity and the lower part of the momentum spectrum are particularly sensitive to the modelling of the particle multiplicity and kinematics by the Monte Carlo generator, as are other distributions such as the number of hits in the innermost pixel layers.
Detector effects that are not modelled by the simulation, such as the dynamic readout inefficiency in the pixel system, can also contribute to the remaining discrepancies.
In Fig.~\ref{fig:TrackSelection} and the following figures, simulated events with gluon splitting to \cPqb\cPaqb\ are shown as a special category.
The \cPqb\ jets in these events tend to be close in space and can be inadvertently merged by the clustering algorithm, resulting in a higher average track multiplicity per jet.

\begin{figure}[htb]
\begin{center}
\includegraphics[angle=0,width=.49\textwidth]{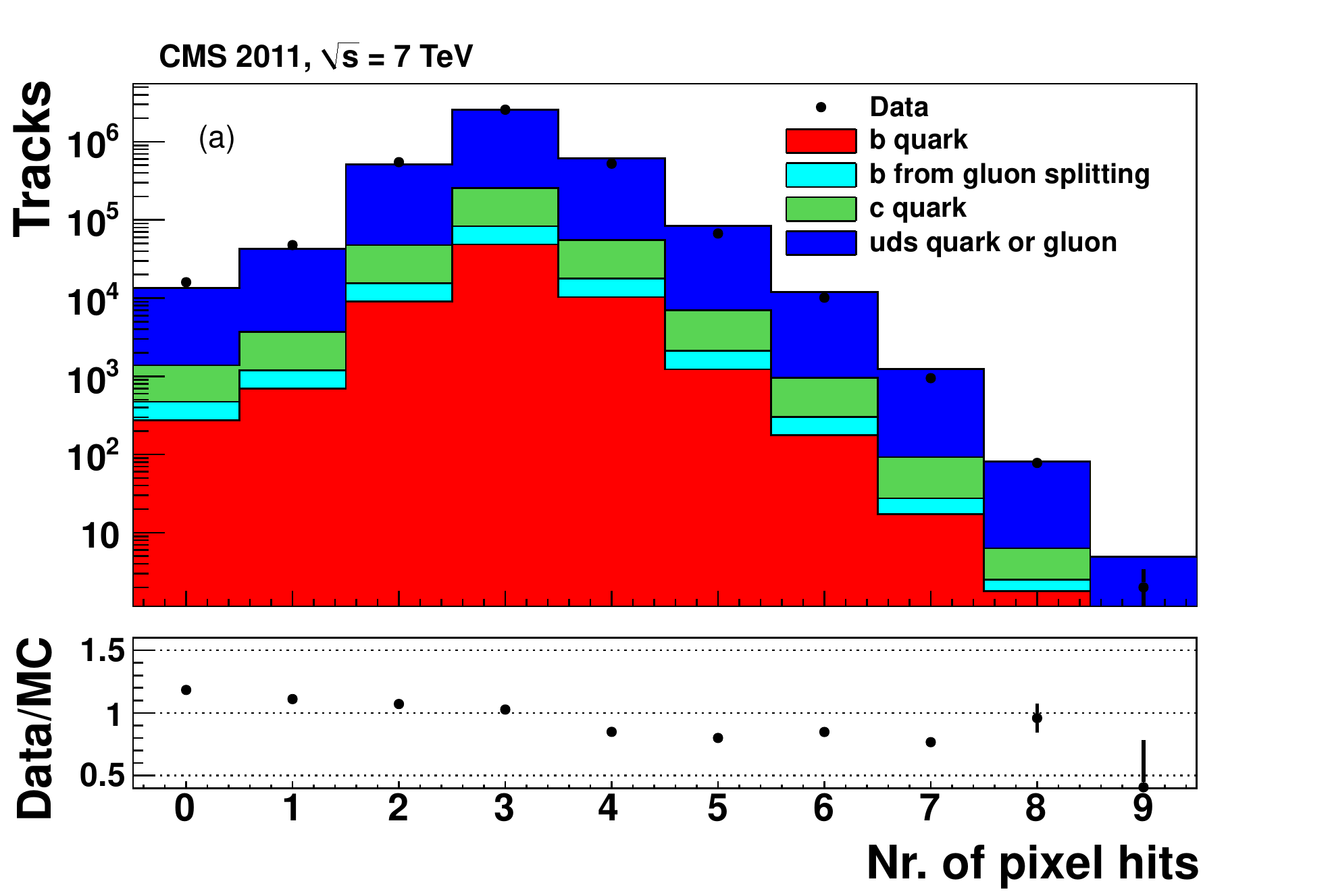} \hfil
\includegraphics[angle=0,width=.49\textwidth]{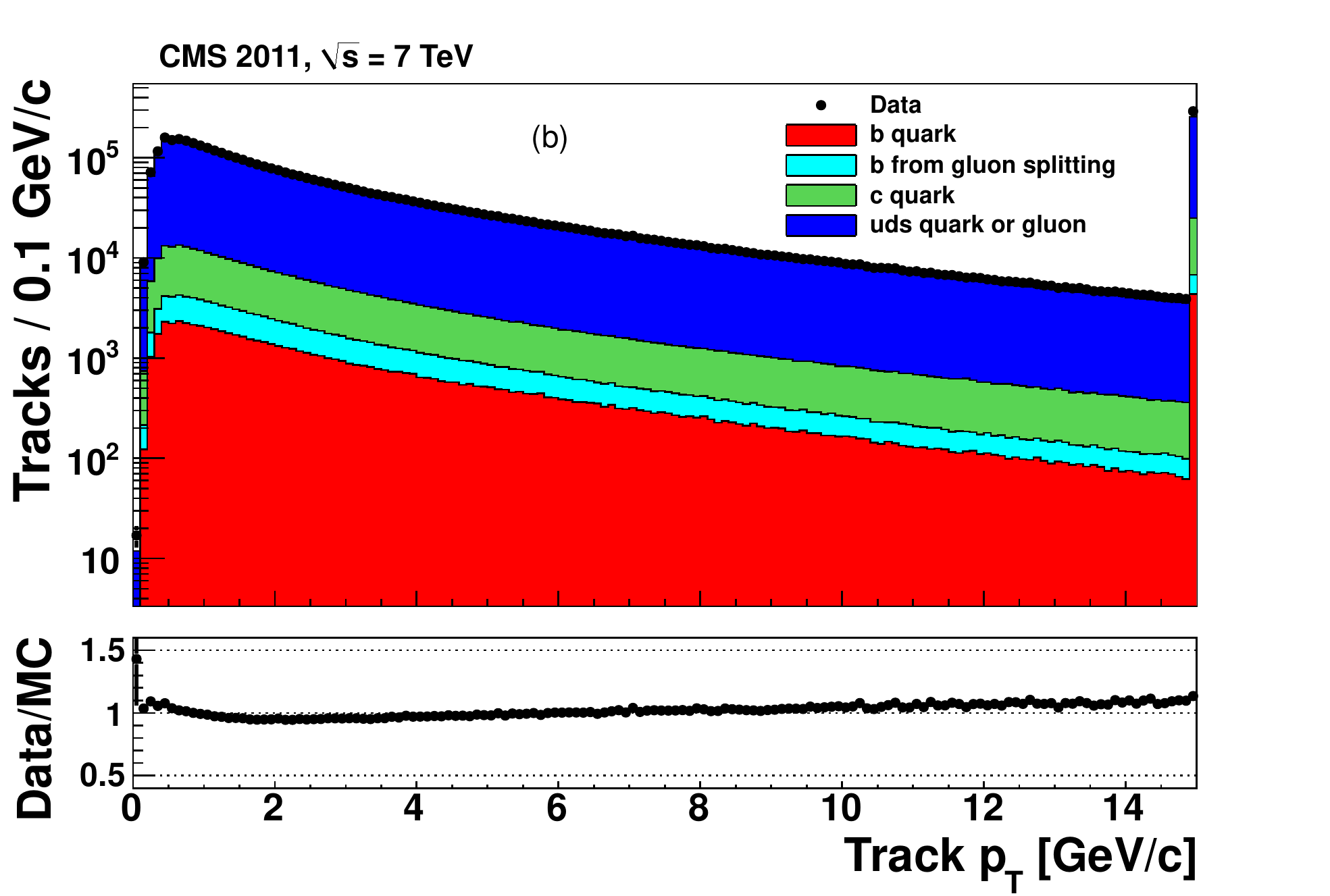} \\
\includegraphics[angle=0,width=.49\textwidth]{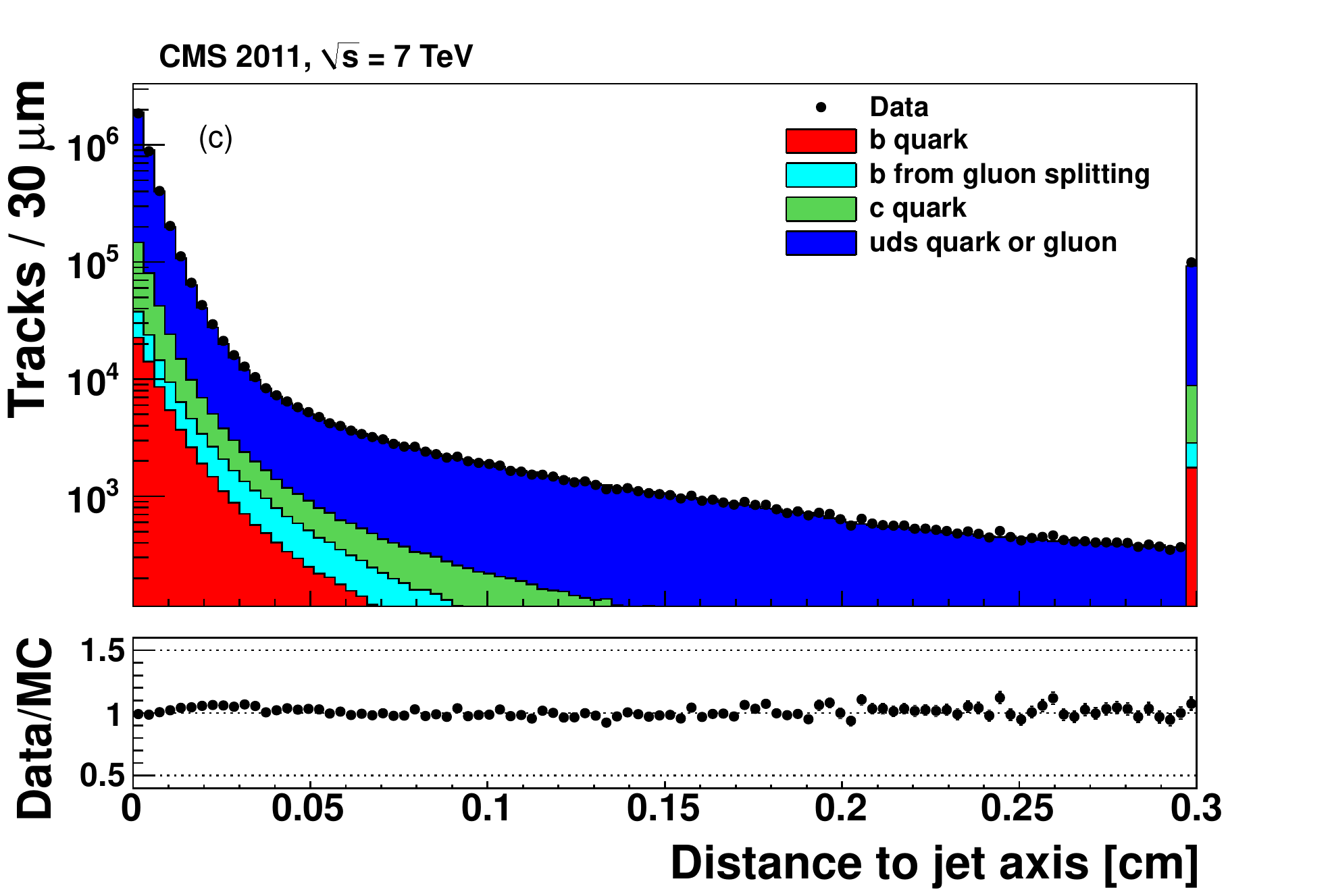} \hfil
\includegraphics[angle=0,width=.49\textwidth]{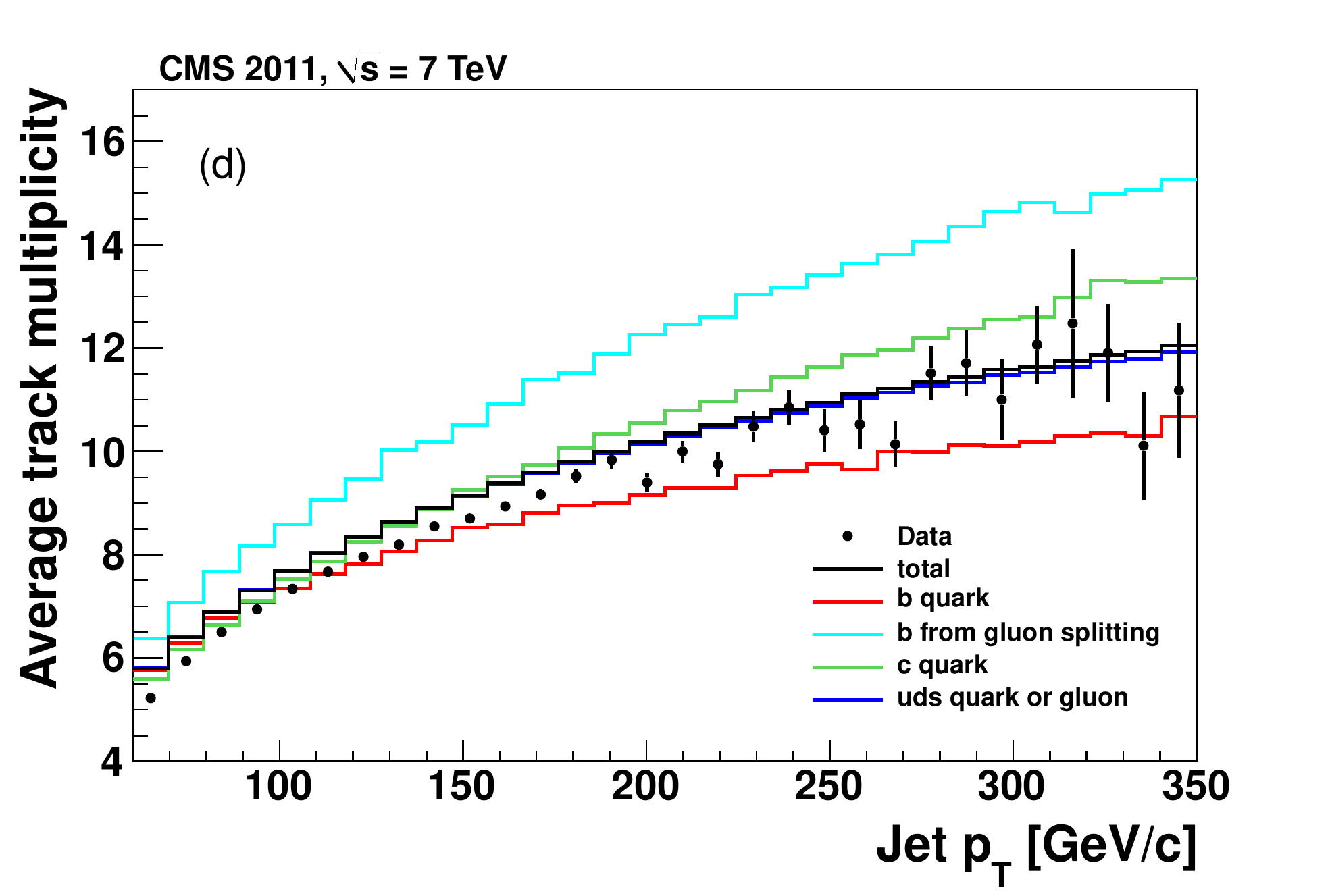}
\caption{Track properties after basic selection (except for the variable plotted):
(a) number of hits in the pixel system,
(b) transverse momentum,
(c) distance to the jet axis.
The average number of tracks passing the basic selection is shown in (d) as a function of the transverse momentum of the jet.
In (a)--(c) the distributions from simulation have been normalized to match the counts in data.
The filled circles correspond to data.
The stacked, coloured histograms indicate the contributions of different components from simulated multijet (``QCD'') samples.
Simulated events involving gluon splitting to \cPqb\ quarks (``\cPqb\ from gluon splitting'') are indicated separately from the other \cPqb\ production processes (``\cPqb\ quark'').
In each histogram, the rightmost bin includes all events from the overflow.
The sample corresponds to a trigger selection with jet $\pt > 60 \GeVc$.
 }\label{fig:TrackSelection}
\end{center}
\end{figure}

The combinatorial complexity of the reconstruction of the decay points (secondary vertices) of \cPqb\ or \cPqc\ hadrons is more challenging in the presence of multiple proton-proton interactions.
In order to minimize this complexity a different track selection is applied.
Tracks must be within a cone of $\DR = 0.3$ around the jet axis with a maximal distance to this axis of 0.2\cm and pass a ``high-purity'' criterion~\cite{TRK-11-001}. The ``high-purity'' criterion uses the normalized $\chi^2$ of the track fit, the track length, and impact parameter information to optimize the purity for each of the iterations in track reconstruction.
The vertex finding procedure begins with tracks defined by this selection and proceeds iteratively.
A vertex candidate is identified by applying an adaptive vertex fit~\cite{AVFitter}, which is robust in the presence of outliers.
The fit estimates the vertex position and assigns a weight between 0 and 1 to each track based on its compatibility with the vertex.  All tracks with weights $> 0.5$ are then removed from the sample.
The fit procedure is repeated until no new vertex candidate can be found.  In the first iteration the interaction region is used as a constraint in order to identify the prompt tracks in the jet. The subsequent iterations produce decay vertex candidates.

\subsection{Identification using track impact parameters}\label{sec:ImpactParameter}

\begin{figure}[htb]
\centering
\includegraphics[width=0.49\textwidth]{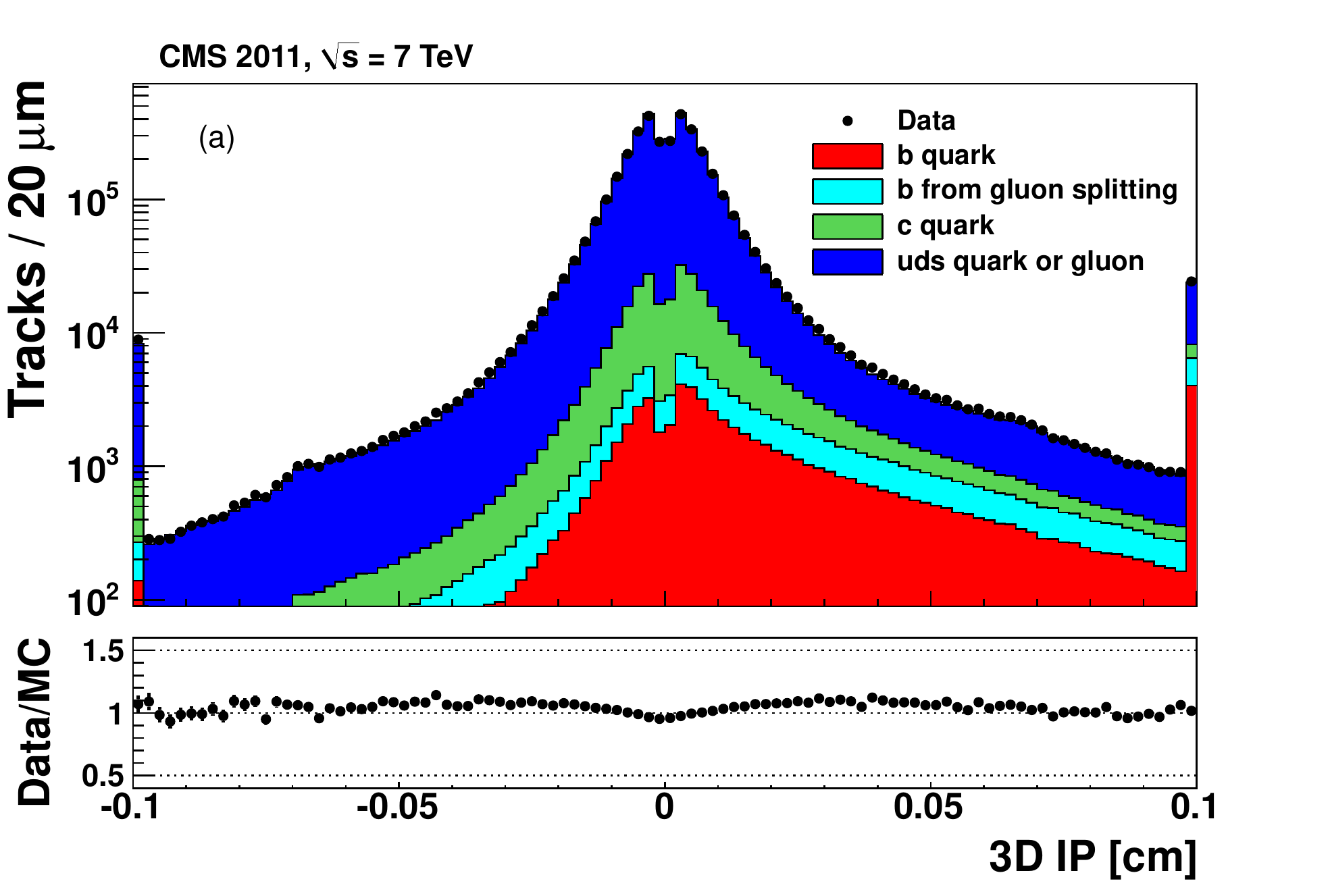} \hfil
\includegraphics[width=0.49\textwidth]{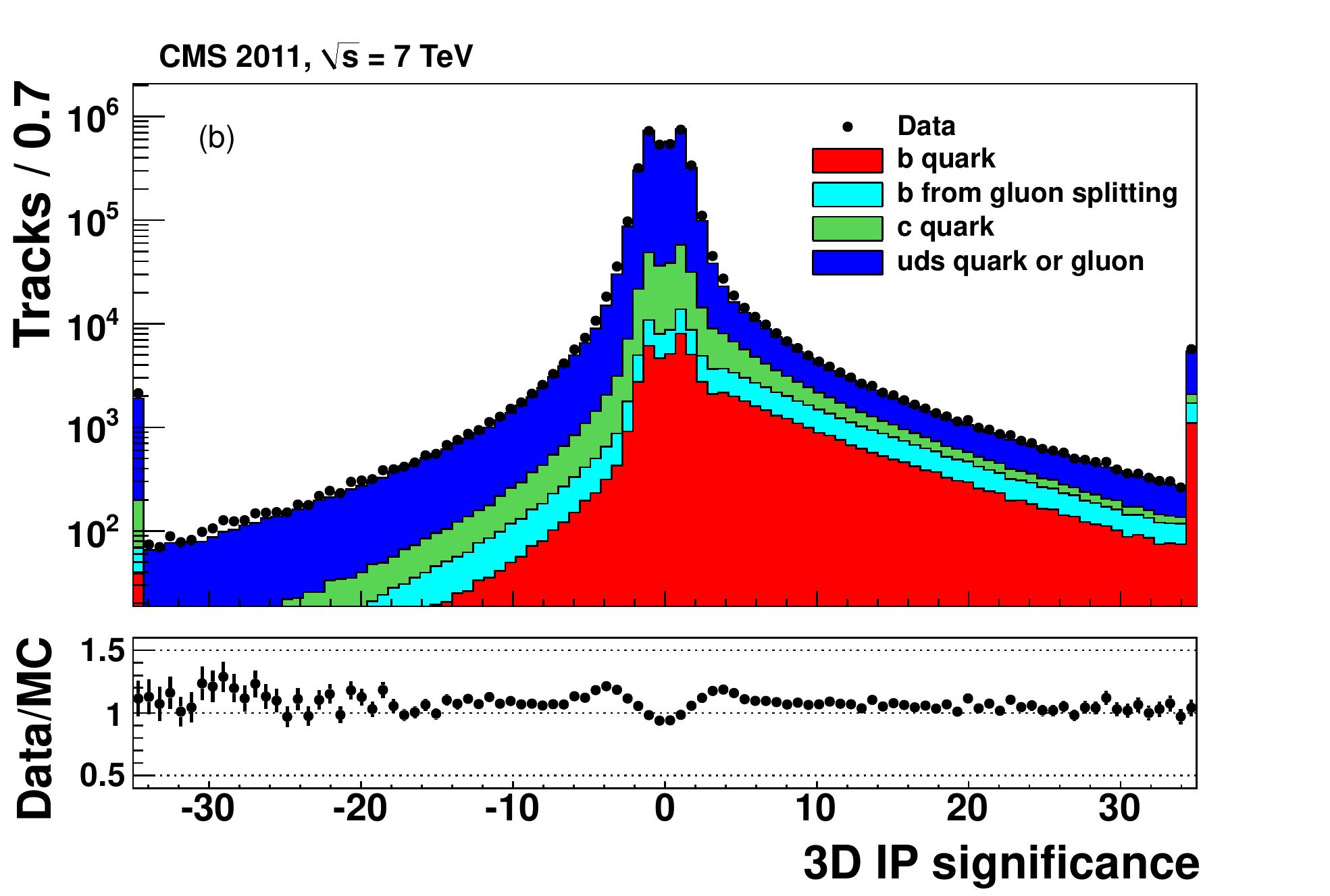}
\caption{Distributions of (a) the 3D impact parameter and (b) the significance of the 3D impact parameter for all selected tracks.
Selection and symbols are the same as in Fig.~\ref{fig:TrackSelection}.
Underflow and overflow are added to the first and last bins, respectively.
}

\label{fig:IPallTracks}
\end{figure}

The impact parameter of a track with respect to the primary vertex can be used to distinguish the decay products of a \cPqb\ hadron from prompt tracks.  The IP is calculated in three dimensions by taking advantage of the excellent resolution of the pixel detector along the $z$ axis.  The impact parameter has the same sign as the scalar product of the vector pointing from the primary vertex to the point of closest approach with the jet direction.
Tracks originating from the decay of particles travelling along the jet axis will tend to have positive IP values.
In contrast, the impact parameters of prompt tracks can have positive or negative IP values.  The resolution of the impact parameter depends strongly on the \pt and $\eta$ of a track.  The impact parameter significance $S_\mathrm{IP}$, defined as the ratio of the IP to its estimated uncertainty, is used as an observable.  The distributions of IP values and their significance are shown in Fig.~\ref{fig:IPallTracks}. In general, good agreement with simulation is observed with the exception of a small difference in the width of the core of the IP significance distribution.

\begin{figure}[htb]
\centering
\includegraphics[width=0.49\textwidth]{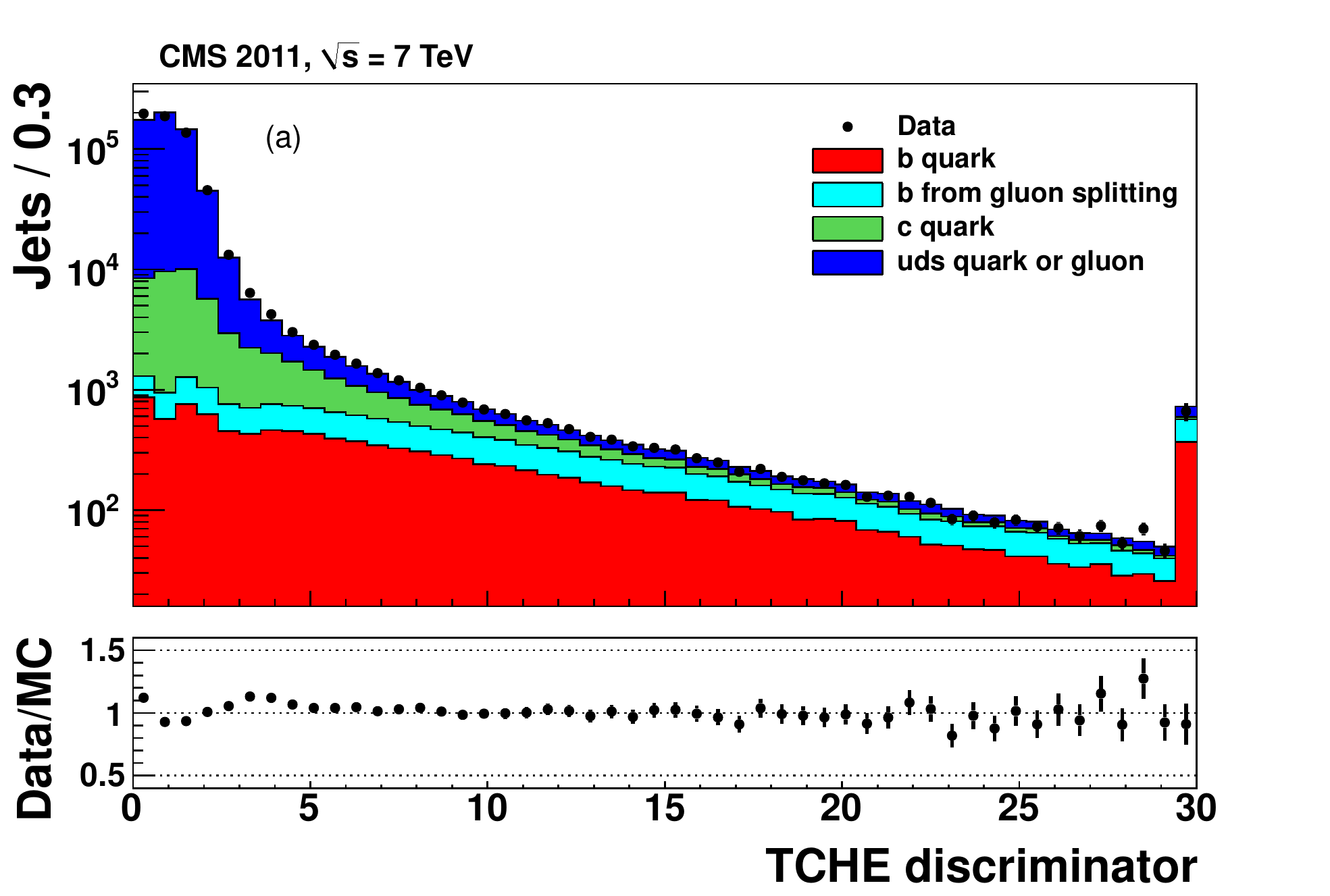} \hfil
\includegraphics[width=0.49\textwidth]{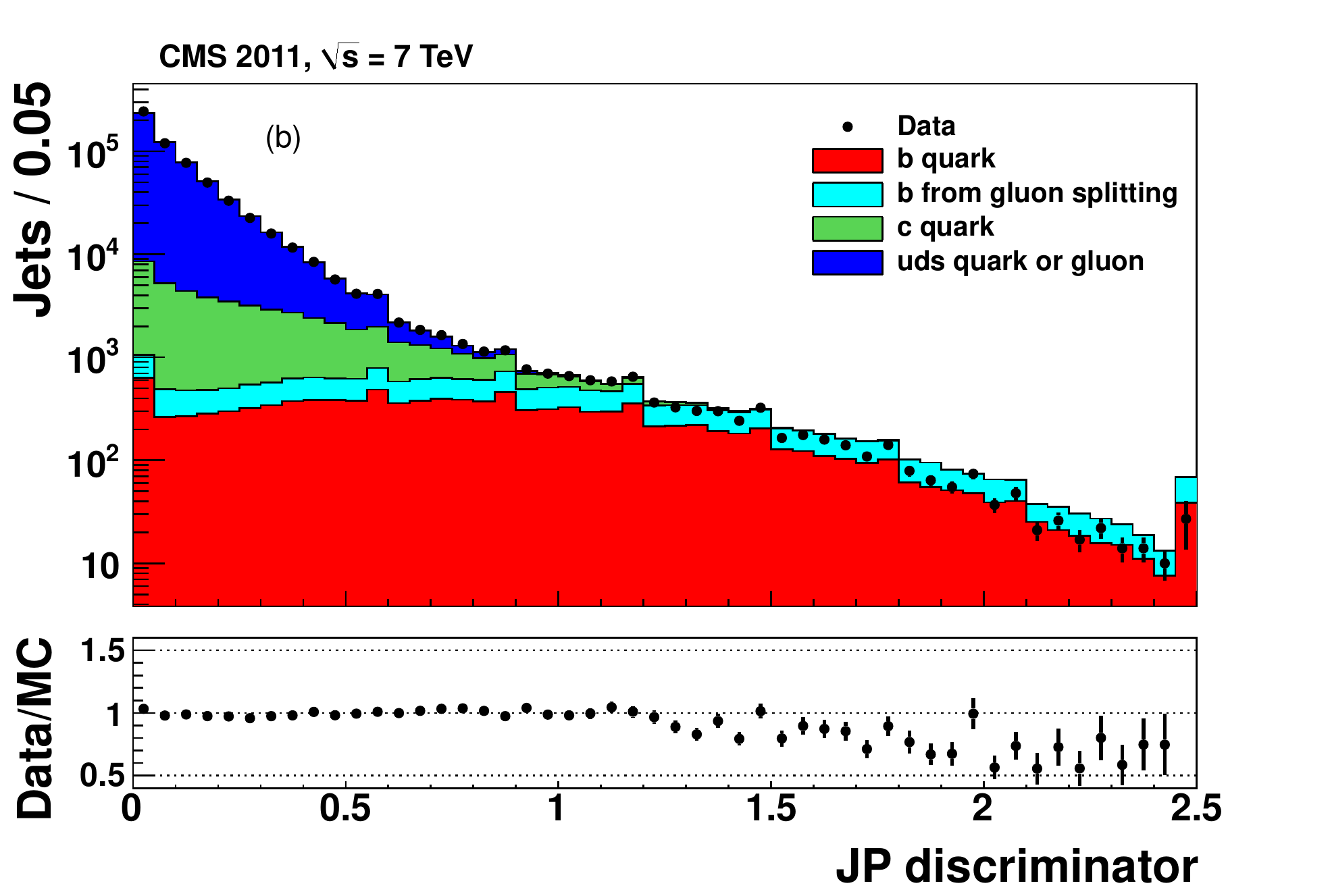}
\caption{Discriminator values for (a) the TCHE and (b) the JP algorithms.
Selection and symbols are the same as in Fig.~\ref{fig:TrackSelection}.
The small discontinuities in the JP distributions are due to the single track probabilities which are required to be greater than 0.5\%.
}
\label{fig:tcjpDisc}
\end{figure}

By itself the impact parameter significance has discriminating power between the decay products of \cPqb\ and non-\cPqb\ jets.  The {\it Track Counting} (TC) algorithm sorts tracks in a jet by decreasing values of the IP significance.  Although the ranking tends to bias the values for the first track to high positive IP significances, the probability to have several tracks with high positive values is low for light-parton jets.  Therefore the two different versions of the algorithm use the IP significance of the second and third ranked track  as the discriminator value.
These two versions of the algorithm are called {\it Track Counting High Efficiency} (TCHE) and {\it Track Counting High Purity} (TCHP), respectively.
The distribution of the TCHE discriminator is shown in Fig.~\ref{fig:tcjpDisc} (a).

A natural extension of the TC algorithms is the combination of the IP information of several tracks in a jet.  Two discriminators are computed from additional algorithms.
The {\it Jet Probability} (JP) algorithm uses an estimate of the likelihood that all tracks associated to the jet come from the primary vertex.
The {\it Jet B Probability} (JBP) algorithm gives more weight to the tracks with the highest IP significance, up to a maximum of four such tracks, which matches the average number of reconstructed charged particles from \cPqb-hadron decays.  The estimate for the likelihood, $P_\mathrm{jet}$, is defined as \begin{eqnarray}
P_\mathrm{jet}=\Pi \cdot \sum^{N-1}_{i=0}\frac{(-\ln\Pi)^i}{i!} \quad \mathrm{with} \quad \Pi = \prod^{N}_{i=1}\mathrm{max}(P_i , 0.005) \; , \label{eq:jetProb}
\end{eqnarray}
where $N$ is the number of tracks under consideration and $P_i$ is the estimated probability for track $i$ to come from the primary vertex~\cite{Buskulic:1993,Borisov:1996}.
The $P_i$ are based on the probability density functions for the IP significance of prompt tracks.
These functions are extracted from data for different track quality classes, using the shape of the negative part of the $S_\mathrm{IP}$ distribution.
Eight quality classes are defined for tracks with $\chi^2$/n.d.o.f $< 2.5$, depending on the momentum ($< 8$ or $> 8\GeVc$) and pseudorapidity ($|\eta|$ within 0-0.8, 0.-1.6, 1.6-2.4 if there are at least three pixel hits or $|\eta| < 2.4$ if there are only two pixel hits). A ninth quality class is defined for tracks with $\chi^2$/n.d.o.f $> 2.5$.
The cut-off parameter for $P_i$ at 0.5\% limits the effect of single, poorly reconstructed tracks on the global estimate.
The discriminators for the jet probability algorithms have been constructed to be proportional to $-\ln P_\mathrm{jet}$.
The distribution of the JP discriminator in data and simulation is shown in Fig.~\ref{fig:tcjpDisc} (b).

\subsection{Identification using secondary vertices}\label{sec:SecondaryVertex}

The presence of a secondary vertex and the kinematic variables associated with this vertex can be used to discriminate between $\cPqb$ and non-$\cPqb$ jets.
Two of these variables are the flight distance and direction, using the vector between primary and secondary vertices.
The other variables are related to various properties of the system of associated secondary tracks such as the multiplicity, the mass (assuming the pion mass for all secondary tracks), or the energy.
Secondary-vertex candidates must meet the following requirements to enhance the \cPqb\ purity:
\begin{itemize}
\item secondary vertices must share less than 65\% of their associated tracks with the primary vertex and the significance of the radial distance between the two vertices has to exceed  $3 \sigma$;
\item secondary vertex candidates with a radial distance of more than 2.5\cm with respect to the primary vertex, with masses compatible with the mass of \PKz\ or exceeding $6.5 \GeVcc$ are rejected, reducing the contamination by vertices corresponding to the interactions of particles with the detector material and by decays of long-lived mesons;
\item the flight direction of each candidate has to be within a cone of $\DR < 0.5$ around the jet direction.
\end{itemize}

\begin{figure}
\centering
\includegraphics[width=0.49\textwidth]{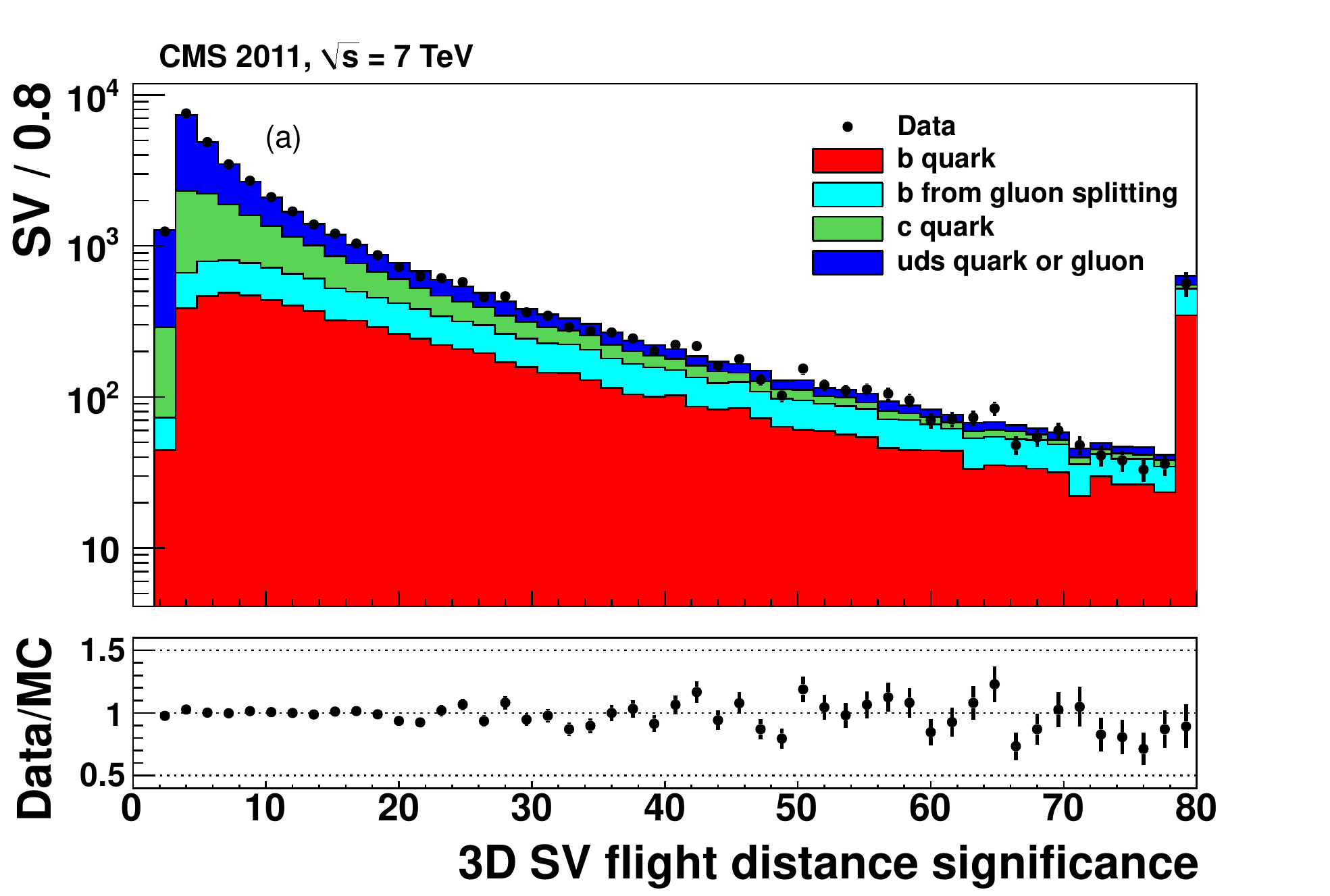} \hfil
\includegraphics[width=0.49\textwidth]{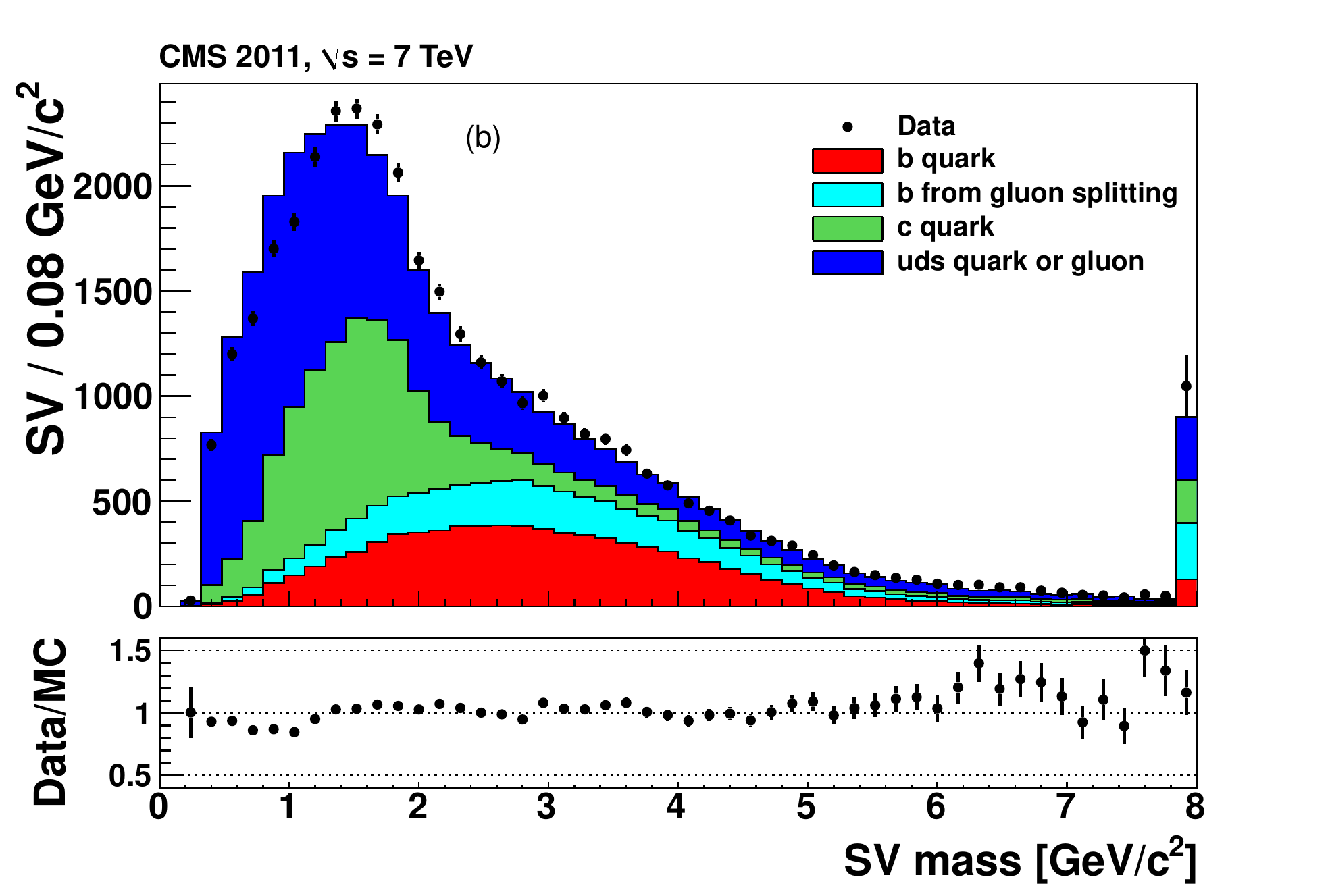}
\caption{Properties of reconstructed decay vertices:  (a) the significance of the 3D secondary vertex (3D SV) flight distance and (b) the mass associated with the secondary vertex.
Selection and symbols are the same as in Fig.~\ref{fig:TrackSelection}.
}
\label{fig:ssvQuantities}
\end{figure}

The {\it Simple Secondary Vertex} (SSV) algorithms use the significance of the flight distance (the ratio of the flight distance to its estimated uncertainty) as the discriminating variable.
The algorithms' efficiencies are limited by the secondary vertex reconstruction efficiency to about 65\%.
Similar to the {\it Track Counting} algorithms, there exist two versions optimized for different purity: the {\it High Efficiency} (SSVHE) version uses vertices with at least two associated tracks, while for the {\it High Purity} (SSVHP) version at least three tracks are required. In Fig.~\ref{fig:ssvQuantities} the flight distance significance and the mass associated with the secondary vertex are shown.

A more complex approach involves the use of secondary vertices, together with track-based lifetime information.  By using these additional variables, the {\it Combined Secondary Vertex} (CSV) algorithm provides discrimination also in cases when no secondary vertices are found, increasing the maximum efficiency with respect to the SSV algorithms.  In many cases, tracks with an $S_\mathrm{IP}> 2$ can be combined in a ``pseudo vertex'', allowing for the computation of a subset of secondary-vertex-based quantities even without an actual vertex fit.  When even this is not possible, a ``no vertex'' category reverts to track-based variables that are combined in a way similar to that of the JP algorithm.

The following set of variables with high discriminating power and low correlations is used (in the ``no vertex'' category only the last two variables are available):
\begin{itemize}
\item the vertex category (real, ``pseudo,'' or ``no vertex'');
\item the flight distance significance in the transverse plane (``2D'');
\item the vertex mass;
\item the number of tracks at the vertex;
\item the ratio of the energy carried by tracks at the vertex with respect to all tracks in the jet;
\item the pseudorapidities of the tracks at the vertex with respect to the jet axis;
\item the 2D IP significance of the first track that raises the invariant mass above the charm threshold of $1.5 \GeVcc$ (tracks are ordered by decreasing IP significance and the mass of the system is recalculated after adding each track);
\item the number of tracks in the jet;
\item the 3D IP significances for each track in the jet.
\end{itemize}

\begin{figure}
\centering
\includegraphics[width=0.49\textwidth]{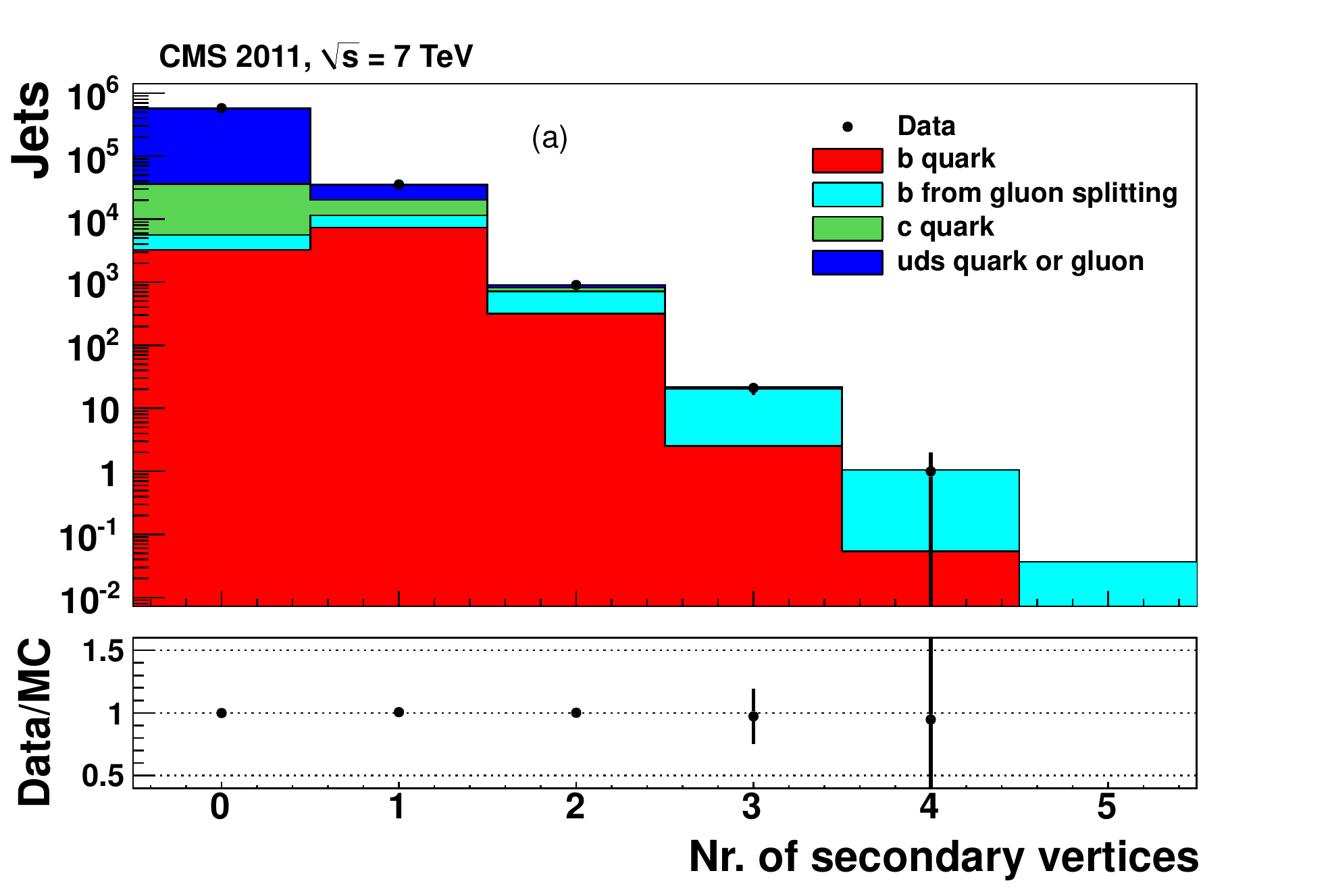} \hfil
\includegraphics[width=0.49\textwidth]{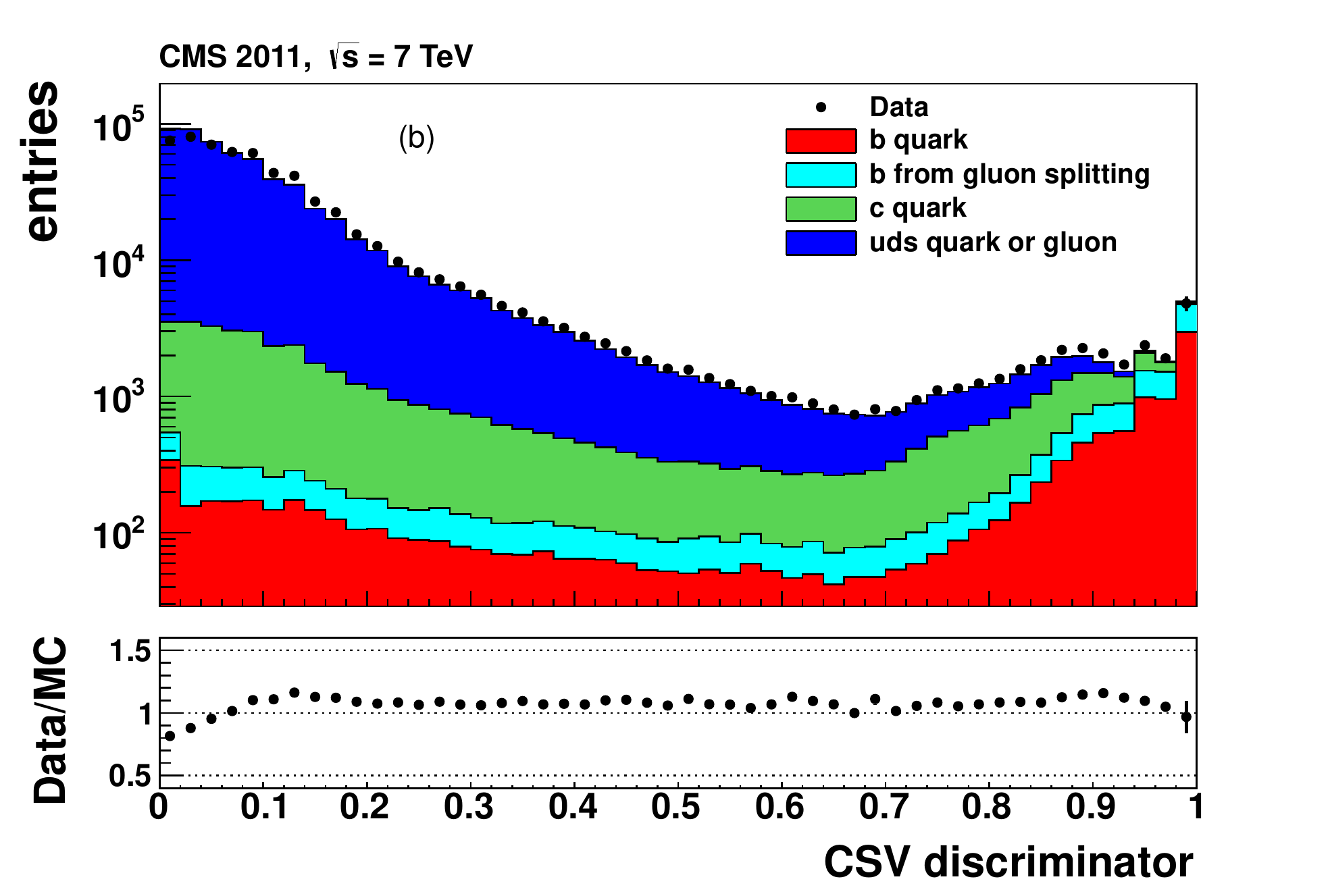}
\caption{Distributions of (a) the secondary vertex multiplicity and (b) the CSV discriminator.
Selection and symbols are the same as in Fig.~\ref{fig:TrackSelection}.
 }\label{fig:csvQuantities}
\end{figure}

Two likelihood ratios are built from these variables.
They are used to discriminate between \cPqb\  and \cPqc\ jets and between \cPqb\ and light-parton jets.
They are combined with prior weights of $0.25$ and $0.75$,  respectively.
The distributions of the vertex multiplicity and of the CSV discriminator are presented in Fig.~\ref{fig:csvQuantities}.

\subsection{Performance of the algorithms in simulation}\label{sec:MCPerformance}

\begin{figure}
\centering
\subfigure{\includegraphics[width=0.49\textwidth]{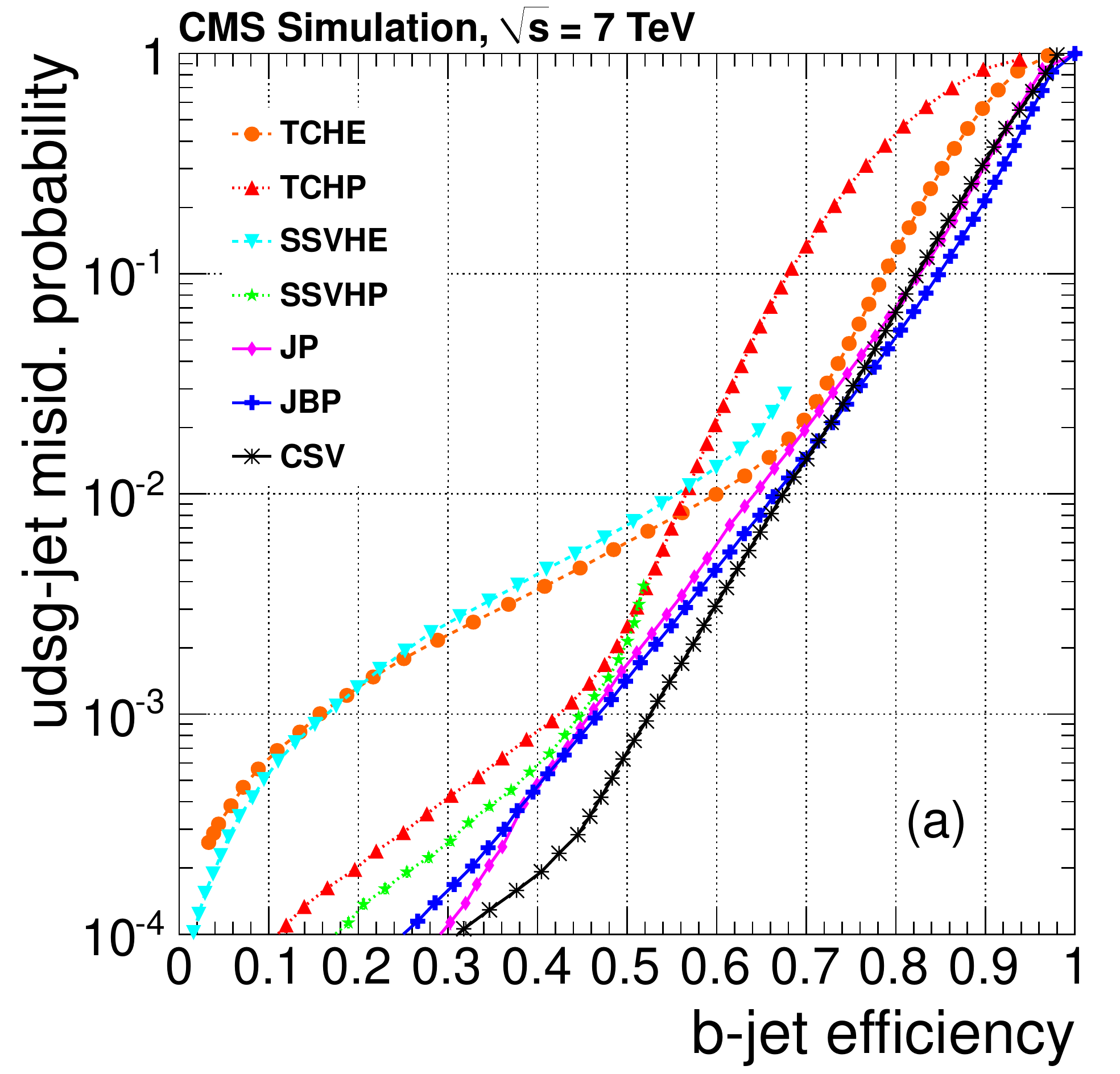}}
\subfigure{\includegraphics[width=0.49\textwidth]{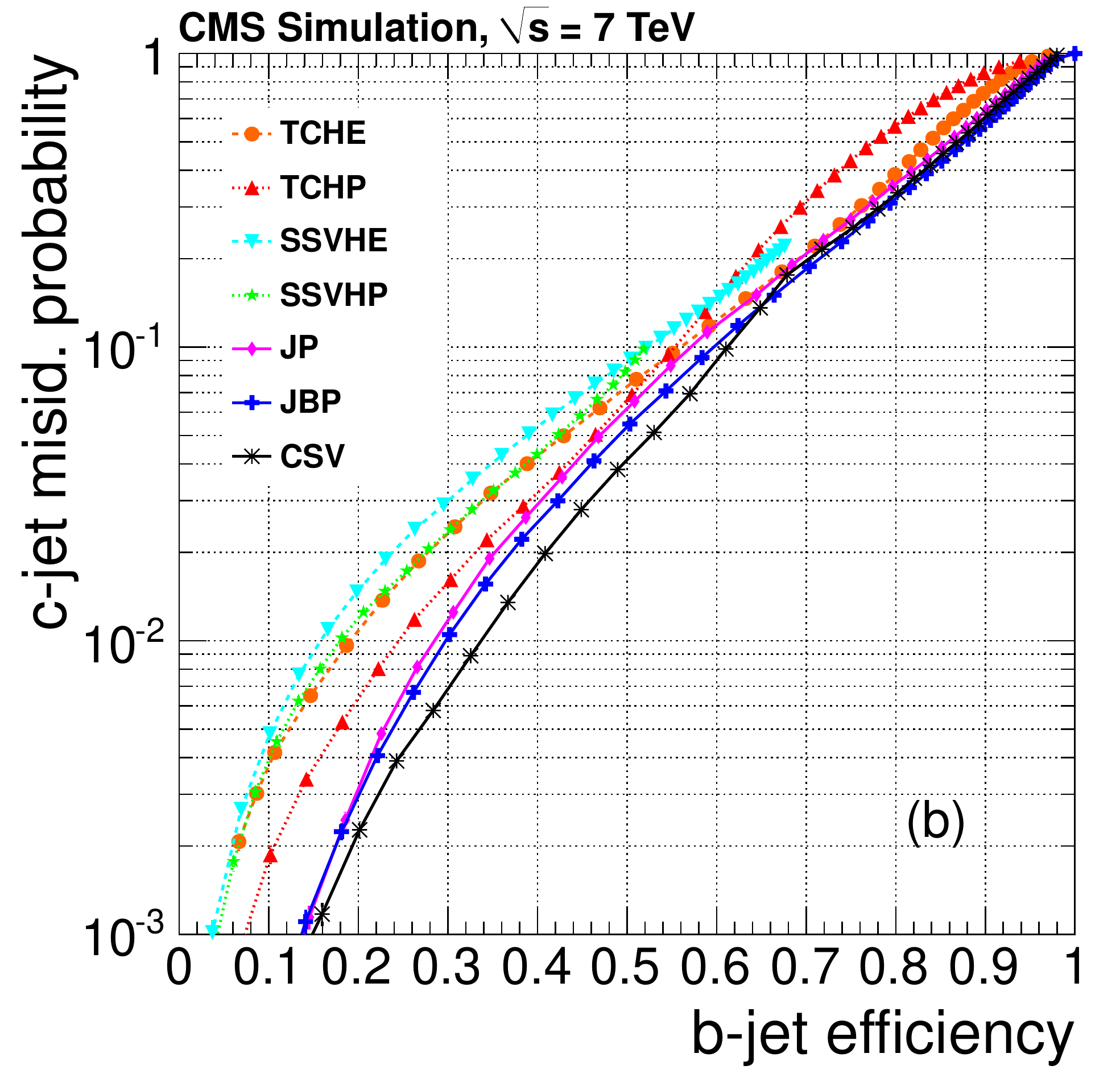}}
\caption{
Performance curves obtained from simulation for the algorithms described in the text.
(a) light-parton- and (b) \cPqc-jet misidentification probabilities as a function of the \cPqb-jet efficiency.  Jets with $\pt > 60\GeVc$ in a sample of simulated multijet events are used to obtain the efficiency and misidentification probability values.
}
\label{fig:mcPerformance}
\end{figure}

The performance of the algorithms described above is summarized in Fig.~\ref{fig:mcPerformance} where the predictions of the simulation for the misidentification probabilities (the efficiencies to tag non-\cPqb\ jets) are shown as a function of the \cPqb-jet efficiencies.
Jets with $\pt > 60\GeVc$ in a sample of simulated multijet events are used to obtain the efficiencies and misidentification probabilities.
For loose selections with 10\% misidentification probability for light-parton jets a \cPqb-jet tagging efficiency of $\sim 80 \text{--} 85$\% is achieved.
In this region the JBP has the highest \cPqb-jet tagging efficiency.
For tight selections with misidentification probabilities of 0.1\%, the typical \cPqb-jet tagging efficiency values are $\sim 45 \text{--} 55\%$.
For medium and tight selections the CSV algorithm shows the best performance.
As can be seen in Fig.~\ref{fig:mcPerformance}, the TC and SSV algorithms cannot be tuned to provide good performance for the whole range of operating points.
Therefore two versions of these algorithms are provided, with the ``high efficiency'' version to be used for loose to medium operating points and the ``high purity'' version for tighter selections.
Because of the non-negligible lifetime of \cPqc\ hadrons the separation of \cPqc\ from \cPqb\ jets is naturally more challenging.
Due to the explicit tuning of the CSV algorithm for light-parton- and \cPqc-jet rejection it provides the best \cPqc-jet rejection values in the high-purity region.

\begin{sidewaysfigure}[p]
\centering 
\includegraphics[width=0.49\textwidth]{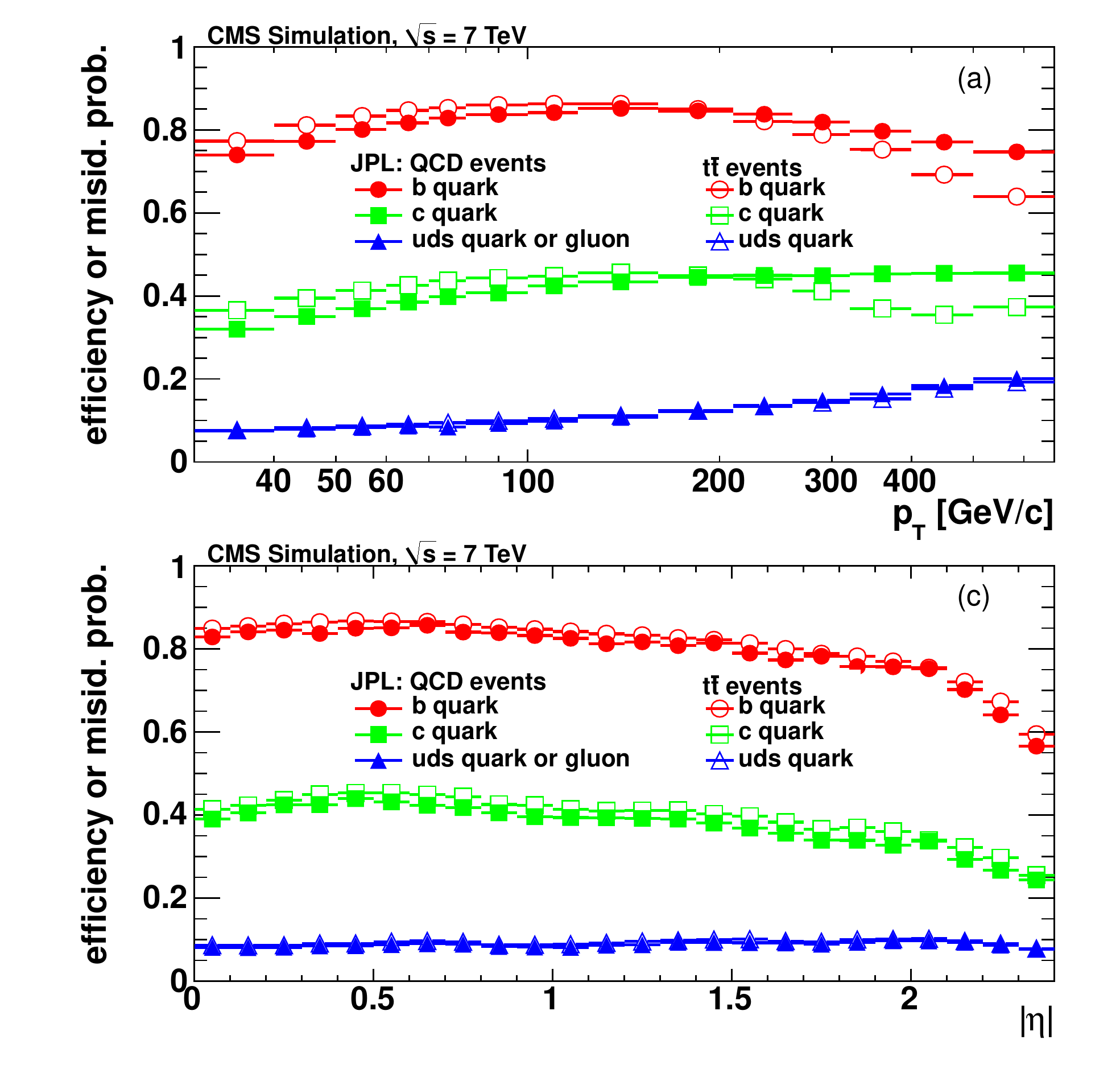}  \hfil
\includegraphics[width=0.49\textwidth]{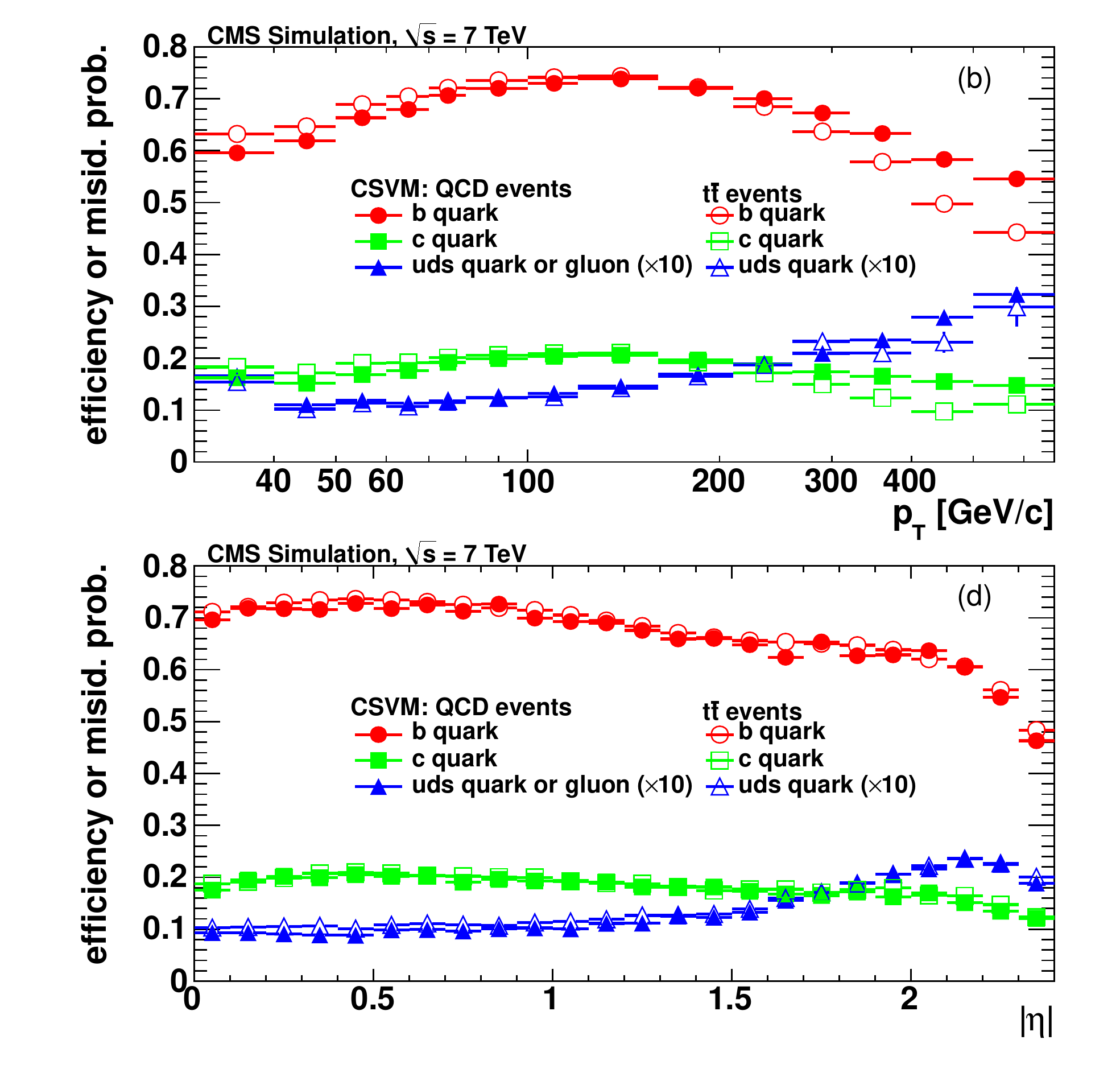}
\caption{
  Efficiency for \cPqb-jets and misidentification probabilities for \cPqc\ and light-parton jets of the (a, c) JPL and (b, d) CSVM taggers as a function of (a, b) jet \pt\ and (c, d) jet pseudorapidity in QCD multijet events (filled symbols) and \ttbar\ events (open symbols).  
  A trigger threshold of $\pt > 60\GeVc$ is applied to the leading jet in the QCD events. 
  Jets with $\pt > 30\GeV/c$ and $|\eta|<2.4$ are used in both samples. In (a) and (b), the rightmost bins includes all jets with $\pt > 500\GeVc$.  
  For the CSVM tagger, the misidentification probability for light partons is scaled up by a factor of ten.
}
\label{fig:MCeffi_JPL_CSVM}
\end{sidewaysfigure}

Figure~\ref{fig:MCeffi_JPL_CSVM} presents the efficiencies and misidentification probabilities as a function of jet \pt\ and pseudorapidity for the JPL and CSVM taggers. 
Two simulated samples are used: a QCD multijet sample with a jet \pt\ trigger threshold of 60\GeVc\ applied to the leading jet, and a \ttbar\ sample.  
Jets with $\pt > 30\GeVc$ and $|\eta|<2.4$ are considered in both cases.  
The \cPqb-jet identification efficiency is slightly larger in \ttbar\ events at small jet \pt ($< 100 \GeV/c$) due to the presence of more central jets.
At large jet \pt\ ($> 200\GeVc$), the presence of \cPqb\ and \cPqc\ jets from gluon splitting explains the apparent higher identification efficiency in the QCD multijet sample. 
The \cPqb-jet efficiency and the \cPqc-jet misidentification probability rise with jet \pt\ for values below 100\GeVc\ and decrease above 200\GeVc. 
This dependence is due to a convolution of the track impact parameter resolution (which is larger at low \pt), of the heavy-hadron decay lengths (which scale with jet \pt) and of the track-selection criteria. 
The misidentification probability for light-parton jets rises continuously with jet \pt\ due to the logarithmic increase of the number of particles in jets and the higher fraction of merged hits in the innermost layers of the tracking system.  
However, both the identification efficiencies and misidentification probabilities stay roughly constant over most of the pixel detector acceptance.

\subsection{Impact of running conditions on b-jet identification}\label{sec:RunCond}

All tagging algorithms rely on a high track identification efficiency and a reliable estimation of the track parameters and their uncertainties.
These are both potentially sensitive to changes in the running conditions of the experiment.
The robustness of the algorithms with respect to the misalignment of the tracking system and an increase in the density of tracks due to pile up, which are the most important of the changes in conditions, has been studied.

The alignment of the CMS tracker is performed using a mixture of tracks from cosmic rays and minimum bias collisions~\cite{TkAl_VBlobel, TkAl_Millepede},
and is regularly monitored.
During the 2011 data taking, the most significant movements were between the two halves of the pixel barrel detector, where discrete changes in the relative $z$ position of up to 30\mum were observed.
The sensitivity of \cPqb-jet identification to misalignment was studied on simulated ${\rm t\bar{t}}$ samples.
With the current estimated accuracy of the positions of the active elements, no significant deterioration is observed with respect to a perfectly aligned detector.
The effect of displacements between the two parts of the pixel barrel detector was studied by introducing artificial separations of 40, 80, 120, and 160\mum in the detector simulation.
The movements observed in 2011 were not found to cause any significant degradation of the performance.

\begin{figure} \centering
\subfigure{\includegraphics[width=0.49\textwidth]{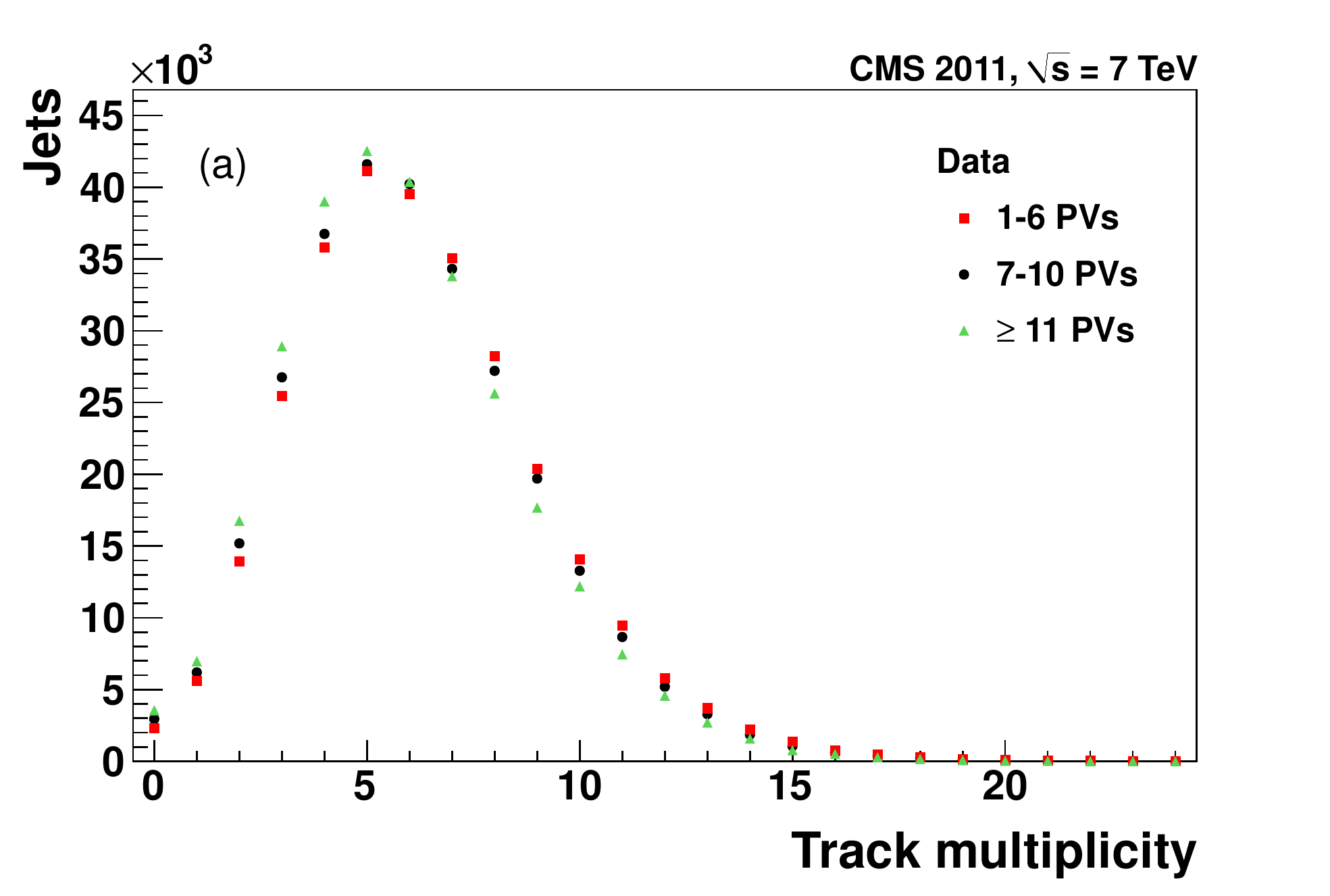} }
\subfigure{\includegraphics[width=0.49\textwidth]{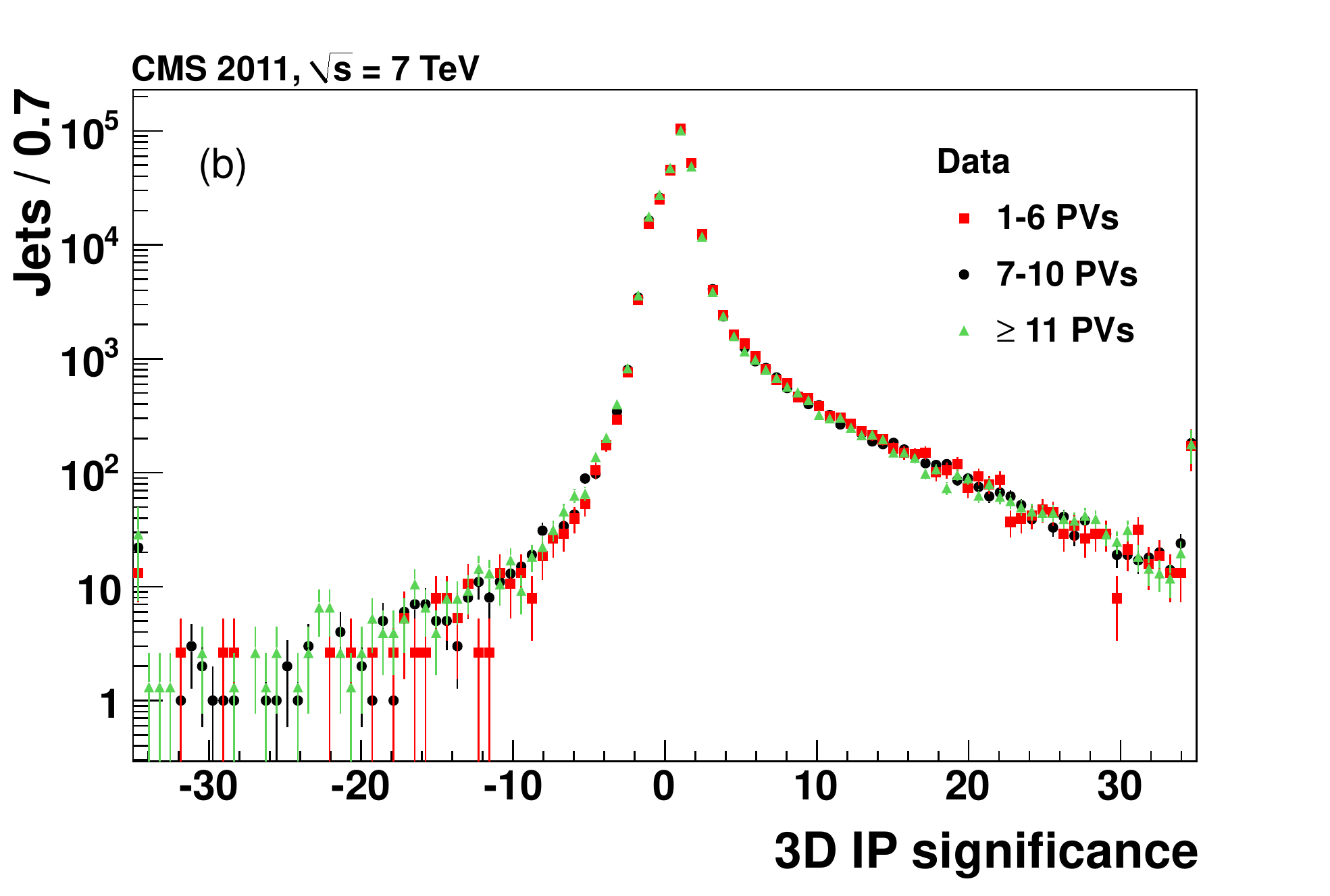} }
\caption{(a) the number of tracks associated with the selected jets for three ranges of primary vertex (PV) multiplicity. (b) the IP significance of the second-most significant track, for the three ranges of primary vertex multiplicity.
The selection is the same as in Fig.~\ref{fig:TrackSelection}.
The distributions are normalized to the event count for 1--6 PV range.
Underflow and overflow entries are added to the first and last bins, respectively.}
\label{fig:pile-upObservables}
\end{figure}

\begin{figure}
\centering
\subfigure{\includegraphics[width=0.49\textwidth]{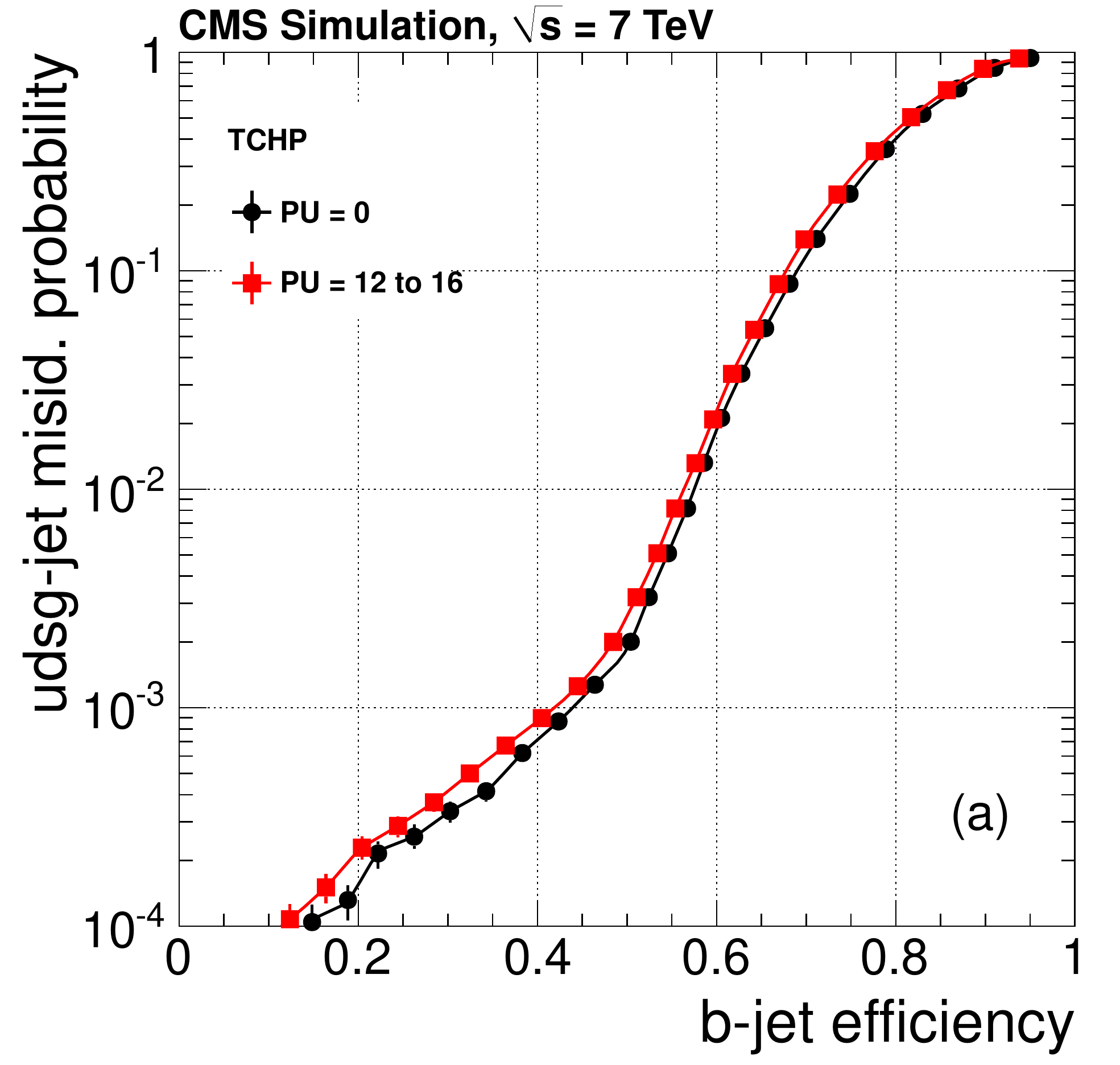}}
\subfigure{\includegraphics[width=0.49\textwidth]{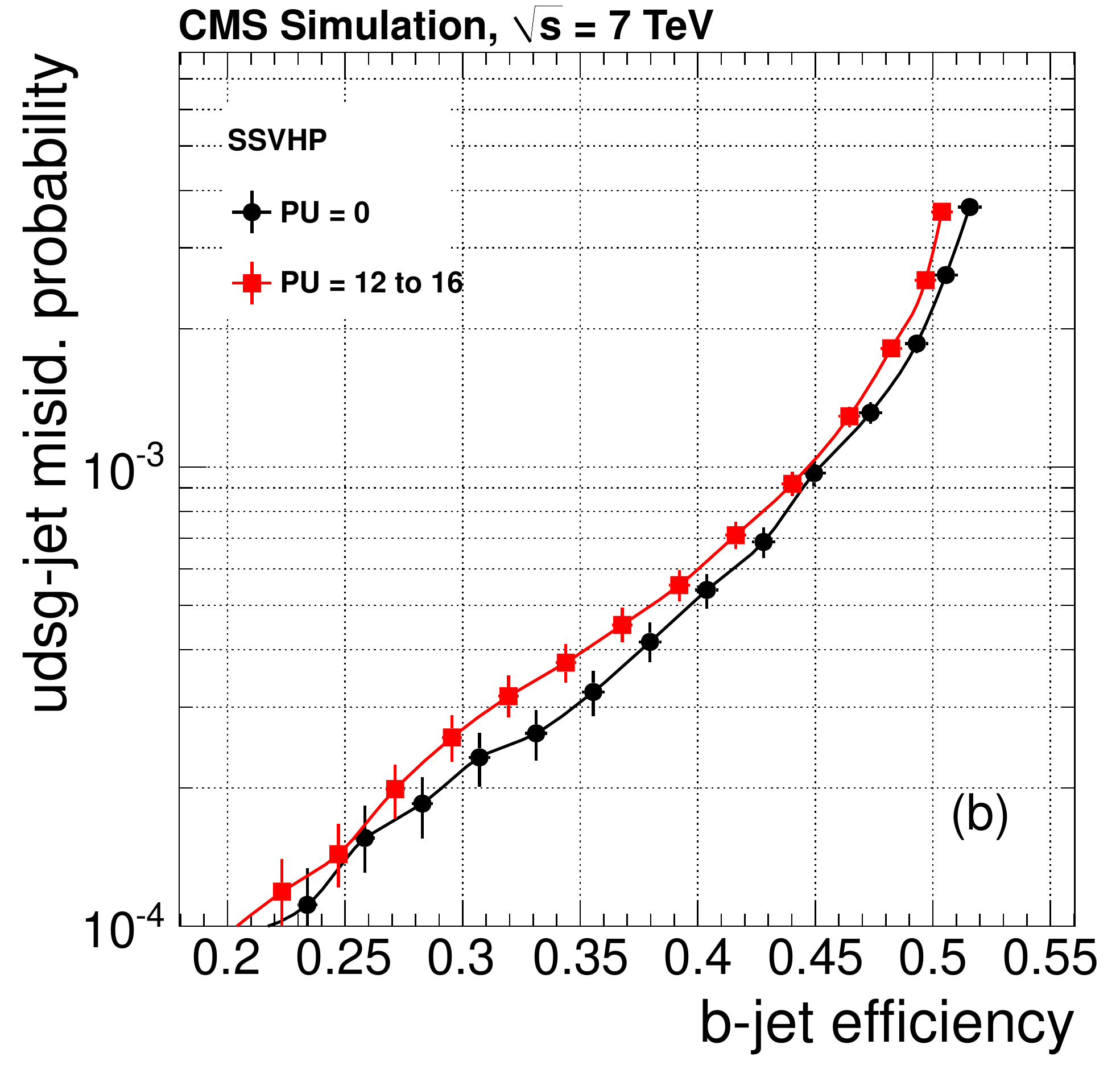}}
\caption{Light-parton misidentification probability versus \cPqb-jet tagging efficiency for jets with $\pt > 60\GeVc$ at generator level
for the (a) TCHP and (b) SSVHP algorithms for different pileup (PU) scenarios.}
\label{fig:pile-upBTagPerformance}
\end{figure}

Because of the luminosity profile of the 2011 data, the number of proton collisions taking place simultaneously in one bunch crossing was of the order of 5 to 20 depending on the time period.
Although these additional collisions increase the total number of tracks in the event, the track selection is able to reject tracks from nearby primary vertices.
The multiplicity distribution of selected tracks is almost independent of the number of primary vertices, as shown in Fig.~\ref{fig:pile-upObservables} (a).
There is an indication of a slightly lower tracking efficiency in events with high pileup.
The rejection of the additional tracks is mainly due to the requirement on the distance of the tracks with respect to the jet axis.
This selection criterion is very efficient for the rejection of tracks from pileup.
The reconstruction of track parameters is hardly affected.
The distribution of the second-highest IP significance is stable, as shown in Fig.~\ref{fig:pile-upObservables} (b).
The impact of high pileup on the \cPqb-jet tagging performance is illustrated in Fig.~\ref{fig:pile-upBTagPerformance}.
This shows the light-parton misidentification probability versus the \cPqb-jet tagging efficiency for the TCHP and SSVHP algorithms.
In order to focus on the changes due to the \cPqb-jet tagging algorithms, the performance curves have been compared using a jet \pt threshold of 60\GeVc at the generator level.
The changes are small and concentrated in the regions of very high purity.

%% file: EfficienciesMuonJets.tex
\section{Efficiency measurement with multijet events}\label{sec:EfficienciesMuonJets}

\input{EfficienciesMu}

\input{EfficienciesIP}

\input{EfficienciesSum}

%% file: EfficienciesMu.tex
For the \cPqb-jet tagging algorithms to be used in physics analyses, it is crucial
to know the efficiency for each algorithm to select genuine \cPqb\ jets.  There are
a number of techniques that can be applied to CMS data to measure the
efficiencies {\it in situ}, and thus reduce the reliance on simulations.
If event distributions from MC simulation match those observed in
data reasonably well, then the simulation can be used for a wide range of topologies after applying corrections determined from specific data samples.
Corrections can be applied to simulated events using a scale factor \SFb, defined as the ratio of the efficiency measured with collision data to the efficiency found in the equivalent simulated samples, using MC generator-level information to identify the jet flavour.
Furthermore, the measurement techniques used for data are also applied to
the simulation in order to validate the different algorithms.

Some efficiency measurements are performed using samples that include a jet
with a muon within $\Delta R = 0.4$ from the jet axis (a ``muon jet'').
Because the semileptonic branching fraction of \cPqb\ hadrons is significantly
larger than that for other hadrons (about 11\%, or 20\% when $\mathrm{b}
\rightarrow \mathrm{c} \rightarrow \ell$ cascade decays are included), these
jets are more likely to arise from \cPqb\ quarks than from another flavour.
Muons are identified very efficiently in the CMS detector, making it
straightforward to collect samples of jets with at least one muon.
These muons can be used to
measure the performance of the lifetime-based tagging algorithms, since the
efficiencies of the muon- and lifetime-based \cPqb-jet identification techniques are largely
uncorrelated.  Sections~\ref{sec:EffPtRel} and~\ref{sec:EffS8} describe
efficiency measurements that use muon jets, while the technique of
Section~\ref{sec:EffIP} makes use of a more generic dijet sample.  The
results are given in Section~\ref{sec:Results}.

\subsection{Efficiency measurement with kinematic properties of muon jets}\label{sec:EffPtRel}

Due to the large \cPqb-quark mass, the momentum component of the muon
transverse to the jet axis, \ptrel, is larger for muons from \cPqb-hadron
decays than for muons in light-parton jets or from charm hadrons. 
This component is used as the discriminant for the ``PtRel'' method. In
addition, the impact parameter of the muon track, calculated in three
dimensions, is also larger for \cPqb\ hadrons than for other hadrons. 
This parameter is used as the discriminant for the ``IP3D'' method.
Both of
these variables can thus be used as a discriminant in the \cPqb-jet tagging
efficiency determination.  
In both cases, the discriminating power of the variable depends on the muon jet \pt.
The muon \ptrel\ (IP) distributions provide better separation for jets with \pt smaller (greater) than about 120\GeVc.
The PtRel and IP3D methods rely on fits to the \ptrel~\cite{ptrel}
and muon IP distributions in the data with respect to simulated spectra for the \cPqb\ signal and charm+light-parton background.

In the two methods, the \ptrel\ and IP spectra for muon jets are modelled
using simulated distributions that represent the spectra expected for
different jet flavours to obtain the \cPqb-jet content of the sample.  The
efficiency for a particular tagger is obtained by measuring the fraction of
muon jets that satisfy the requirements of the tagger.  To make the
treatment of the statistical uncertainty more straightforward, the muon jet
sample is separated into those jets that satisfy and those that fail the
requirements of the tagger.
These jets are referred to as ``tagged'' and ``untagged.''

\begin{figure}[hbtp]
\begin{center}
  \includegraphics[width=.49\textwidth]{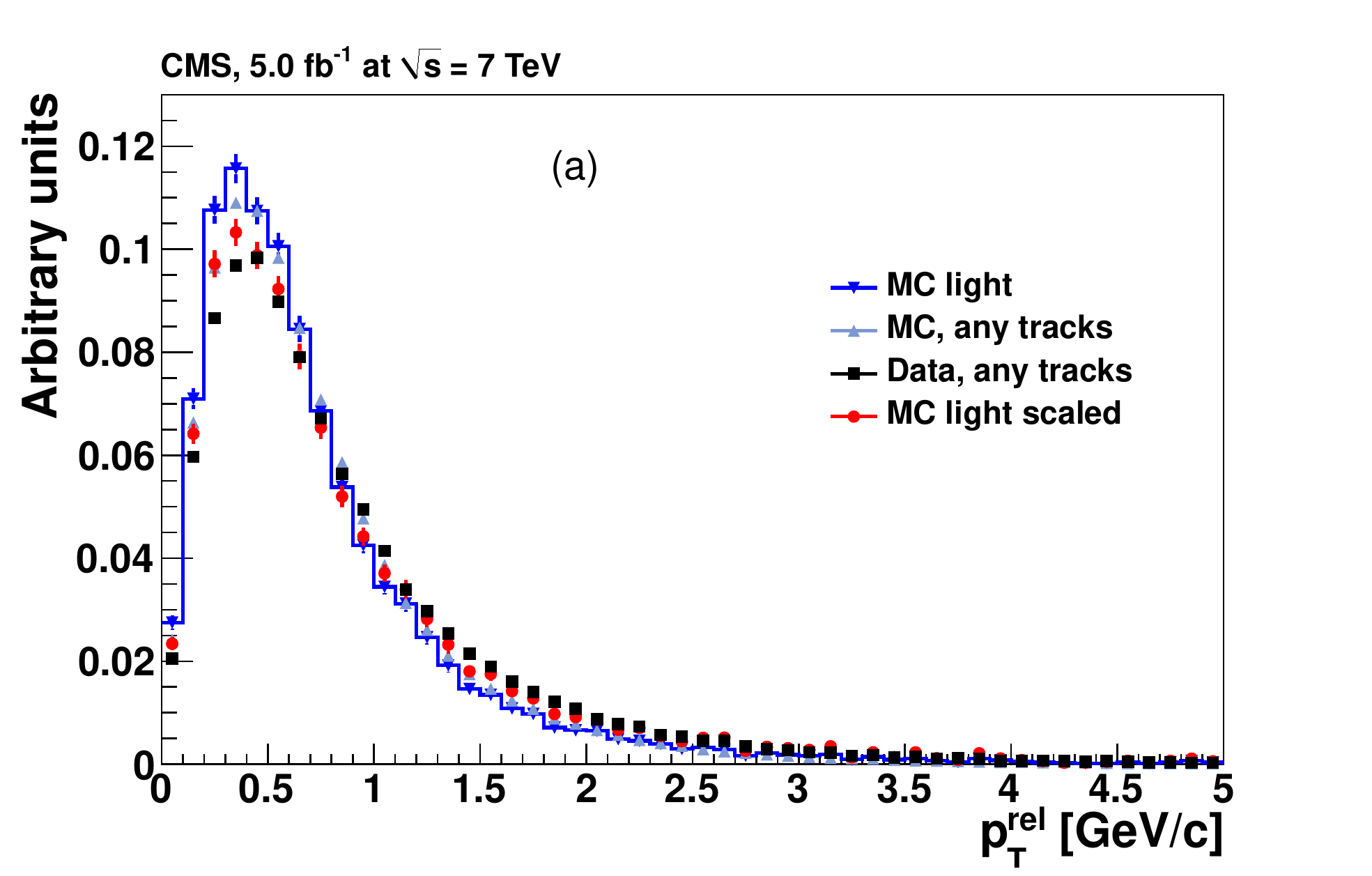} \hfil
  \includegraphics[width=.49\textwidth]{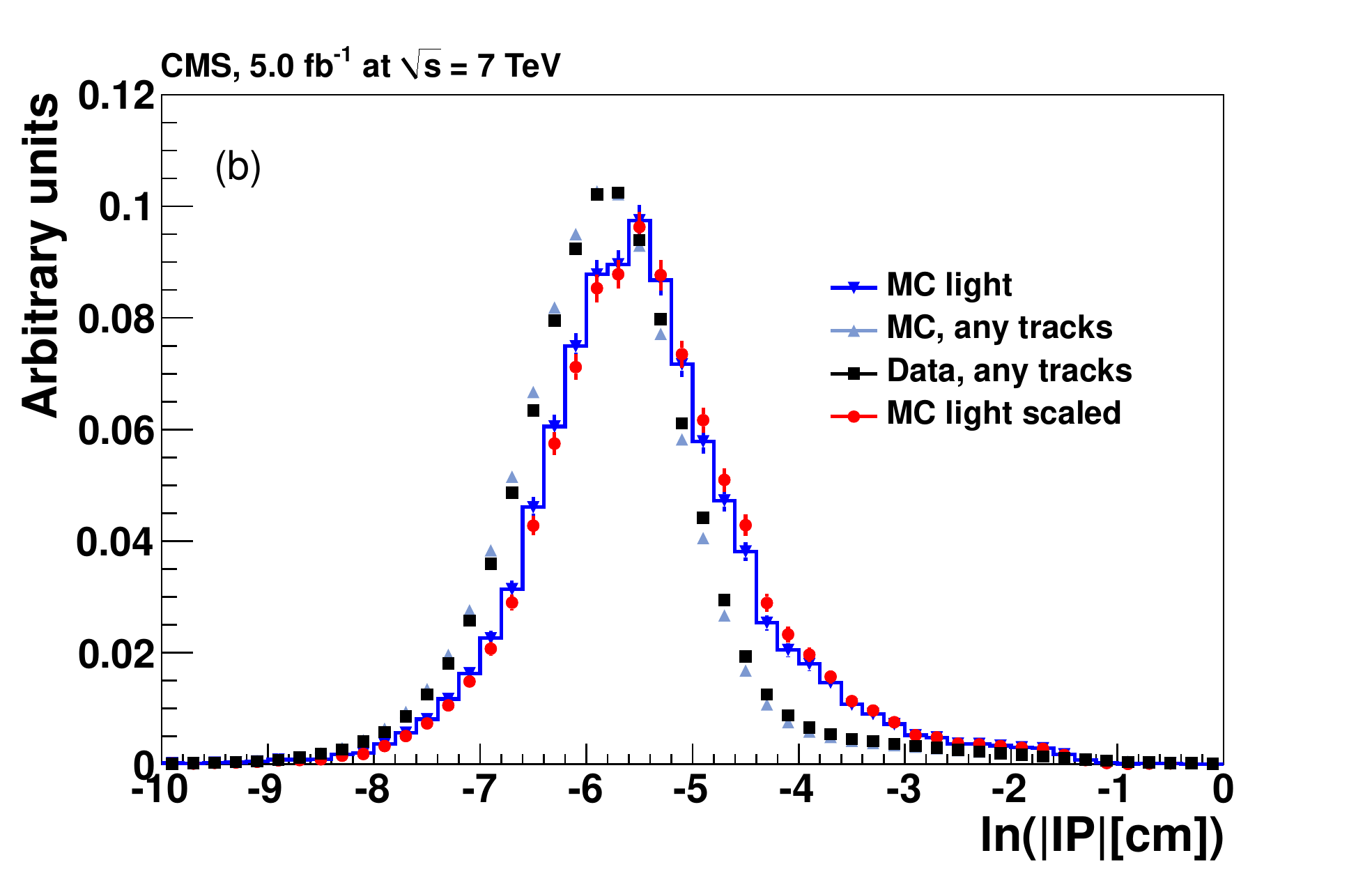}
\caption{Comparison of distributions of (a) muon \ptrel\ for jets with \pt between 80 and 120\GeVc and (b) $\ln(|\mathrm{IP}|\mathrm{[cm]})$ for jets with \pt between 160 and 320\GeVc  for muons in simulated light-parton jets (``MC light''), tracks from simulated inclusive jet events (``any tracks''), tracks from data, and muons in simulated light-parton jets after corrections based on data (``MC light scaled'').}
\label{fig:ptrel_lightTemplate}
\end{center}
\end{figure}

A dijet sample with high \cPqb-jet purity is obtained by requiring that events
have exactly two reconstructed jets: the muon jet as defined above and
another jet fulfilling the TCHPM \cPqb-jet tagging criterion (the ``medium''
operating point for the TCHP algorithm).  Simulated MC events are
used to establish \ptrel\ and IP spectra for muon jets resulting from the
fragmentation of \cPqb, \cPqc, and light partons.  Muons in light-parton
jets mostly arise from the decay of charged pions or kaons and from misidentified
muons or hadronic punch-through in the calorimeters, effects that might not
be modelled well in the simulation.  The spectra for light-parton jets
from simulation can be validated against control samples of collision
data. 
In Fig.~\ref{fig:ptrel_lightTemplate} the distributions of \ptrel\ and $\ln(|\mathrm{IP}|\mathrm{[cm]})$ derived from the simulation are compared to the ones obtained for tracks in inclusive jet data by applying the same kinematic selection and track reconstruction quality requirements as for the muon candidates. 
In order to measure the ability of the simulation to model the investigated spectra, we apply the same procedure to a sample of simulated inclusive jet events.
The spectra derived for low-\pt muons from light-parton jets in simulation are corrected by multiplying them with the ratio of shapes of the inclusive distributions obtained in data and simulation on a bin-by-bin basis.

\begin{figure}[h!]
  \begin{center}
    \includegraphics[width=.45\textwidth]{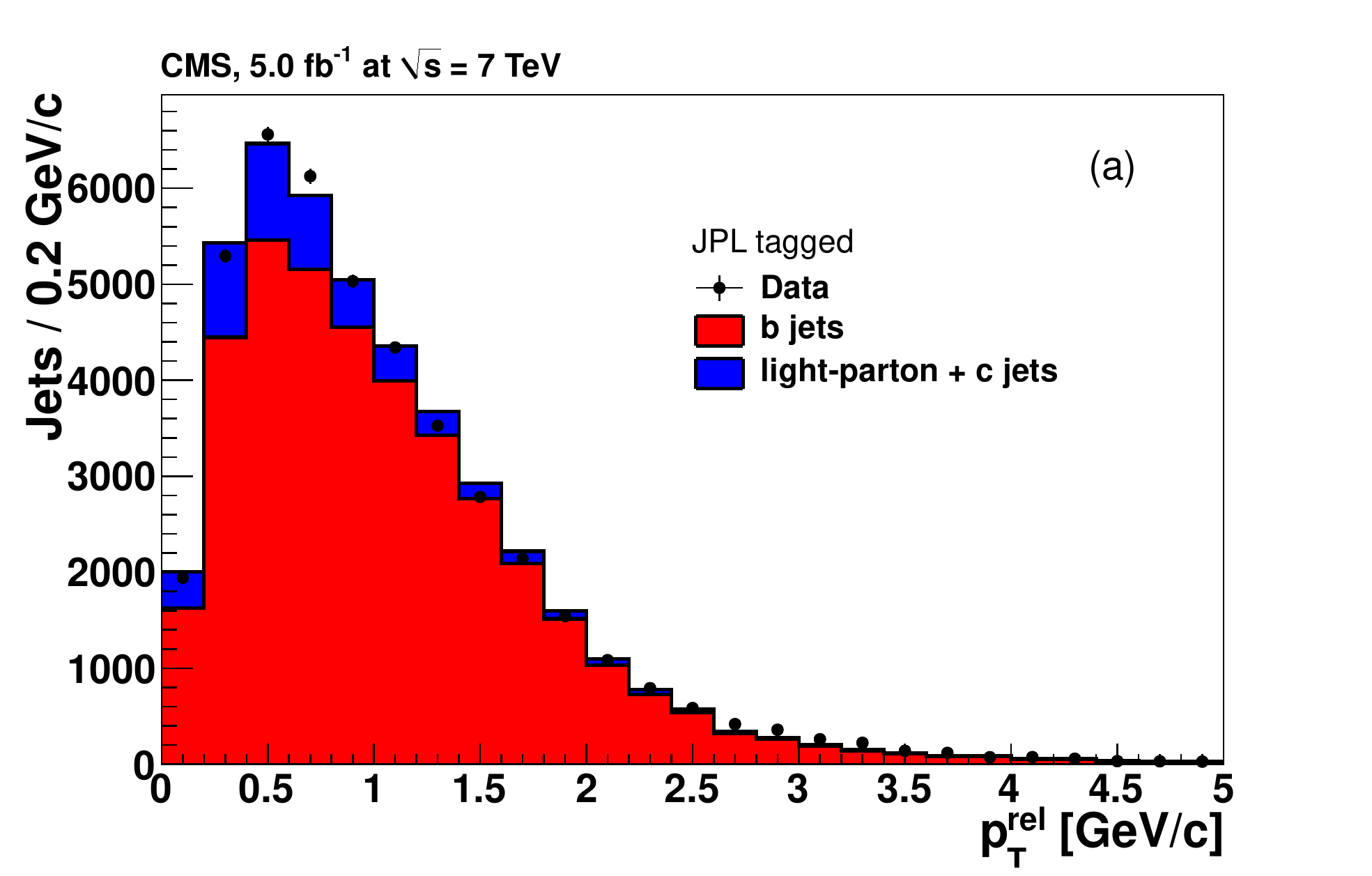} \hfil
    \includegraphics[width=.45\textwidth]{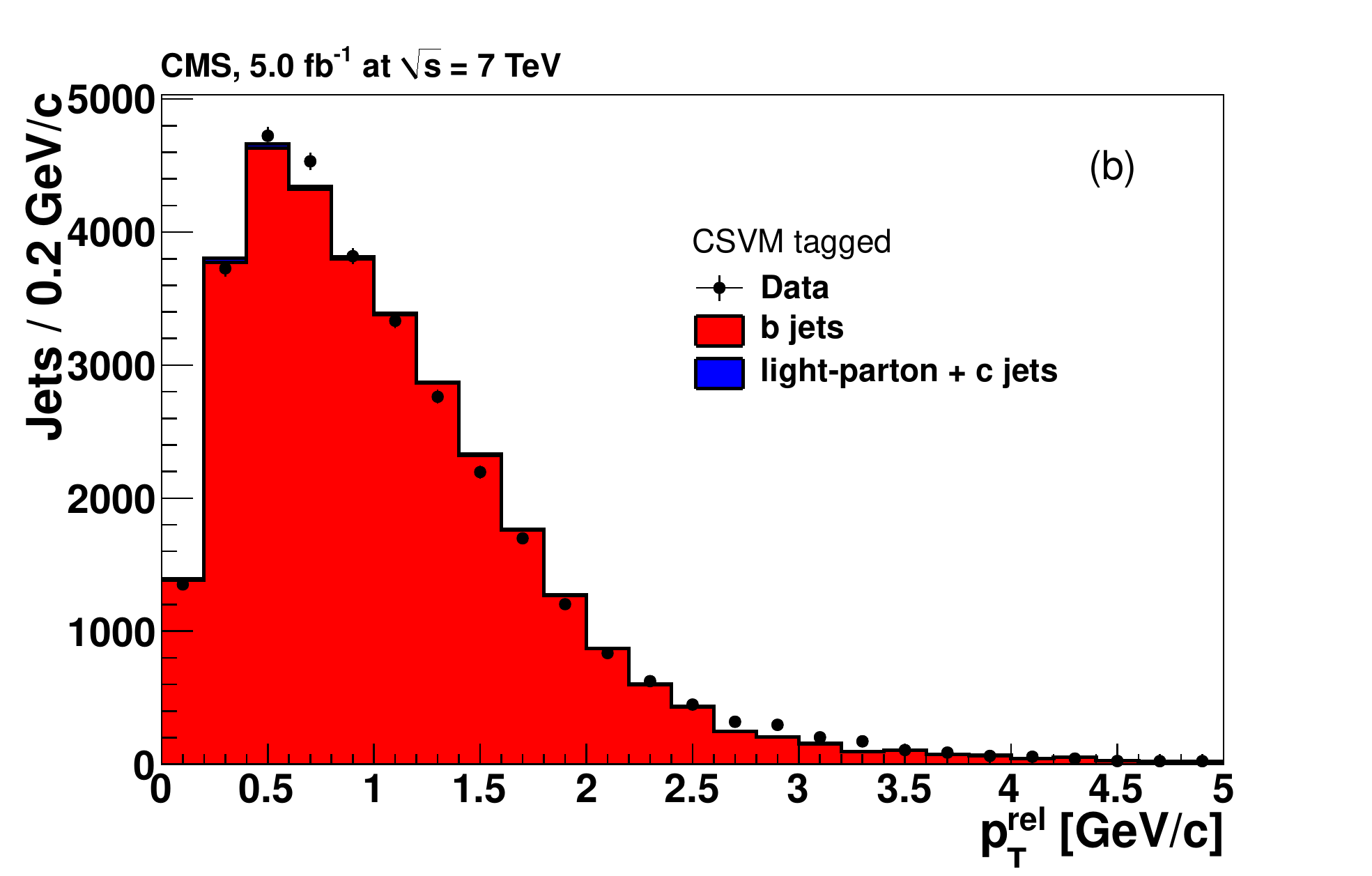} \\
    \includegraphics[width=.45\textwidth]{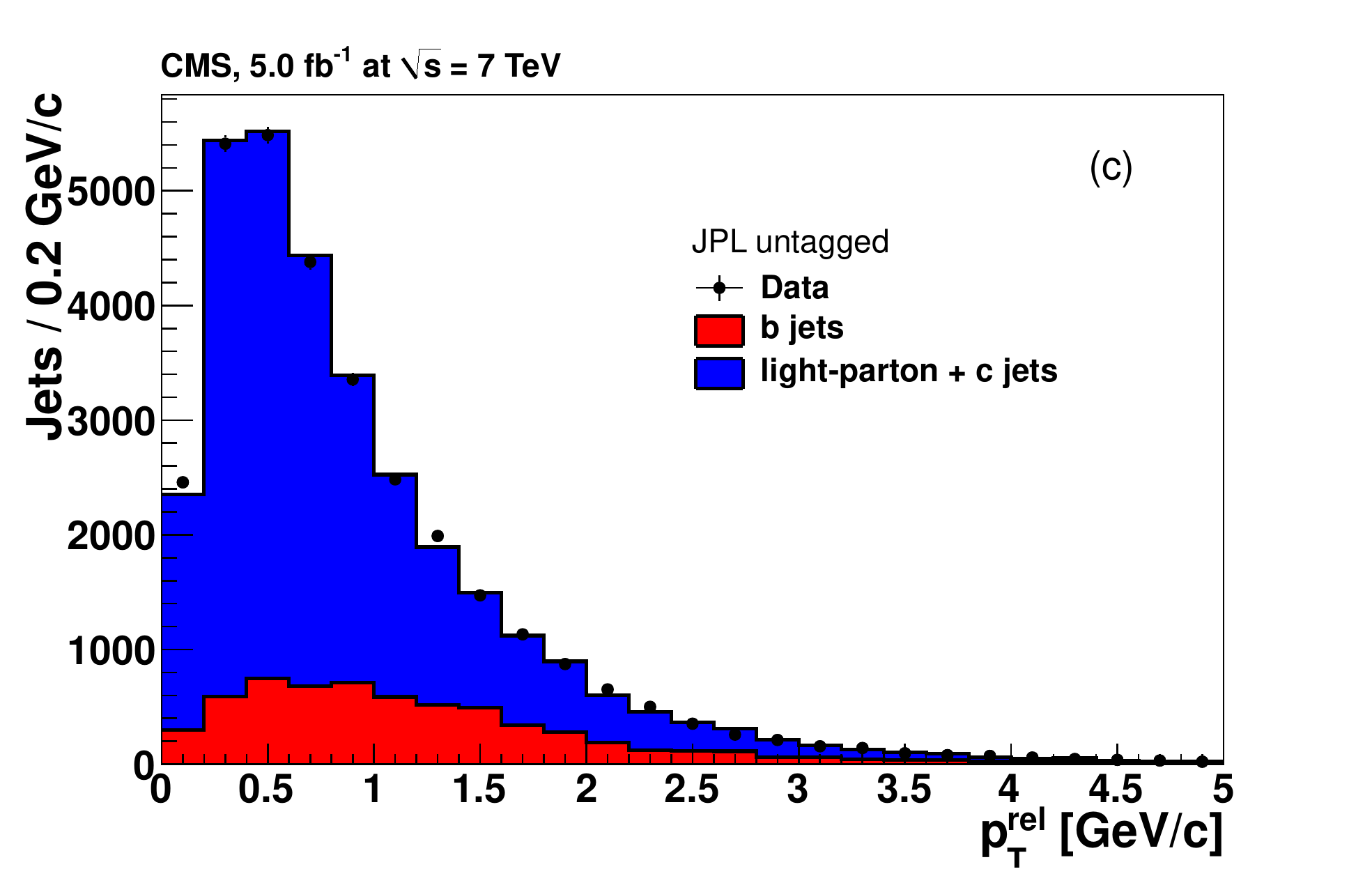} \hfil
    \includegraphics[width=.45\textwidth]{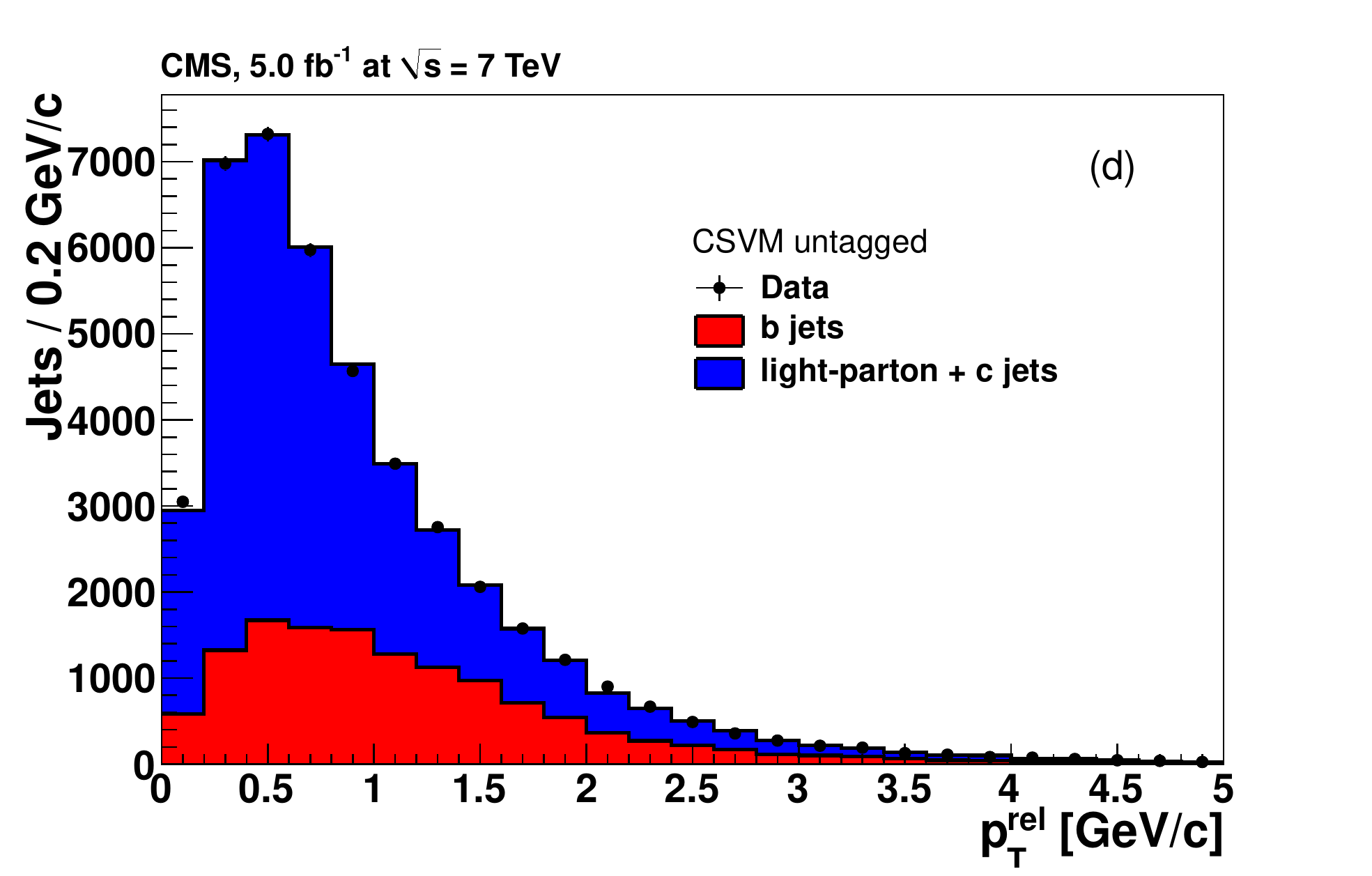}
\caption{Fits of the summed \cPqb\ and non-\cPqb\ templates, for simulated muon jets, to the muon \ptrel\ distributions from data. (a) and (c) show the results for muon jets that pass (tagged) or fail (untagged) the \cPqb-jet tagging criteria of the JPL method, respectively. (b) and (d) are the equivalent plots for the CSVM method.
      The muon jet \pt is between 80 and 120\GeVc.}
    \label{fig:ptrel_fit}
  \end{center}
\end{figure}

\begin{figure}[h!]
  \begin{center}
    \includegraphics[width=.45\textwidth]{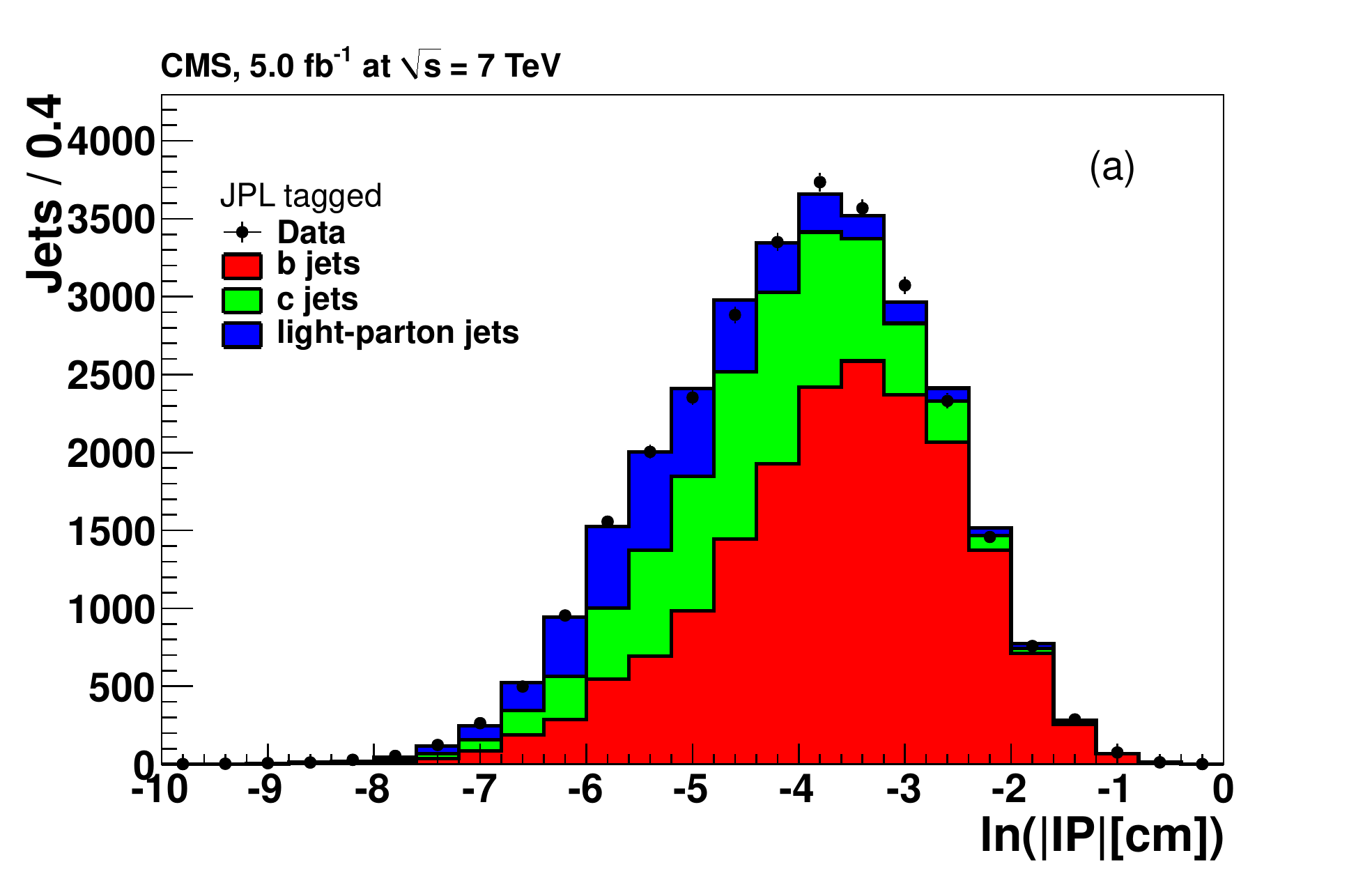} \hfil
    \includegraphics[width=.45\textwidth]{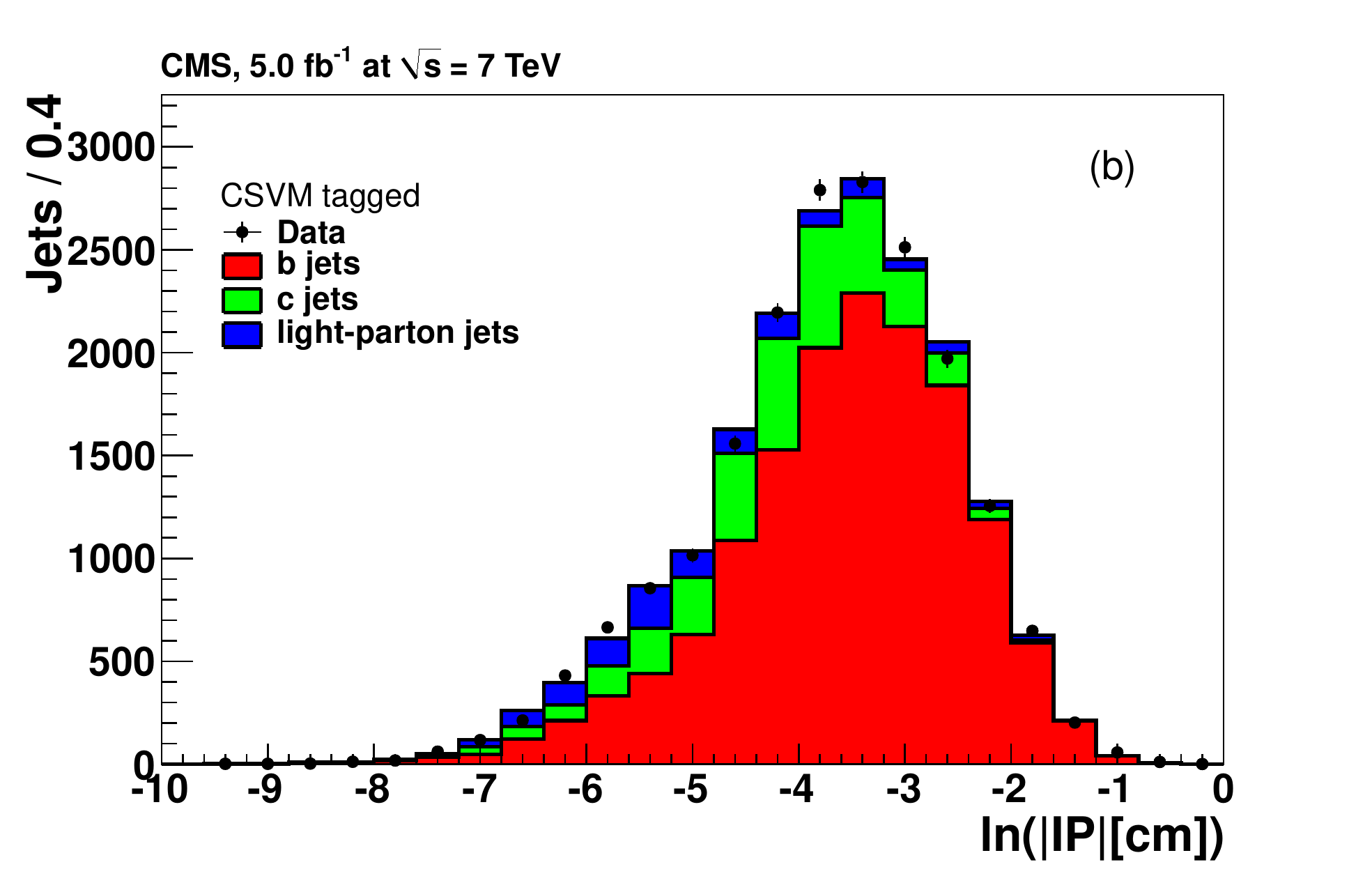} \\
    \includegraphics[width=.45\textwidth]{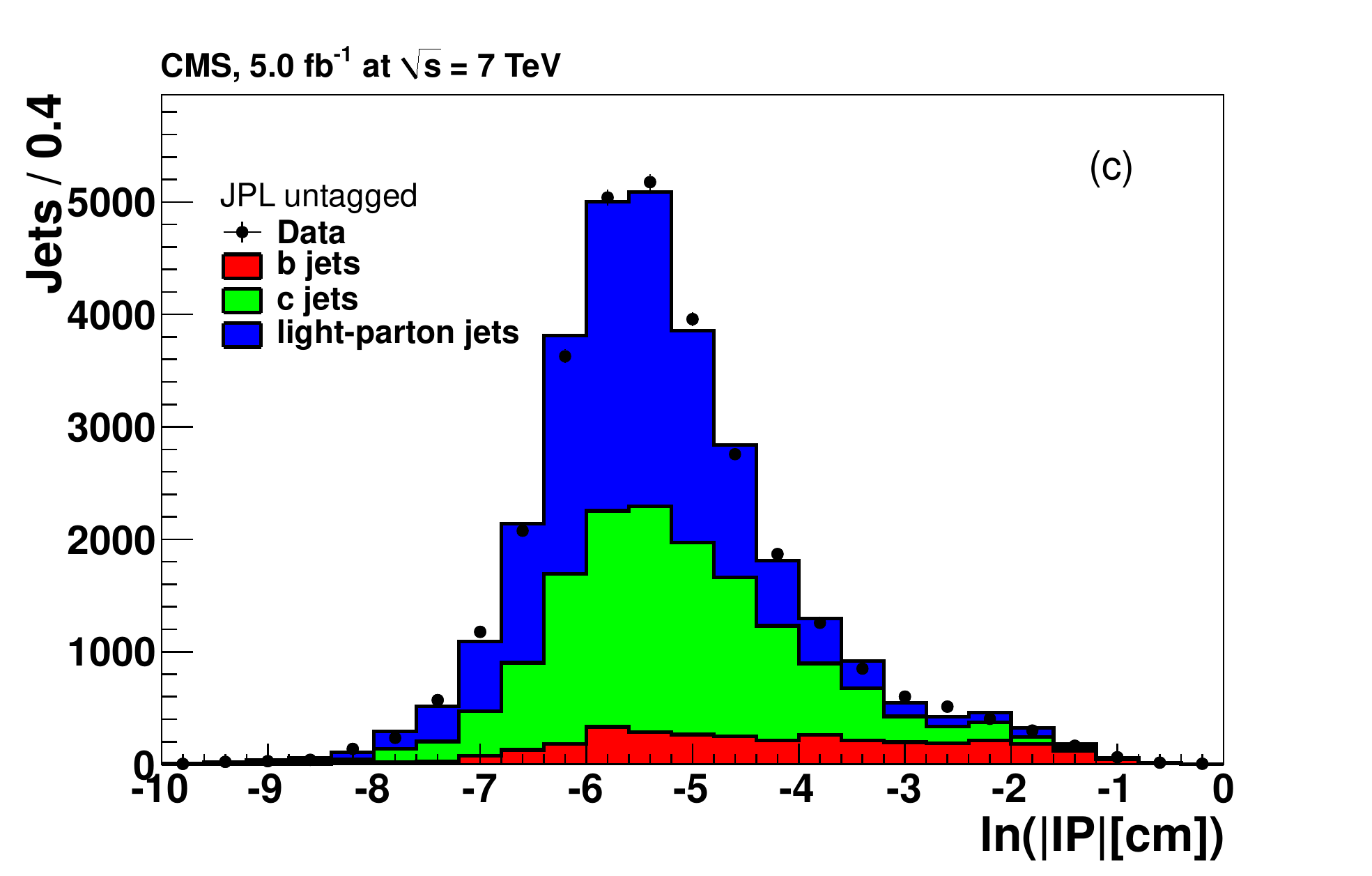} \hfil
    \includegraphics[width=.45\textwidth]{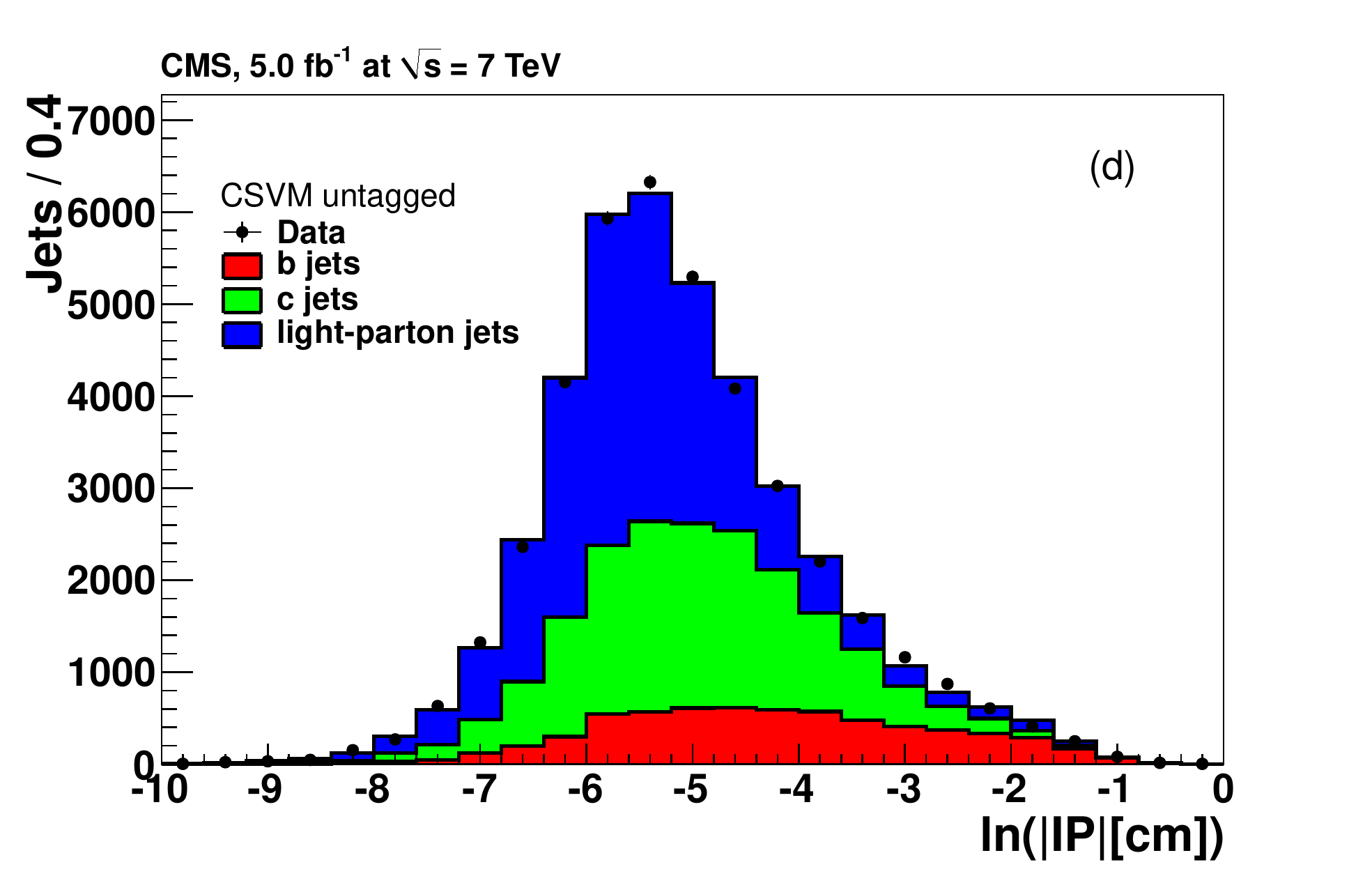}
    \caption{Same as Fig.~\ref{fig:ptrel_fit} using the $\ln(|\mathrm{IP}|\mathrm[cm])$ distributions.  The muon jet \pt is between 160 and 320\GeVc.}
    \label{fig:ip3d_fit}
  \end{center}
\end{figure}

The fractions of each jet flavour in the dijet sample are extracted with a binned maximum likelihood fit using  \ptrel\ and IP templates for \cPqb, \cPqc\ and light-parton jets derived from simulation or inclusive jet data.
The fits are performed independently in the tagged and
untagged subsamples of the muon jets.  Results of representative fits are
shown in Figs.~\ref{fig:ptrel_fit} and \ref{fig:ip3d_fit}.

From each fit the fractions of \cPqb\ jets ($f_\cPqb^\mathrm{tag}$, $f_\cPqb^\mathrm{untag}$) are
extracted from the data.  With these fractions and the total yields of
tagged and untagged muon jets ($N_\mathrm{data}^\mathrm{tag}$, $N_\mathrm{data}^\mathrm{untag}$), the
number of \cPqb\ jets in these samples are calculated, and the efficiency \effb\
for tagging \cPqb\ jets in the data is inferred:
\begin{equation}
\effb =
\frac{f_\cPqb^\mathrm{tag} \cdot N_\mathrm{data}^\mathrm{tag}}
     {f_\cPqb^\mathrm{tag} \cdot N_\mathrm{data}^\mathrm{tag} + f_\cPqb^\mathrm{untag} \cdot N_\mathrm{data}^\mathrm{untag}} \; .
\label{eq:eff_b_data}
\end{equation}

To obtain \SFb, the efficiency for tagging \cPqb\ jets in the simulation is
obtained from jets that have been identified as \cPqb\ jets with MC
generator-level matching.

\subsection{Efficiency measurement with the System8 method}\label{sec:EffS8}

The ``System8'' method~\cite{Benoit,Abazov:2010}  is applied to events with a muon jet and at least one other, ``away-tag'', jet. The muon jet is used as a probe. The reference lifetime tagger and a supplementary \ptrel-based selection are tested on this jet. The away-tag jet is tested with a separate lifetime tagger. 
There are eight quantities that can be counted from the full data sample. 
The quantities depend on the number of passing or failing tags. 
A set of equations correlates these eight quantities with the tagging efficiencies.

A muon jet can be tagged as a \cPqb\ jet using either a lifetime tagger, or by
requiring that the muon has large \ptrel.  In this analysis, the
requirement is $\ptrel\ > 0.8\GeVc$.  These two tagging criteria have
efficiencies \effb\ and $\varepsilon_\cPqb^\mathrm{PtRel}$, respectively, for \cPqb\ jets.
The third tagging criterion is the requirement that another jet in the event passes also a lifetime-based tagger.  This last requirement defines the ``away-tag sample''.
It enriches the \cPqb\ content of the events, and thus makes it more likely that
the muon jet is a \cPqb\ jet.  
Correlations between the efficiencies of the two
tagging criteria are estimated from simulation.  As \ptrel\ provides less
discrimination between jet flavours at higher jet energies, the System8
method loses sensitivity for jet $\pt > 120\GeVc$.

With these criteria eight quantities are measured. 
The four quantities for the muon jets are: the total number of muon jets in the sample $n$, the number of muon jets that pass the lifetime-tagging criterion $n^\mathrm{tag}$, the number of muon jets that pass the \ptrel\ requirement $n^\mathrm{PtRel}$, and the number of muon jets that pass both criteria $n^{\mathrm{tag},\mathrm{PtRel}}$. 
Likewise, the four quantities for the away-tag sample are labelled $p$,
$p^\mathrm{tag}$, $p^\mathrm{PtRel}$, $p^{\mathrm{tag},\mathrm{PtRel}}$. 
The away-tag jets are tagged with the TCHPL criterion.

The full muon jet sample, $n$, and the away-tag sample, $p$, are each composed of an unknown mix of \cPqb\ and non-\cPqb\ jets. 
The non-\cPqb\ jets are labelled ``$\cPqc\ell$''. 
The muon sample thus comprises $n_\cPqb$ and $n_{\cPqc\ell}$, and the away-tag sample, $p_\cPqb$ and $p_{\cPqc\ell}$.
The efficiencies
of the two tagging criteria on \cPqb\ jets ($\varepsilon_\cPqb^\mathrm{tag}$,
$\varepsilon_\cPqb^\mathrm{PtRel}$) and on non-\cPqb\ jets ($\varepsilon_{\cPqc\ell}^\mathrm{tag}$,
$\varepsilon_{\cPqc\ell}^\mathrm{PtRel}$) are also unknown, for a total of eight
unknown quantities.  Thus, a system of eight equations can be written that
relates the measurable quantities to the unknowns:
\begin{eqnarray}
n &=& n_\cPqb + n_{\cPqc\ell}\;, \nonumber\\
p &=& p_\cPqb+ p_{\cPqc\ell}\;, \nonumber\\
n^\mathrm{tag} &=&
\varepsilon_\cPqb^\mathrm{tag} n_\cPqb + \varepsilon_{\cPqc\ell}^\mathrm{tag} n_{\cPqc\ell}\;, \nonumber\\
p^\mathrm{tag} &=&
\beta^\mathrm{tag} \; \varepsilon_\cPqb^\mathrm{tag} p_\cPqb + \alpha^\mathrm{tag} \; \varepsilon_{\cPqc\ell}^\mathrm{tag} p_{\cPqc\ell}\;, \\
n^\mathrm{PtRel} &=&
\varepsilon_\cPqb^\mathrm{PtRel} n_\cPqb + \varepsilon_{\cPqc\ell}^\mathrm{PtRel} n_{\cPqc\ell}\;, \nonumber \\
p^\mathrm{PtRel} &=& \beta^\mathrm{PtRel} \; \varepsilon_\cPqb^\mathrm{PtRel} p_\cPqb + \alpha^\mathrm{PtRel} \; \varepsilon_{\cPqc\ell}^\mathrm{PtRel} p_{\cPqc\ell}\;, \nonumber\\
n^{\mathrm{tag},p_\mathrm{Trel}} &=&
\beta^{n} \; \varepsilon_\cPqb^\mathrm{tag} \varepsilon_\cPqb^\mathrm{PtRel} n_\cPqb +
\alpha^{n} \; \varepsilon_{\cPqc\ell}^\mathrm{tag} \varepsilon_{\cPqc\ell}^\mathrm{PtRel} n_{\cPqc\ell}\;, \nonumber\\
p^{\mathrm{tag},p_\mathrm{Trel}} &=&
\beta^{p} \; \varepsilon_\cPqb^\mathrm{tag} \varepsilon_\cPqb^\mathrm{PtRel} p_\cPqb +
\alpha^{p} \; \varepsilon_{\cPqc\ell}^\mathrm{tag} \varepsilon_{\cPqc\ell}^\mathrm{PtRel} p_{\cPqc\ell} \;.\nonumber
\end{eqnarray}

The method assumes that the efficiencies for a combination of tagging
criteria are factorizable.  Thus eight correlation factors are introduced
to solve the system of equations: $\alpha^\mathrm{tag}$, $\beta^\mathrm{tag}$,
$\alpha^\mathrm{PtRel}$, $\beta^\mathrm{PtRel}$, $\alpha^{n}$, $\beta^{n}$, $\alpha^{p}$,
and $\beta^{p}$. 
These factors are obtained from the simulation as a function of the muon jet \pt and $|\eta|$.  
The factors $\alpha$ and $\beta$ are determined for non-\cPqb\ and \cPqb\ jets, respectively.
The superscripts ``tag'' and ``PtRel'' of $\alpha$ and $\beta$ indicate the efficiency ratio of the $p$ to the $n$ samples for the lifetime and \ptrel\ criteria.
The superscripts ``${n}$'' and ``${p}$'' refer to the correlation between the two tagging efficiencies, ``tag'' and ``PtRel'', in the $n$ and $p$ samples.

The simulation predicts that the correlation coefficients typically range
between 0.95 and 1.05 for those associated with the \cPqb-jet tagging efficiencies,
and between 0.7 and 1.2 for those associated with the c+$\ell$-tagging
efficiencies.
A numerical computation is applied to solve the system of eight equations
in the data to determine the eight unknowns, thus simultaneously
determining the tagging efficiencies and flavour contents of both the full
and away-tag samples.

%% file: EfficienciesIP.tex
\subsection{Efficiency measurement using a reference lifetime algorithm}\label{sec:EffIP}

While muon \ptrel\ provides less discrimination power between jet flavours at large jet \pt, the lifetime-based algorithms described in Sections~\ref{sec:ImpactParameter} and \ref{sec:SecondaryVertex} (TCHE, TCHP, JP, JBP, SSVHE, SSVHP and CSV) retain their sensitivity to distinguish different jet flavours.  
In particular, the
discriminant for the jet probability algorithm has different distributions for
different jet flavours for jet momenta in the range $30<\pt<700 \GeVc$.  
The JP algorithm can be calibrated directly with data. Tracks with negative
impact parameter are used to compute the probability that those tracks 
come from the primary vertex.  The same calibration is performed
separately in simulated samples.  As a result, the JP algorithm serves as a
reference for estimating the fraction of \cPqb\ jets in a data sample,
and also for estimating the fraction of \cPqb\ jets in a subsample that has been
selected by an independent tagging algorithm.  In this manner
the efficiency of the independent algorithm can be measured.  
This method is called the lifetime tagging method (``LT'').
It can be performed on both inclusive and muon jet samples.
The resulting scale factors are compared to obtain an estimate of the systematic uncertainty.

The efficiency measurement is performed in inclusive jet events in which at
least one jet must be above a given \pt threshold, and separately in dijet
events in which at least one jet is a muon jet.  To increase the fraction
of \cPqb\ jets in the inclusive sample, an additional jet tagged by the JPM
algorithm is also required.  The sample with muon jets is already
sufficiently enriched in \cPqb\ jets by the muon requirement.  The same set of
samples can be established with simulated events, so that the true tagging
efficiency can be measured there and a scale factor computed.

Because a value of the JP discriminant can be defined for jets that have as
few as one track with a positive impact parameter significance, the
discriminant can be calculated for most \cPqb\ jets, regardless of their \pt.
The fraction of \cPqb\ jets that have JP information, $C_\cPqb$, rises from about
0.91 at $\pt = 20\GeVc$ to more than 0.98 for $\pt > 50\GeVc$.

\begin{figure}[hbtp]
  \begin{center}
  \includegraphics[width=0.9\textwidth]{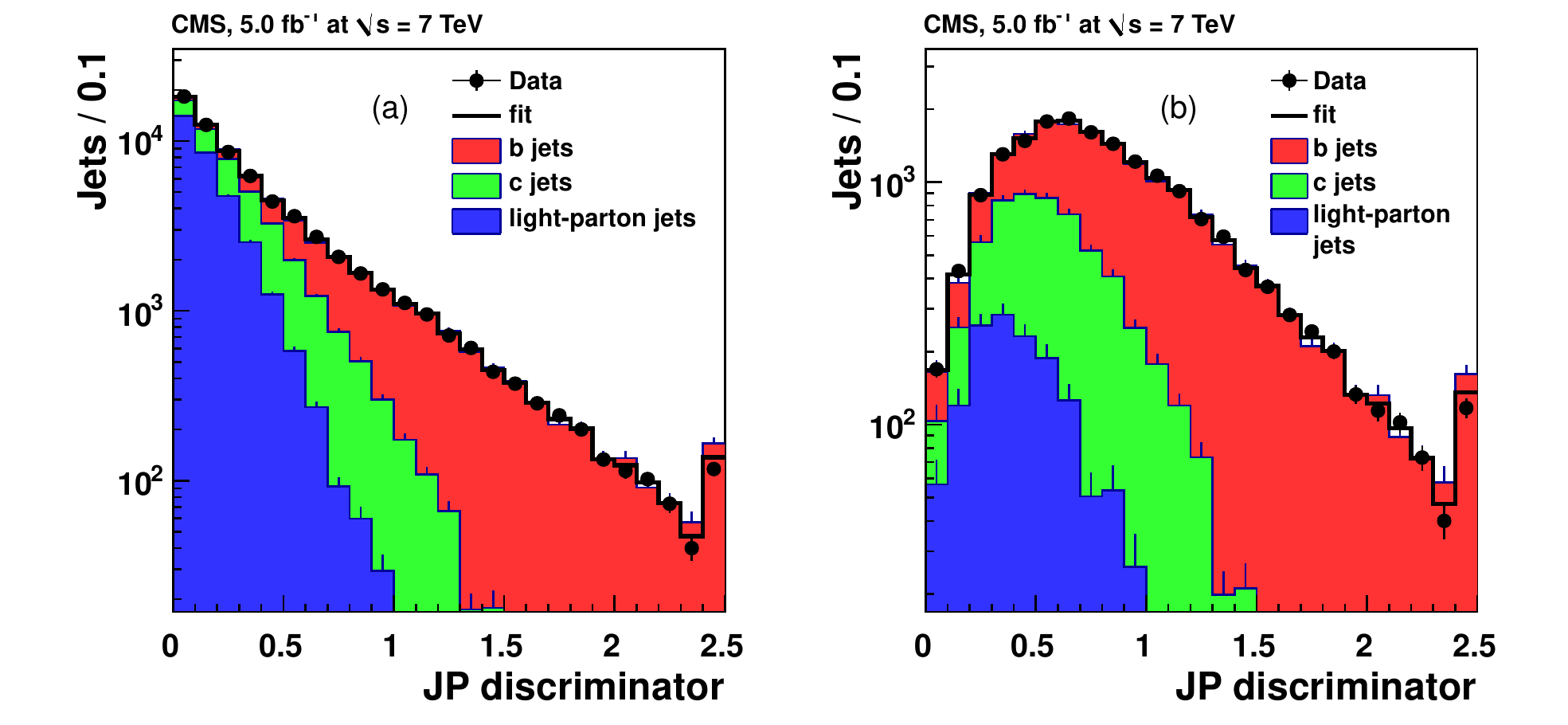} \\
  \includegraphics[width=0.9\textwidth]{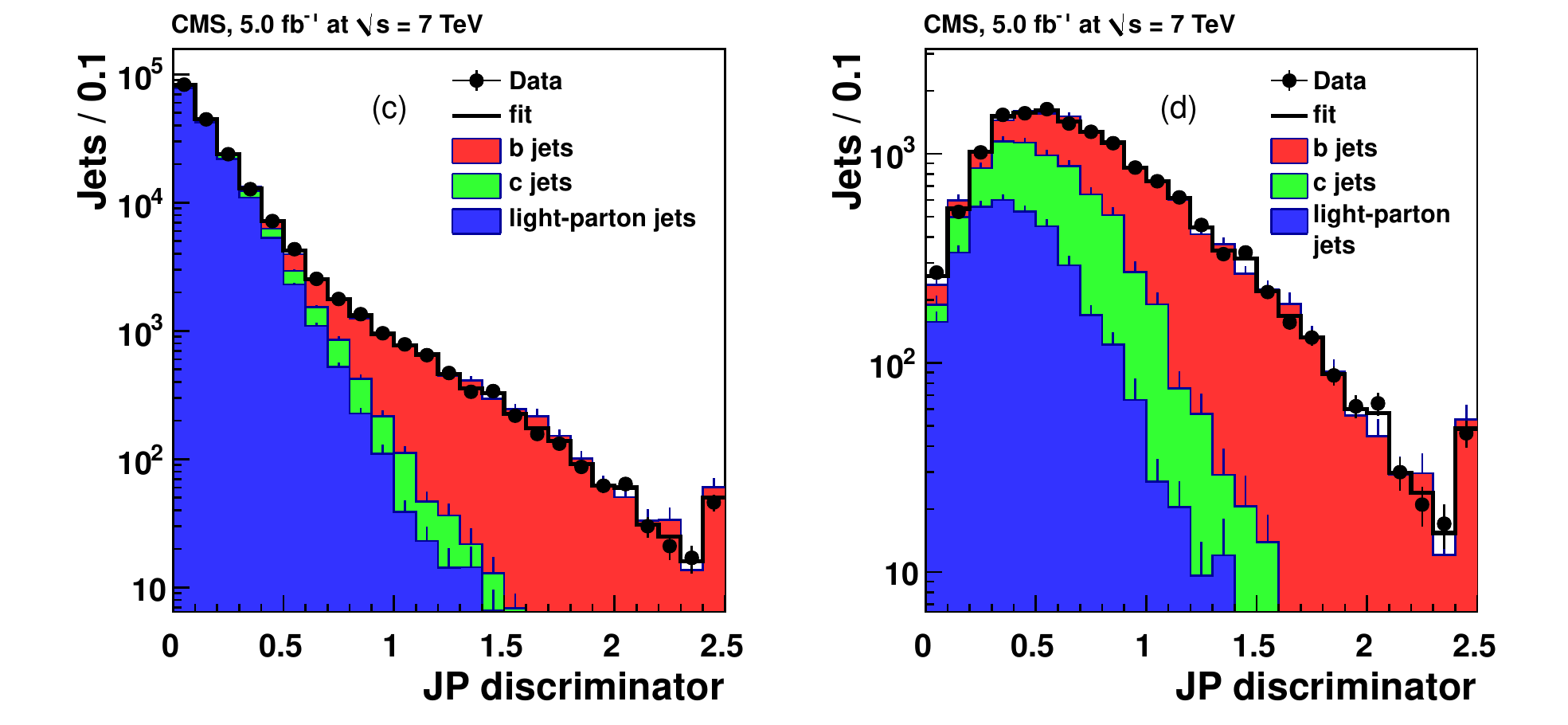} \\
  \caption{Fits of the summed \cPqb, \cPqc\ and light-parton templates, for
    simulated jets, to the JP-discriminant distributions from data. (a) and
    (b) show the results for muon jets before and after identification with
    the CSVM tagger, respectively. (c) and (d), the equivalent plots for
    inclusive jets.  
The black line is the sum of the contributions from the templates.
The jet \pt is in the range $260 < \pt < 320$\GeVc. 
    Overflows are displayed in the rightmost bins.}
    \label{fig:fit-jp}
  \end{center}
\end{figure}

Figure~\ref{fig:fit-jp} shows the JP discriminant distributions in the
muon jet sample and the inclusive sample, before and after tagging the
jets with an independent tagger, in this case the CSVM tagger.  Also shown
is a fit to the distributions using JP-discriminant templates
derived from simulations of \cPqb, \cPqc, and light-parton jets.  The normalization
of the relative flavour fractions $f_\cPqb$, $f_\cPqc$ and $f_\mathrm{light}$ is left free,
with the constraint that $f_\cPqb + f_\cPqc + f_\mathrm{light} = 1$.  
The \cPqb-jet tagging efficiency is the ratio of the number of \cPqb\ jets that are tagged by the independent tagger to the number of \cPqb\ jets before the tagging. 
The numbers are calculated using the fit. 
The \cPqb-jet tagging efficiency is corrected for the fraction of jets that have JP information.

\begin{equation}
\varepsilon_\cPqb^\mathrm{tag} = \frac{C_\cPqb \cdot f_\cPqb^\mathrm{tag} \cdot N_\mathrm{data}^\mathrm{tag}}
{f_\cPqb^{\mathrm{before \; tag}} \cdot N_\mathrm{data}^{\mathrm{before \; tag}}},
\end{equation}
where the superscripts ``before tag'' and ``tag'' refer to the samples before and after application of the tagging criterion.

\begin{figure}[hbtp]
  \begin{center}
    \includegraphics[width=.62\textwidth]{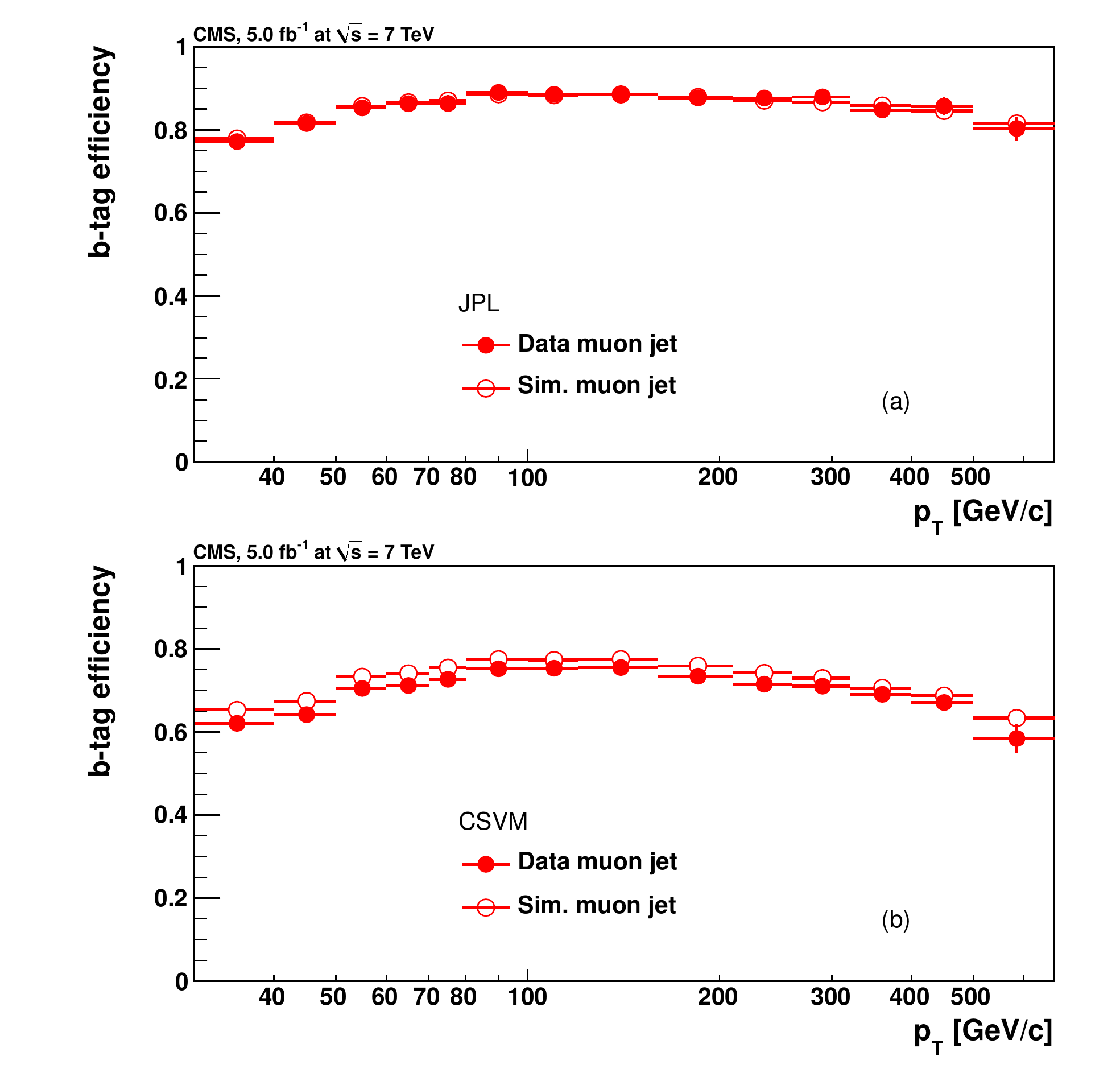} 
    \caption{Efficiencies for the identification of \cPqb-jets measured for (a) the JPL and (b) the CSVM tagger with the LT method in the muon jet sample.
      Filled and open circles correspond to data and simulation, respectively.
    }\label{fig:effi_jp}
  \end{center}
\end{figure}

Examples of the efficiencies measured for the JPL and CSVM taggers are shown in Fig.~\ref{fig:effi_jp}.
In both cases the results from simulation are close to those obtained from data.

This technique cannot be used to measure the efficiency of the JP algorithm
itself, as the JP discriminant is used in the fit to determine the \cPqb-jet
content of the sample.  However, the CSV discriminant, which is mostly
based on information from secondary vertices, can be used in its place to
determine the flavour content.  More than 90\% of jets have CSV information,
as is the case with the JP discriminant.  But unlike the JP discriminant,
the CSV discriminant cannot be calibrated solely with the data.  To remedy
this, the CSV discriminant is used to estimate the tagging efficiency of
the TC algorithms.  By comparing these results to those using
the JP discriminant, the bias due to using the CSV discriminant is
determined to be (0--2\%, 4--6\%, 6--9\%) for the (loose, medium, tight)
operating points.  The efficiencies and scale factors for the JP algorithm are
corrected for these biases.

%% file: EfficienciesSum.tex
\subsection{Systematic uncertainties on efficiency measurements\label{sec:effi_syst}}

Several systematic uncertainties affect the measurement of the \cPqb-jet tagging
efficiency.  Some are common to all four methods (PtRel, IP3D, System8,
LT), some are common to a subset of them, and some are unique to a
particular method.

Common systematic uncertainties for all methods:
\begin{itemize}
\item {\bf Pileup:} The measured \cPqb-jet tagging efficiency depends on the
  number of \Pp\Pp\ collisions superimposed on the primary interaction of
  interest.
  The systematic uncertainty is computed
  by varying the average value of the pileup in data by 10\%  and calculating
  the difference in the values of \SFb\ after reweighting the simulation
  with the modified distribution.

\item {\bf Gluon splitting:} Studies of angular correlation between \cPqb\ jets
  at the LHC~\cite{BPH-10-010} indicate that QCD events may have a larger
  fraction of gluon splitting into \bbbar\ pairs than is assumed in the generation
  of the simulation.  A study was carried out with the MC sample
  where the number of events with gluon splitting was artificially changed
  by 50\%.  Results obtained with this modified gluon splitting MC
  sample are then compared to those with the original sample.  The observed
  deviation is quoted as a systematic uncertainty.

\item {\bf Muon \ptmu:} The central value of the \cPqb-jet tagging efficiency is
  extracted from data with muon $\ptmu > 5$\GeVc.  The choice of the
  selection affects the shape of the template distributions used in
  fits, and also the number of events used to measure the tagging
  efficiencies.  The \ptmu\ threshold is varied up to 9\GeVc\ to test the
  sensitivity to this choice.

\end{itemize}

Common uncertainty for the PtRel, IP3D and System8 methods:

\begin{itemize}
\item {\bf Away-jet tagger:} The dependency of the calculated \cPqb-jet tagging
  efficiency on the away-jet tagger is studied by comparing the results
  obtained by tagging the away jets with different variants of the TC algorithm
  (TCHEL, TCHEM, TCHPM).
  The measured \SFb\ tends to increase when the away tag is
  tighter. The maximum deviation from the default away-jet tagger is taken
  as a systematic uncertainty.
\end{itemize}

Uncertainty unique to the PtRel method:
\begin{itemize}
\item {\bf Ratio of light-parton to charm jets in simulation:}
The shapes of the \ptrel\ and IP spectra for light-parton jets have been obtained from control samples in data, which minimizes the bias due to a mismodelling of the muon kinematics in the simulation.
However, since the \ptrel\ distribution in data is fitted with a sum of templates for \cPqb\ jets and for \cPqc +\cPqu\cPqd\cPqs\cPg\ jets, uncertainties on the ratio between light-parton and charmed jets in the simulation must be considered.
To do so, the predicted ratio is varied by $\pm 20\%$, and the fit is repeated, taking the variation in the results as a systematic uncertainty.
This uncertainty does not apply to the IP method, where a three-component fit is performed that determines the light-parton and charm contributions independently.

\end{itemize}

Uncertainties unique to the System8 method:
\begin{itemize}
\item {\bf Selection on \ptrel:} One of the System8 criteria is a
  selection on the muon $\ptrel >0.8$\GeVc.  In order to test the
  sensitivity to the \cPqb\ purity in the muon jet sample and the relative
  charm/light-parton fraction in the non-\cPqb\ background, this selection was changed
  from 0.5 to 1.2\GeVc in the data.  The correlation factors were
  recomputed accordingly in the simulation and the System8 method was
  applied again to the data in order to compute the \cPqb-jet tagging efficiency.
  The largest deviation observed from the central value is quoted as a
  systematic uncertainty.

\item {\bf MC closure test:} The \cPqb-jet tagging efficiency can be
  directly calculated from the simulated QCD muon-enriched sample, as the
  flavour of the jets at generator level is known.  In this case, the
  efficiency can be measured by taking the number of identified true \cPqb\ jets
  over all true \cPqb\ jets.  The resulting value is denoted as the MC
  truth \cPqb-jet tagging efficiency.  The System8 method is also applied to this
  MC sample.  The resulting \cPqb-jet tagging efficiencies are in good
  agreement with the MC truth, giving a negligible systematic
  uncertainty.  (This systematic uncertainty does not appear for the other
  methods as they rely on template fits, making such a test trivial.)
\end{itemize}

Uncertainties unique to the LT method:

\begin{itemize}
\item {\bf Fraction of \cPqb\ jets with JP information:} The fraction of
  inclusive jets with JP information is well described by the
  simulation. As explained above, the number of \cPqb\ jets before tagging is
  measured by a fit to the JP distribution and corrected by the fraction
  $C_\cPqb$ of \cPqb\ jets with JP information.  A systematic uncertainty of half
  the residual correction, $(1 - C_\cPqb) / (2 C_\cPqb)$, is estimated from the
  simulation as a function of the \cPqb-jet \pt.  A corresponding factor with a
  similar uncertainty is needed for measuring the efficiency of the JP and
  JBP taggers with the CSV discriminator spectrum.

\item {\bf Difference between muon jets and inclusive jets:} In the fits to
  the Jet Probability discriminator, the shape for the light-parton
  contribution is mostly calibrated from the data. However, as the LT
  method relies on a lifetime discriminator, a systematic effect may arise
  from some mismodelling of correlations for \cPqb\ jets between the JP
  discriminator and the other tagging criterion under study. This effect is
  specific to the LT method.  In order to estimate the uncertainty due to
  this effect, two independent samples with different \cPqb-jet fractions
  are considered: the muon-jet sample and an inclusive jet sample (where
  another jet is tagged by the JPM criterion).  The difference between the
  measured \SFb\ in muon jets and in inclusive \cPqb\ jets is taken as a
  systematic uncertainty. This is the largest contribution to the
  systematic uncertainty on \SFb\ with the LT method.  Due to the large
  statistical uncertainty on \SFb\ for inclusive jets with $\pt < 80$\GeVc,
  the same systematic uncertainty is used for $\pt < 80$\GeVc and for the
  range 80--210\GeVc.  If the difference on \SFb\ between muon jets and
  inclusive jets is smaller than the statistical error on \SFb\ for
  inclusive jets, this uncertainty is used for the systematic uncertainty
  estimate.

\item {\bf Bias for the JP and JBP taggers:} The uncertainty on the
  measurement of the bias, when using the CSV discriminant to measure the
  efficiency of the JP and JBP taggers as estimated for the TC taggers, is
  propagated into the uncertainty on the scale factors for these taggers.
\end{itemize}

The systematic uncertainties on the data/MC scale factors for different tagging criteria are detailed in Tables~\ref{tab:syst_ptrel}--\ref{tab:syst_jp} for the PtRel and System8 methods at low jet \pt\ ($80 < \pt < 120 \GeVc$) and for the IP3D and LT methods at higher jet \pt\ ($160 < \pt < 320 \GeVc$).
In these momentum ranges the average uncertainty is about 3\% for the PtRel method, 6--10\% for the System8 method, 3--4\% for the IP3D method, and 2--7\% for the LT method.

\begin{table}[!htb]
\begin{center}
\topcaption{Relative systematic uncertainties on \SFb\ for the PtRel method in the muon jet \pt range 80--120 \GeVc, using the medium operating point of the \cPqb-jet tagging algorithms.}
\begin{tabular}{@{\extracolsep{\fill}}ccccccc}
\hline
 \cPqb\ tagger  & pileup & $\cPg \to \cPqb\cPaqb$ & $\pt^{\mu}$ & away jet & light / charm & total \\
\hline
JPM & 1.6\% & 1.6\% & 1.1\% & 1.9\% & 0.4\% & 3.2\%  \\
JBPM & 0.9\% & 0.5\% & 1.8\% & 1.2\% & 0.5\% & 2.5\%  \\
TCHEM & 1.5\% & 0.5\% & 1.4\% & 1.6\% & 0.4\% & 2.7\%  \\
TCHPM & 1.1\% & 0.1\% & 2.3\% & 1.1\% & 0.6\% & 2.8\%  \\
SSVHEM & 0.7\% & 1.3\% & 2.0\% & 0.4\% & 0.4\% & 2.5\%  \\
CSVM & 1.4\% & 1.3\% & 1.3\% & 1.2\% & 0.5\% & 2.6\% \\
\hline
\label{tab:syst_ptrel}
\end{tabular}
\end{center}
\end{table}

\begin{table}[!htb]
\begin{center}
  \topcaption{Relative systematic uncertainties on \SFb\ for the System8
    method in the muon jet \pt range 80--120 \GeVc, using the medium operating point of the \cPqb-jet tagging algorithms.}
    \begin{tabular}{@{\extracolsep{\fill}}cccccccc}
\hline
    \cPqb\ tagger & pileup & $\cPg \to \cPqb\cPaqb$ & $\pt^{\mu}$ & away jet & \ptrel & MC closure  & total \\ \hline
  JPM   & 1.4\% & 0.6\% & 4.2\% & 3.9\% & 1.6\% & \phantom{$<$}0.1\% & 6.1\%  \\
  JBPM  & 1.5\% & 1.9\% & 6.5\% & 1.5\% & 4.0\% & $<$0.1\% & 8.2\%  \\
  TCHEM & 1.3\% & 1.3\% & 6.6\% & 2.1\% & 2.4\% & $<$0.1\% & 7.5\%  \\
  TCHPM & 1.3\% & 2.7\% & 8.2\% & 1.9\% & 4.0\% & \phantom{$<$}0.1\% & 9.7\%  \\
  SSVHEM& 1.3\% & 0.1\% & 3.7\% & 2.8\% & 3.0\% & $<$0.1\% & 5.6\%  \\
  CSVM  & 1.5\% & 0.4\% & 4.3\% & 1.3\% & 4.5\% & \phantom{$<$}0.1\% & 6.5\%  \\
\hline
\label{tab:syst_s8}
\end{tabular}
\end{center}
\end{table}

\begin{table}[!htb]
\begin{center}
\topcaption{Relative systematic uncertainties on \SFb\ for the IP3D method in the muon jet \pt range 160--320 \GeVc, using the medium operating point of the \cPqb-jet tagging algorithms.}\label{tab:syst_ip3d}
\begin{tabular}{@{\extracolsep{\fill}}cccccc}
\hline
\cPqb\ tagger & pileup &  $\cPg \to \cPqb\cPaqb$ & $\pt^{\mu}$ & away jet & total \\
\hline
JPM        & 0.1\% & 0.7\% & 0.1\% & 3.2\% & 3.2\% \\
JBPM       & 0.2\% & 1.7\% & 0.4\% & 2.8\% & 3.3\% \\
TCHEM      & 0.1\% & 1.2\% & 0.4\% & 2.4\% & 2.7\% \\
TCHPM      & 0.2\% & 2.3\% & 0.7\% & 2.2\% & 3.3\% \\
SSVHEM     & 0.6\% & 2.2\% & 0.3\% & 2.9\% & 3.6\% \\
CSVM       & 0.6\% & 2.3\% & 0.1\% & 3.2\% & 4.0\% \\
\hline
\end{tabular}
\end{center}
\end{table}

\begin{table}[bth]
\begin{center}
 \topcaption[]{Relative systematic uncertainty on \SFb\ with the LT method in the
 muon jet \pt range 160--320\GeVc, using the medium operating point of the \cPqb-jet tagging algorithms.}
 \label{tab:syst_jp}
\begin{tabular}{cccccccc}
\hline
 \cPqb\ tagger & pileup & $\cPg \to \cPqb\cPaqb$ & $\pt^{\mu}$ & $C_\cPqb$ & inc. jets & bias  & total \\
\hline
 JPM      & $0.1\%$ & $0.8\%$ & $0.5\%$ & $0.1\%$ & $4.4\%$ & $4.0\%$ & $6.0\%$ \\
 JBPM     & $0.1\%$ & $0.4\%$ & $0.8\%$ & $0.1\%$ & $4.3\%$ & $4.0\%$ & $5.9\%$ \\
 TCHEM    & $0.1\%$ & $1.6\%$ & $0.3\%$ & $0.2\%$ & $2.8\%$ &   ---   & $3.2\%$ \\
 TCHPM    & $0.2\%$ & $0.5\%$ & $0.5\%$ & $0.2\%$ & $1.7\%$ &   ---   & $1.9\%$ \\
 SSVHEM   & $0.1\%$ & $2.3\%$ & $0.8\%$ & $0.2\%$ & $6.6\%$ &   ---   & $7.0\%$ \\
 CSVM     & $0.2\%$ & $2.3\%$ & $0.7\%$ & $0.2\%$ & $5.2\%$ &   ---   & $5.7\%$ \\
\hline
\end{tabular}
\end{center}
\end{table}

%% file: EfficienciesTTbar.tex
\section{Efficiency measurement with \texorpdfstring{\ttbar}{t t-bar} events}\label{sec:EfficienciesTTbar}

\input{introductionTT}

\input{samplesTT}

\input{systematicsTT}

\input{method_profile}

\input{method_flavourTag}

\input{method_RB}

\input{method_bSample}

%% file: introductionTT.tex
In the framework of the standard model, the top quark is expected to decay to a
\PW~boson and a b~quark about 99.8\% of the time~\cite{PDG}.  
Experimentally, the measurement of the heavy-flavour content of \ttbar
events can provide  either a direct measurement of the branching fraction of the
decay of the top quark to a \PW\ boson and a b~quark, $B(\mathrm{t}\rightarrow \PW \mathrm{b})$, or, assuming
$B(\mathrm{t}\rightarrow \PW \mathrm{b})=1$,  the \bdtag efficiency.
The \cPqb\ jets in \ttbar events have an average \pt of about 80\GeVc and cover a \pt range relevant for many processes both within the standard model and for many models beyond the standard model.

In this Section, we present several methods to study the heavy-flavour content of \ttbar\ events.  
The \plr (PLR) method, described in Section~\ref{sec:plr}, and the \ftm (FTM) method, described in Section~\ref{sec:ftmm}, use \ttbar\ events in the dilepton channel in which both \PW\ bosons decay into leptons.
The \ftc (FTC) method, described in Section~\ref{sec:ftcm}, and the bSample method (Section~\ref{sec:method-b-sample}) use \ttbar\ events in the lepton+jets
channel, in which one \PW\ boson decays into quarks and the other into
a charged lepton and a neutrino.
These methods are used to measure the
efficiency of tagging \cPqb\ jets in the data and the simulation  over the average \pt and $\eta$ range of jets in the
top-quark events. The differences in efficiencies observed between the
data and MC simulation are provided as a data/MC scale
factor $SF_{b}$ similar to the techniques described in  Section~\ref{sec:EfficienciesMuonJets}.

%% file: samplesTT.tex
\subsection{Event selection}
\label{sec:evtSelTT}

The event reconstruction used herein follows closely the event selection performed for the \ttbar production cross section measurements~\cite{CMS-PAS-TOP-11-003,CMS-PAS-TOP-11-005}, with the exception of the \bdtag requirements.
All objects are reconstructed using a particle-flow algorithm.

In the  lepton+jets channel, the final state is composed of four jets, one energetic
 isolated muon and missing transverse energy.
Events are required to pass a single-muon trigger.
After offline reconstruction, events are selected requiring exactly one isolated muon with $\pt>30\GeVc$ and $|\eta|<2.1$ and at least four jets with $\pt > 30\GeVc$ and $| \eta | < 2.4$.
The \FTCm further requires that the two leading jets have transverse momenta greater than 70\GeVc and 50\GeVc respectively, and that the transverse momentum of the muon is greater than 35\GeVc.
The reconstructed missing transverse energy ($\mymet$) is required to be above 20\GeV.

In the  dilepton channel, the final state is composed of two jets, two energetic
 isolated leptons (electron or muon) and missing transverse energy.
Events are required to pass dilepton triggers in which two muons, two electrons, or one electron and one muon are required to be present.
After offline reconstruction, events are selected with two isolated, oppositely charged leptons with $\pt>20\GeVc$ and $|\eta|<2.5$ (2.4) for electrons (muons), at least
two jets with $\pt>30\GeVc$ and $|\eta|<2.4$, and $\mymet>30\GeV$ for $\Pe \Pe /\mu\mu$ events. The selected leptons and jets are required to originate from
the same primary interaction vertex. Events with same-flavour lepton pairs in the dilepton mass window ($76<m_{\ell\ell}<106\GeVcc$)
are removed to suppress the dominant Z+jet background.
Dilepton pairs from heavy-flavour resonances and low-mass Drell--Yan production are also removed by requiring a minimum dilepton invariant mass of $12\GeVcc$.

The numbers of observed and predicted events in the lepton+jets channel and the dilepton channel are given in Tables~\ref{tab:cut_flow_LJ} and~\ref{tab:CutFlow_DL}, respectively.
The uncertainties include the uncertainties on the luminosity measurement and the cross sections.
For all MC predictions, events are reweighted to take into account differences in trigger and lepton selection efficiencies between data and simulation~\cite{CMS-PAS-TOP-11-003,CMS-PAS-TOP-11-005}.
The lepton selection efficiency scale factors are estimated from data using \Z events.
For dilepton events, the trigger efficiencies are estimated on a data sample using a trigger that is weakly correlated to the dilepton triggers.
The dilepton trigger selection efficiency is estimated on events which contain two leptons that fulfil the complete dilepton event selection.

The Drell--Yan  background is measured using data.
Two different methods are used, and the two estimates are compatible.
In the PLR method, for the $\Pe \Pe$ and $\mu\mu$ channels, the ratio of Drell--Yan events outside and inside the dilepton invariant mass
 window, $R_\text{out/in}$, is estimated from the simulation. This is used to
 estimate the Drell-Yan background using the number of data events inside the
 dilepton invariant mass window~\cite{CMS-PAS-TOP-11-005}.
A contamination from other backgrounds can still be present in the \Z-mass window, and this contribution is subtracted using the \Pe$\mu$ channel scaled
 according to the event yields in the \Pe\Pe\ and $\mu\mu$ channels.
For the \Pe$\mu$ channel, the DY  background yield is estimated
after performing a binned maximum-likelihood 
fit to the dilepton invariant mass distribution.
In the \FTMm, the number of Drell-Yan events \Pe\Pe\ and $\mu\mu$ channels is estimated from the shape of the distribution of the angle between the momentum of the two leptons. For the \Pe$\mu$ channel, the predictions are taken from simulation.

\begin{table}
\begin{center}
\small
\topcaption{
Number of observed and predicted events in the lepton+jets sample after applying all selection requirements of the \FTCm. All MC samples have been scaled to an integrated luminosity of $2.3\fbinv$.
The uncertainties include the uncertainties on the luminosity and the cross sections.
The CSVM operating point has been used for the \bdtag requirement.
}
\label{tab:cut_flow_LJ}
 \begin{tabular}{lccc}
      \hline
      &  no tagging & $\geq 1$ b-tagged jets & $\geq 2$ b-tagged jets \\
      \hline
      \ttbar & 8504  $\pm$  1275 & 7425 $\pm$ 1113 & 3744 $\pm$ 561  \\
      Single top & 477 $\pm$ 82  & 394 $\pm$  118  & 162 $\pm$ 49  \\
      W+jets  & 6170  $\pm$ 1851  & 1367 $\pm$ 410   & 214 $\pm$ 64  \\
      Z+jets & 459  $\pm$ 138  & 83 $\pm$ 25   & 15 $\pm$ 5  \\
      QCD &  23   $\pm$ 7  & 3 $\pm$ 1  & 0.20 $\pm$  0.06 \\
      \hline
      Total prediction & 15633  $\pm$ 2253  &  9272 $\pm$ 1921   &  4134 $\pm$ 566  \\
      \hline
      Data & 14391 &  8781  & 3897  \\
      \hline
\end{tabular}
\normalsize
   \end{center}
\end{table}

\begin{table}[hbtp!]
  \begin{center}
\topcaption{
Number of observed and predicted events in the dilepton sample after applying all selection requirements of the \PLRm.
All MC samples have been scaled to a luminosity of $2.3\fbinv$.
The uncertainties include the uncertainties on the luminosity and the cross sections.
The TCHEL operating point has been used for the \bdtag requirement.
The component ``\textit{\ttbar\ signal}" stands for the dilepton events.
The component ``\textit{\ttbar\ other}" contains the events in all other decay channels.
}
\label{tab:CutFlow_DL}
\begin{tabular}{lccc}
\hline
Processes & Channel \Pe\Pe & Channel $\mu\mu$ & Channel \Pe$\mu$ \\
\hline
\hline
\multicolumn{4}{c}{Without \bdtag requirement}\\
\hline
\ttbar signal &971 $\pm$ 147 &1275 $\pm$ 182 &3453 $\pm$ 521    \\
\ttbar other &11.5 $\pm$ 5.7 &3.3 $\pm$  1.7 &23.6 $\pm$ 11.8     \\
Single top &48.7 $\pm$ 14.6 &62.7 $\pm$  18.9 &163.7 $\pm$ 49.0    \\
Di-bosons &22.3 $\pm$ 6.7 &29.2 $\pm$  8.8 &49.4 $\pm$ 14.8    \\
Z+jets &409 $\pm$ 204 &545 $\pm$  273 &200 $\pm$ 100    \\
W+jets &12.0 $\pm$ 6.0 & $<0.5$ &11.4 $\pm$ 5.7    \\
\hline
Total prediction & 1475 $\pm$ 259 &1915 $\pm$ 343 &3902 $\pm$ 512    \\
Data &1442 &1773 &3898    \\
\hline
\hline
\multicolumn{4}{c}{With $\geq 1$ b-tagged jets}\\
\hline
Total prediction  & 1088 $\pm$ 170 &1429 $\pm$ 218 &3390 $\pm$ 475  \\
Data    &1080 &1364 &3375  \\
\hline
\hline
\multicolumn{4}{c}{With $\geq 2$ b-tagged jets}\\
\hline
Total prediction  & 529 $\pm$ 73 &697 $\pm$ 97 &1827 $\pm$ 263  \\
Data    &554 &686 &1854 \\
\hline
\end{tabular}
  \end{center}
\end{table}

%% file: systematicsTT.tex
\subsection{Systematic uncertainties}
\label{sec:systTT}

Most of the sources of systematic uncertainties are common to all methods, and several methods have specific additional contributions.
A description of the common systematic uncertainties is given in this section.
The description of the procedure to estimate the systematic uncertainties in each analysis and the influence of the different sources will be given separately for each analysis in its relevant section.

There are different sources of uncertainties originating from detector
knowledge or  related to the theory and the simulation. 
These uncertainties can affect the normalization factor for each process or they can distort the distributions themselves.

The dominant sources of uncertainty arise from the MC simulation.
The uncertainty due to the modelling of the underlying event is estimated by comparing results between the main sample generated with the Z2 tune to that with the D6T tune~\cite{Field:2009zz}.
The effect due to the scale used to match clustered jets to partons (\ie, jet-parton matching) is estimated with dedicated samples generated
by varying the nominal matching \pt\ thresholds by factors of 2 and 1/2.
Effects due to the definition of the renormalization and factorization scales 
used in the simulation of the signal are
studied with dedicated MC samples with the scales varied 
simultaneously by factors of 2 and 1/2.
The uncertainties related to the parton distribution function (PDF) used to model the hard scattering of the proton-proton collisions are estimated by varying the parameters of the PDF by $\pm 1 \sigma$ with respect to their nominal values and  using the PDF4LHC prescription~\cite{Lai:2010nw, Botje:2011sn}.
Variations in the relative composition of the simulated samples are studied by varying the contributions of each background with respect to the signal and each other.

Several systematic uncertainties pertain to the modelling of the CMS detector in the MC simulations.
Important uncertainties are the energy scales of the jets and, to a lesser extent, of the leptons, as they shift the momenta of the reconstructed objects.
Similarly, the uncertainty in jet energy resolution  has also been considered.
The effects of the jet energy scale are taken into account by varying the energy scale of the jets according to its uncertainty~\cite{Chatrchyan:2011ds}. A further source comes from the uncertainties associated with the measurement of the trigger and lepton selection efficiencies.
The uncertainty due to pileup is evaluated by varying the mean value of the measured pileup distribution by $\pm 10\%$.

%% file: method_profile.tex
\subsection{Profile likelihood ratio method}
\label{sec:plr}

In this method, the data/MC scale factor of the \cPqb-jet tagging
efficiency is measured with the PLR method
using the 2-dimensional distribution of the jet multiplicity versus
the \cPqb-tagged jet multiplicity in dilepton events.  The uncertainties in the event yield
and in the shape of the distribution are considered as nuisance
parameters in the likelihood function and are then fitted during the
minimization procedure.  This leads to combined statistical and
systematic uncertainties associated  with the measurement of the scale
factor.

\begin{figure}[t]
\centering
\includegraphics[width=0.7\textwidth]{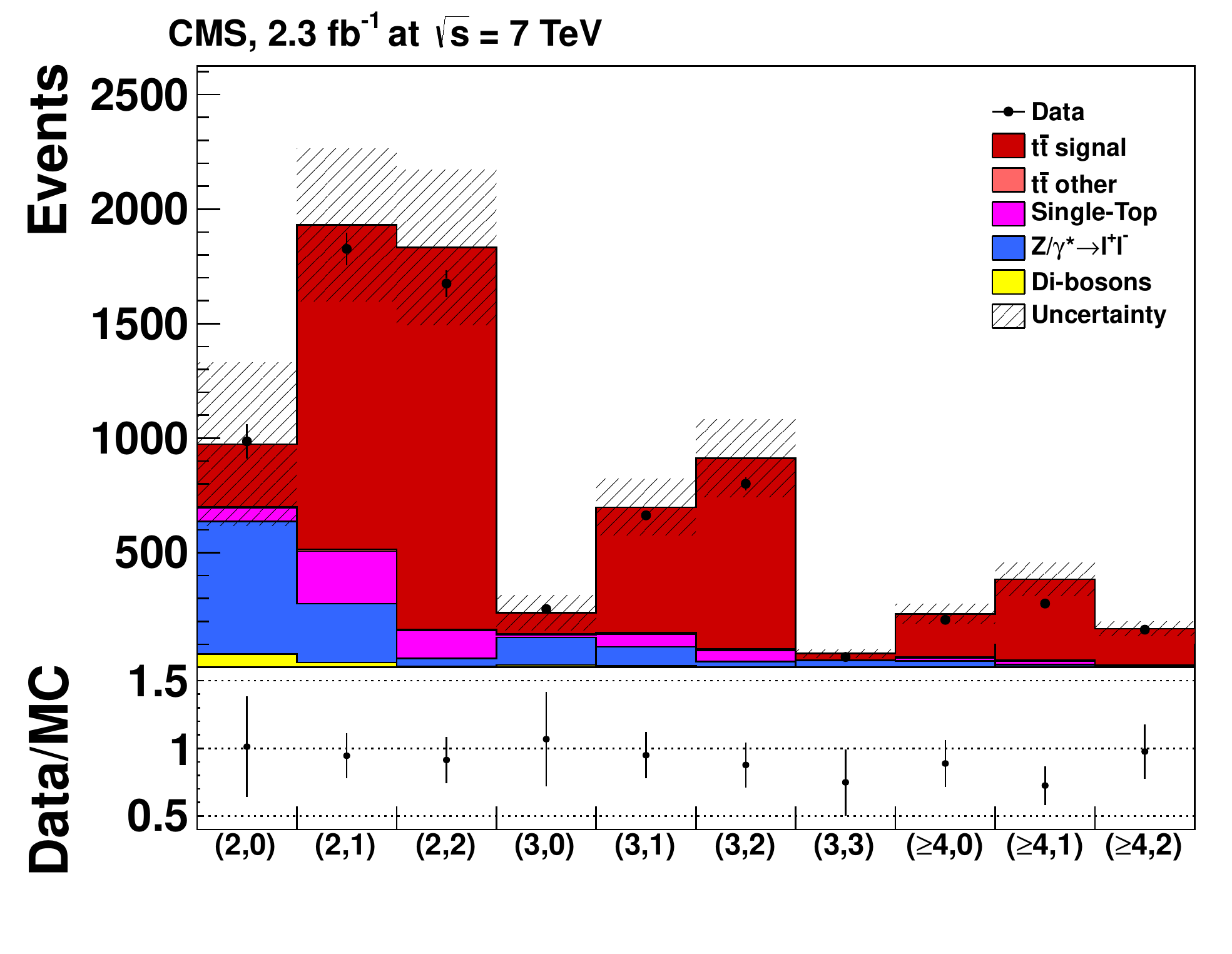}
\hfill
\caption{Number of \cPqb-tagged jets per event in the dilepton channel for the  TCHEL operating point, in the data (filled circles) and the simulation (solid histograms), before the fit. The simulated distribution
is normalized to an integrated luminosity of 2.3\fbinv.
The bin labels $(m,n)$ refer to the number of events with $m$ jets in the event of which $n$ are tagged.
The component ``{\em \ttbar\ signal}" are the dilepton events, and 
the component ``{\em \ttbar\ other}" contains the events in all other decay channels.
The hatched area corresponds to the combined statistical and systematic uncertainty.
}
\label{fig:tagged-jets-plr}
\end{figure}

The likelihood function for a given dilepton channel $j$ (ee, e$\mu$ or $\mu\mu$) and a given bin $i$ of the 2-dimensional distribution (corresponding to $n$ jets and $m$ \cPqb-tagged jets) is written  as~\cite{STATREF}:
\begin{equation}
\mathcal{L}_{i,j}(\SFb, N_{i,j}^\mathrm{obs},\{U_{k}\}) = \mathcal{P}oisson(N_{i,j}^\mathrm{obs},\mu_{i,j}(\SFb,\{U_{k}\})) \times \prod_{k} \mathcal{G}auss(U_{k},0,1)\;,
\label{eq:un}
\end{equation}
where $N_{i,j}^\mathrm{obs}$ is the number of observed events, $\mu_{i,j}$ the number of expected events, and ${U_{k}}$ the nuisance parameters.
The distribution of the number of \cPqb-tagged jets observed in data and predicted in the simulation for the TCHEL operating point 
for \ttbar and background events is shown in Fig.~\ref{fig:tagged-jets-plr}.
The likelihood function for a given channel $j$ is then the product of the likelihood functions over all the bins of the distribution:
\begin{equation}
\mathcal{L}_{j}(\SFb, \{N_{i,j}^\mathrm{obs}\},\{U_{k}\}) = \prod_{i} \mathcal{L}_{i,j}(\SFb, N_{i,j}^\mathrm{obs},\{U_{k}\}) \;.
\end{equation}
Since the decay channels are statistically independent, the overall
likelihood function is then simply the product of the individual
channel likelihoods:
\begin{equation}
\mathcal{L}(\SFb, \{N_{i,j}^\mathrm{obs}\},\{U_{k}\}) = \prod_{j} \mathcal{L}_{j} \;.
\end{equation}

The expression of the profile likelihood ratio $LR$ is then
\begin{equation}
LR(\SFb) = \frac{\mathcal{L}({\SFb},\{{\hat{\hat{U_{i}}}\})}}{\mathcal{L}(\hat{\SFb}, \{\hat{{U_{i}}}\})}\;,
\end{equation}
where $\hat{\hat{U_{i}}}$ represents the conditional
maximum likelihood estimates of ${U_{i}}$ obtained with the
scale factor \SFb\ fixed while $\hat{\SFb}$ and  $\hat{{U_{i}}}$ are
the estimates obtained with  \SFb\ free.

The distribution of $-2\ln(LR(\SFb))$ is asymptotically distributed
as a $\chi ^2$ distribution with one degree of freedom (Wilk's theorem~\cite{wilks}).
An $LR$ curve is obtained by scanning the values of \SFb\ in a given range and used to determine a 68\% confidence level interval.
These uncertainties are the combination of the
statistical uncertainty and the systematic uncertainties considered
as nuisance parameters.  
All the nuisance
parameters are common to the three channels except the 
estimation of the backgrounds from data for W+jets and Z+jets.
The Z+jets background is estimated from data as described in Section~\ref{sec:evtSelTT}.
The small W+jets background is estimated from data using the matrix method~\cite{CMS-PAS-TOP-11-005}.

The expected number of \cPqb-tagged jets in events with $n$ jets of a
given dilepton final state, $\mu_{i,j}$ in Eq.(\ref{eq:un}), is derived
from pre-tagged simulated events with $n$ jets. This is carried out by applying per-jet
\cPqb-jet tagging efficiencies, considering all jet tagging
combinations. These efficiencies are derived as a function of \pt and
$\eta$, using simulated \ttbar events for \cPqb\ jets and using data samples dominated by light-parton jets. A constant scale factor \SFb\ is applied to the \cPqb- and \cPqc-jet efficiencies to model the \cPqb-jet tagging efficiency in data. The value of \SFb\ is then extracted by minimizing the $LR$ as described above.
A closure test is performed on simulated signal events to check that, for a unit scale factor, the
\cPqb-tagged jet multiplicity distribution obtained with the reweighting
procedure is the same as the one obtained directly from MC simulation 
using a requirement on the \cPqb-jet tagging discriminant.

Several uncertainties are considered as nuisance
parameters in the likelihood function and are then fitted during the
minimization procedure.
These are the uncertainties on the energy scale of the jets and the leptons, the expected number of events of the different contributions, and the uncertainty on the light-parton jet scale factor.

Further contributions to the systematic uncertainties are estimated outside the PLR procedure.
The expected input distributions to the PLR method are re-derived, using MC samples with varied parameters, and the \cPqb-jet tagging scale factors are re-measured.
The relative differences of \SFb\ with respect to the nominal values are taken as systematic uncertainties, and added in quadrature to the total uncertainty from the fit.
These uncertainties include the uncertainties on the jet-parton matching scale, the parton-shower/matrix-element threshold, and the top mass. 

The factorization scale is the dominant systematic uncertainty, as it affects the jet multiplicity distribution, with a relative uncertainty of approximately 1.7\% on the scale factor of the CSVL operating point.
The second-largest contribution is from the uncertainty on the \ttbar event yield, which is estimated to be 20\%. It includes the uncertainties on the \ttbar cross section, the trigger and lepton selection efficiencies, and  the branching fraction of the decays of the \PW\ bosons. This results in an uncertainty of 1.4\% on the scale factor of the CSVL operating point.
Further, the statistical uncertainty on the \cPqb-jet tagging efficiency in the simulation was found to range between 0.4\% and 1.6\% depending on the operating point considered. 
A 1.6\% systematic uncertainty on the scale factor was therefore chosen for all the operating points.

Finally, to account for a possible uncertainty coming from the fitting algorithm itself, an additional uncertainty is estimated using different choices of the likelihood minimization. This is taken as a 1\% relative uncertainty.

%% file: method_flavourTag.tex
\subsection{Flavour tag consistency method}  
\label{sec:ftcm}

The  \FTCm requires consistency between the observed and expected number of
tags in the lepton+jets events to study the performance of the
heavy-flavour algorithms. 

In a sample of \ttbar\ pair candidates in the lepton+jets channel, the
expected number of events with $n$ \cPqb-tagged jets $\langle
N_{n}\rangle$ can be written as
\begin{equation}  
    \langle N_{n}\rangle = L \cdot \sigma_{\ttbar} \cdot  \varepsilon \cdot  \sum_{i,j,k}F_{ijk}  \sum_{i'+j'+k' = n}^{i'\leq i,j'\leq j, k'\leq k} 
     [C_{i}^{i'}\varepsilon_\cPqb^{i'}(1-\varepsilon_\cPqb)^{(i-i')}C_{j}^{j'}\varepsilon_\cPqc^{j'}(1-\varepsilon_\cPqc)^{(j-j')}C_{k}^{k'}\varepsilon_{\mathrm{l}}^{k'}(1-\varepsilon_{\mathrm{l}})^{(k-k')}],
\label{eq:lighteq}
\end{equation} 
where $L$ is the integrated luminosity, $\sigma_{\ttbar}$ is
the $\ttbar$ cross section, $\varepsilon$ is the pre-tagging selection
efficiency, $C_{a}^{b}$ is the binomial coefficient, and
$\varepsilon_\cPqb, \varepsilon_\cPqc$, and $\varepsilon_{\mathrm{l}}$ are the \cPqb-, \cPqc-,
and light-parton jet tagging efficiencies.  The factors $F_{ijk}$ are the
fractions of events with $i$ \cPqb\ jets, $j$ \cPqc\ jets, and $k$ light-parton jets.
They are derived from the \ttbar\ simulation in which the true flavour of
the jets is known.

As an example, the  $F_{112}$ term contributes to the expected number
of events with 1 \cPqb-tagged jet $\langle N_{1} \rangle$ in the following
way:
\begin{equation}  
    \langle N_{1}\rangle \propto F_{112}\times \left(\underbrace{  1\cdot \varepsilon_\cPqb(1-\varepsilon_\cPqc)(1-\varepsilon_{\mathrm{l}})^{2}}_\text{the \cPqb\ jet} + \underbrace{1\cdot (1-\varepsilon_\cPqb)\varepsilon_\cPqc(1-\varepsilon_{\mathrm{l}})^{2}}_\text{the \cPqc\ jet}
	  + \underbrace{2\cdot (1-\varepsilon_\cPqb)(1-\varepsilon_\cPqc)\varepsilon_{\mathrm{l}}(1-\varepsilon_{\mathrm{l}})}_\text{the light-parton jet} \right).
\end{equation}

To account for the non-negligible amount of background, Eq.~(\ref{eq:lighteq}) is modified to include each background sample:
\begin{eqnarray}
    \langle N_{n}\rangle & = & \langle N_{n}^{\ttbar}\rangle + \langle N_{n}^{\text{background}}\rangle  \nonumber \\
     & = & {L} \cdot \sigma_{\ttbar} \cdot  \varepsilon_{\ttbar} \cdot \Bigg[ \sum_{i,j,k}F_{ijk}^{\ttbar}  \sum_{i'+j'+k' = n}^{i'\leq i,j'\leq j, k'\leq k} ( \cdots )  \nonumber \\
     & + & \frac{\sigma_{\text{background}}}{\sigma_{\ttbar}} \cdot \frac{\varepsilon_{\text{background}}}{\varepsilon_{\ttbar}} \cdot \sum_{i,j,k}F_{ijk}^{\text{background}}  \sum_{i'+j'+k' = n}^{i'\leq i,j'\leq j, k'\leq k} ( \cdots ) \Bigg],
\label{eq:fulleq}
\end{eqnarray}
where $( \cdots )$ stands for the expression in square brackets from
Eq.~(\ref{eq:lighteq}).

\begin{figure}[t]
\centering
\includegraphics[width=0.49\textwidth]{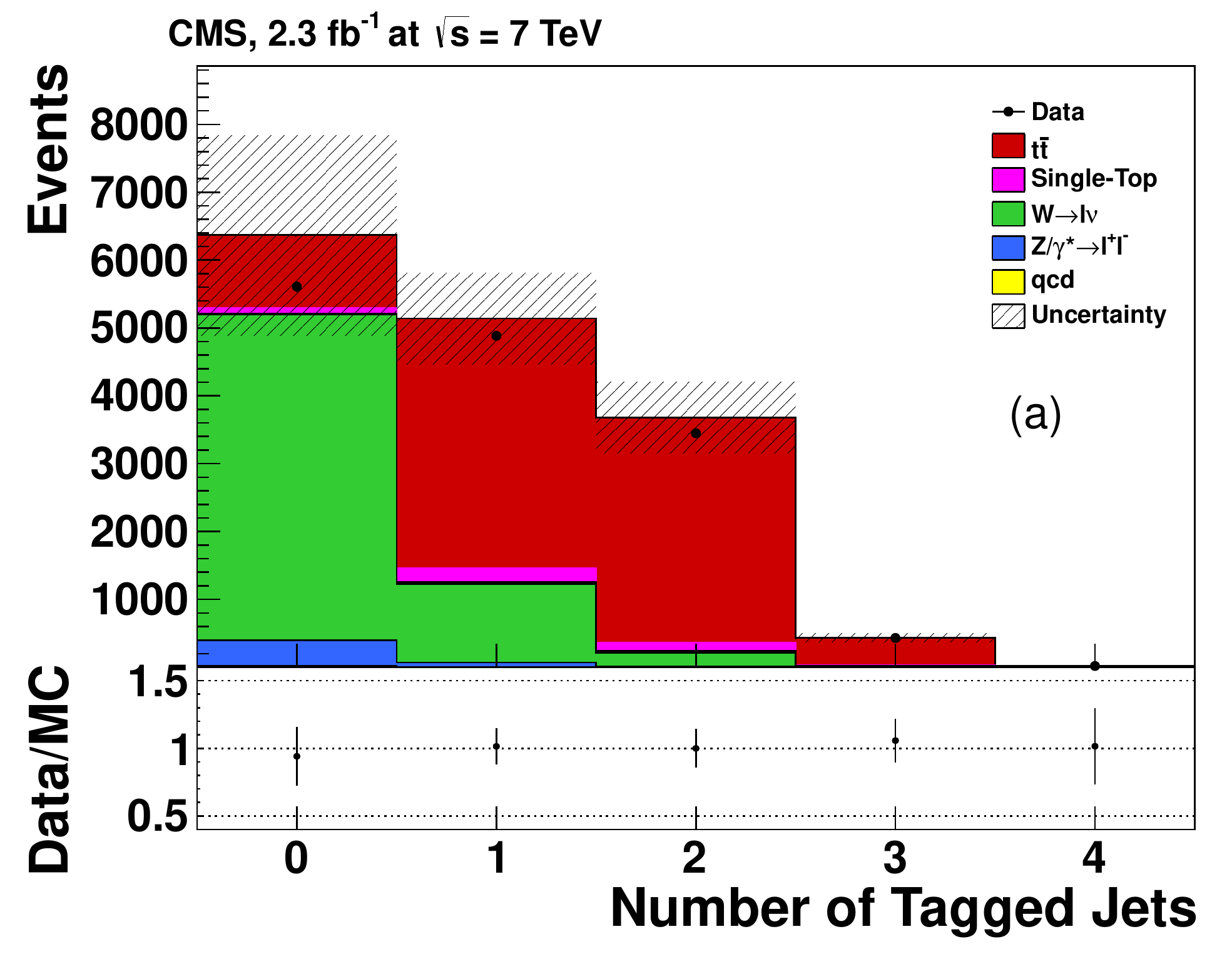}
\includegraphics[width=0.49\textwidth]{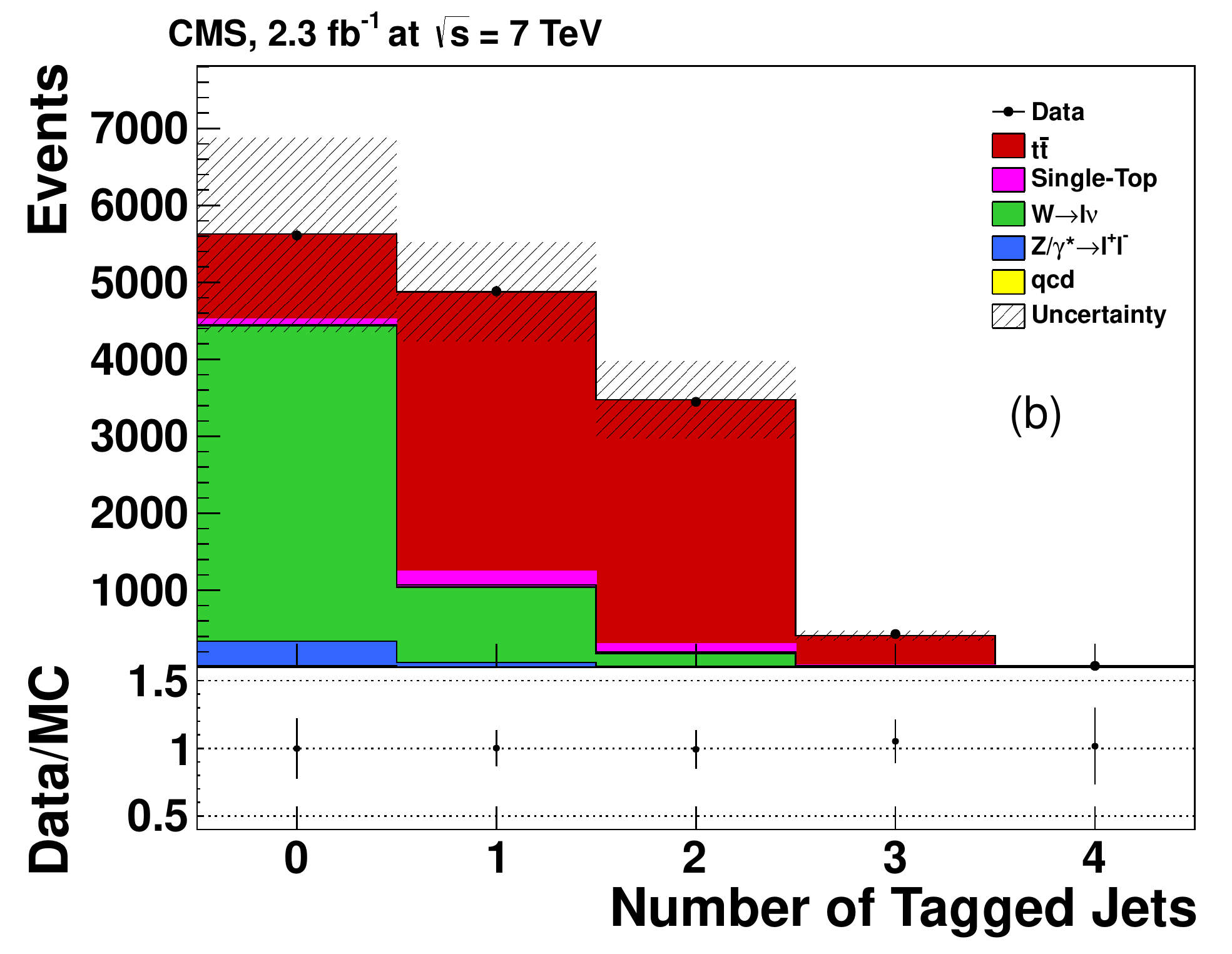}
\hfill
\caption{Number of tagged jets per event in the lepton+jet channel with the \FTCm with the  CSVM operating point, in the data (filled circles) and the simulation (solid histograms), (a) before and (b) after the fit. The simulated distribution
is normalized to an integrated luminosity of 2.3\fbinv.
The hatched area corresponds to the combined statistical and systematic uncertainty.
}
\label{fig:tagged-jets}
\end{figure}

The tagging efficiencies and the \ttbar\ production cross section are
then measured by minimizing the log-likelihood function:
\begin{equation}  
    \mathcal{L} = -2\log \prod_{n} {\mathcal Poisson}(N_{n}, \langle N_{n}\rangle    ), \\
\end{equation} 
where $N_{n}$ is the number of observed events with $n$ \cPqb-tagged jets.
The distribution of the number of \cPqb-tagged jets observed in data and predicted in the simulation before and after the fit
for \ttbar and background events is shown in Fig.~\ref{fig:tagged-jets}.

In the current implementation the likelihood only uses the \cPqb-tagged
jet multiplicity in \ttbar lepton+jets events with between four to
seven reconstructed jets, as it emphasizes the measurement of the
heavy-flavour  \cPqb-jet tagging efficiency.  The \cPqb-jet tagging efficiencies and
$\ttbar$ cross section are treated as free parameters in the fit.
The $\ttbar$ cross section determined in the fits are consistent with the published values.
The  \cPqc- and light-parton-jet tagging efficiencies are taken from the
simulation corrected for the data/MC scale factors.

The systematic uncertainties are determined from ensembles of pseudo-experiments.
In each of these pseudo-experiments, the number of signal and background events are generated using Poisson statistics, using as mean values the number of expected events in each channel.
Events are then randomly chosen in the simulated samples and the 
\cPqb-tagged jet multiplicity distributions are populated according to the simulated jet multiplicity in each event.
The measurement is then performed as described above using the factors $F_{ijk}$ from the nominal simulation. 
The average \bdtag efficiency  is compared to the average \bdtag efficiency
value measured in ensemble tests with the nominal samples.
The difference is taken as a systematic uncertainty.

The dominant contribution is the uncertainty on the jet energy scale, with a relative uncertainty of 2.2\% on the scale factor of the CSVL operating point. 
The second-largest uncertainty arises from the uncertainty on the production cross section of the W+heavy flavour jets, with a relative uncertainty of 0.97\%.
The uncertainties due to the factorization scale and the jet-parton matching are 0.41\% and  0.35\%, respectively, for the CSVL operating point.

%% file: method_RB.tex
\subsection{Flavour tag matching method}
\label{sec:ftmm}

The \FTMm requires consistency between 
the observed and expected number of tags in dilepton events. 
The expected number of events with $n$ \cPqb-tagged jets $\langle N_{n}\rangle$ is written as
\begin{equation}
\langle N_{n}\rangle =\sum_{\text{k~jets}=2}^{\text{all jets}} n_k \cdot P_{n,k}\;,
\label{eq:btagmultmodel}
\end{equation}
where $n_k$ is the observed number of events  with $k$ jets, and $P_{n,k}$ 
is the probability to count $n$ \cPqb-tagged jets in a $k$-jet event.
These probability functions are written in terms of the tagging
 efficiencies and the expected jet composition.

In order to illustrate explicitly the construction of the probability
 functions, the exclusive two-jet multiplicity bin is used and the
 following expression is obtained:
\begin{equation}
P_{n,2} = \mathop{\sum_{\text{i jets}=0}^{2}}_{\text{from~top~decay}} \alpha_{i}\cdot P_{n,2,i}\;,
\label{eq:twojetsprobmodel}
\end{equation}
where $P_{n,2,i}$ is the probability that $n$~\cPqb\ tags are observed in an event 
with two jets of which $i$~jets come from \ttbar decays.

The misassignment probabilities $\alpha_i$ denote the probability in
the sample that $i$ jets from the decay of the \ttbar\ pair have been
reconstructed and selected. These are normalized such that
$\sum_{i}\alpha_{i}=1$.  For example, $\alpha_2$ is the probability
that both \cPqb\ jets from the \ttbar decay have been selected.  They take
into account both the contribution from the background, which is small
in the dilepton channel, and jet misassignment.  Either or
both of the jets from the decays of the two top quarks may not be selected, and jets
from initial- and final- state radiation, or
jets from the proton recoil may enter the selection, further diluting
the sample.

As an example, for the case where two tagged jets are found in a two-jet event, 
the probabilities can be explicitly written as: 
\begin{equation}
\begin{array}{lll}
P_{2,2,0} = \varepsilon_{\mathrm{q}}^2  & 
 & \text{\small if no jets are from \ttbar decays;}\\
& & \\
P_{2,2,1} = 2\varepsilon_\cPqb\varepsilon_{\mathrm{q}} & 
 & \text{\small if 1 jet is from \ttbar decays;}\\
& & \\
P_{2,2,2} = \varepsilon_\cPqb^2 & 
 & \text{\small if 2 jets are from \ttbar decays.}
\end{array}
\label{eq:twotagstwojetbinprob}
\end{equation}
The misidentification probability $\varepsilon_{\mathrm{q}}$ is an effective measurement of the
 probability of tagging gluon, light and charm quark jets
 in the dilepton sample.
Similar expressions can easily be derived for the other jet multiplicity bins.

The misassignment probabilities are determined  
from data, and used in the subsequent likelihood of the \cPqb-tagged jet multiplicity distribution.
In order to estimate the actual fraction of \cPqb\ jets from top-quark
decays in the selected sample, kinematic properties of the top decay topology are used.
The invariant mass of the lepton-jet pairs from a $\cPqt \rightarrow \PW  \cPqb$
decay have a kinematic end-point
at $M_{\ell,\cPqb}^\mathrm{max}\equiv\sqrt{m_{\cPqt}^{2}-m_{\PW}^{2}}\approx 156\GeVcc$.
The invariant mass of misassigned lepton-jet pairs
exhibits a longer tail towards high mass values.
The shape of the misassigned pairs can be modelled by mixing lepton-jet pairs from different events or randomly
changing the lepton momentum direction.
The fraction of jets from $\cPqt \rightarrow \PW  \cPqb$ decays can thus be measured 
normalizing the spectrum obtained from the combinatorial model
to the number of pairs observed in the tail (i.e. $M_{\ell,\cPqb}>180\GeVcc$).
This is estimated  independently for each dilepton channel and for
each jet-multiplicity bin.
The procedure is checked and found to  be unbiased from MC pseudo-experiments.
Taking into account the expected contribution of \ttbar and single-top
events to the final sample, the sample composition in terms of events with $2$, $1$, or $0$  correctly reconstructed and selected \cPqb\ jets is estimated.

The \bdtag\ efficiency $\varepsilon_\cPqb$ can then be measured by maximizing the likelihood function:
\begin{equation}
{\mathcal L}(\varepsilon_\cPqb,\varepsilon_{\mathrm{q}},\alpha_{i}) = \prod_{n=0}^{\text{all~jets}}
{\mathcal Poisson} 
(N_{n}, \langle N_{n}\rangle )\;,
\label{eq:hfcfitlikelihood}
\end{equation}
where $N_{n}$ is the observed number of  events with $n$ \cPqb-tagged jets.

The likelihood only uses the \cPqb\ tagged jet
 multiplicity in \ttbar dilepton events with two and three 
 reconstructed jets.
Gaussian constraints are added for the effective \cPqc- and light-parton jet tagging efficiency $\varepsilon_{\mathrm{q}}$  and the misassignment probabilities:
\begin{equation}
{\mathcal L}= \prod_{n=0}^{\text{all~jets}}
{\mathcal Poisson} 
(N_{n}, \langle N_{n}\rangle )
~\cdot~\prod_{i} {\mathcal Gauss} (\alpha_i,\hat{\alpha}_i,\sigma_{\alpha_i}) \cdot {\mathcal Gauss} (\varepsilon_{\mathrm{q}},\hat{\varepsilon}_{\mathrm{q}},\sigma_{\varepsilon_{\mathrm{q}}})\;.
\label{eq:combinedhfcfitlikelihood}
\end{equation}
The central value $\hat{\varepsilon}_{\mathrm{q}}$ and width $\sigma_{\varepsilon_{\mathrm{q}}}$ of  $\varepsilon_{\mathrm{q}}$ are determined from the simulation. For the misassignment probabilities $\alpha_i$, the central values $\hat{\alpha}_i$ are taken from the measurement described above, and the width $\sigma_{\alpha_i}$ derived from the uncertainty of the expected contribution of \ttbar and single-top events to the final sample.

The systematic uncertainties affect the measurement of the \cPqb-jet tagging
probability through their effect on the parameters of the fit,
namely the measured misassignment probabilities and the misidentification probability for non-\cPqb\ jets.
The effect on the measured misassignment probabilities is determined
from ensembles of pseudo-experiments, where, for each source of
uncertainty, the bias on the probabilities is determined.
Most sources of uncertainties such as jet energy scale and resolution,
and pileup have little effect. This is because the method
used to derive the misassignment probabilities is based on 
templates for the lepton-jet invariant mass obtained from control samples in data.
Other sources, which might affect the contribution from top-quark decays and
from initial- and final-state radiation jets to the final sample, are
evaluated using samples where the QCD factorisation and renormalisation scales and the jet-parton matching
scales are varied.
In the pseudo-experiments the standard \ttbar\ sample is
substituted by each of these samples
and the process is repeated.

This bias is then used to shift the measured misassignment probabilities.
The likelihood fit of the data is repeated with
the modified values.
The difference with respect to the nominal result is taken as the
systematic uncertainty.
The same procedure is applied to evaluate the uncertainty on the misidentification probability for non-\cPqb\ jets.

The final uncertainty is dominated by factors which tend to increase
the contamination of background or alter the jet environment.
The main uncertainties are the factorization scale, and to a
smaller extent, the jet-parton matching
with relative uncertainties on the scale factor of 2.3\% and 1.4\% respectively, for the CSVL operating point.
The second-largest uncertainty arises from the
1.5\% uncertainty on the
light-parton jet tagging efficiency.

%% file: method_bSample.tex
\subsection{Efficiency measurement from a \texorpdfstring{\cPqb-enriched}{b-enriched} jet sample}
\label{sec:method-b-sample}

In this method, the \cPqb-jet tagging efficiency is measured from a
sample enriched with \cPqb\ jets (bSample) in lepton+jets events. The contamination of this sample due to light-parton jets  is
estimated from data and subtracted.

In order to select the correct jets originating from the decay of the top quarks, a $\chi^2$ is calculated for each jet-parton combination based on the masses of the reconstructed \PW\ boson $m_\mathrm{qq}$ and the hadronically decaying top quark $m_\mathrm{bqq}$:
\begin{equation}
\chi^2 = \left( \frac{m_\mathrm{bqq} - m_\mathrm{t}}{\sigma_\mathrm{t}} \right)^2 + \left( \frac{m_\mathrm{qq} - m_\PW}{\sigma_\PW} \right)^2\;.
\end{equation}
The mean masses and widths are obtained from the \ttbar\ simulation using a Gaussian fit to the  mass distributions of the combination with the correct
jet-to-quark assignment.
The mean and width of the reconstructed top-quark mass distribution are $172.5\GeVcc$ and $16.3\GeVcc$, respectively. The mean and width of the reconstructed \PW-boson distribution are $82.9\GeVcc$ and  $9.5\GeVcc$, respectively.
Using the four leading jets, with transverse momenta above
 30\GeVc, there are 12 combinations to pair the four reconstructed jets
 with the quarks from \ttbar decay. The combination with the
 lowest $\chi^2$ is selected to represent the event topology.
The event is rejected if the lowest $\chi^2$ is above 90.

\begin{figure}[tbh]
\begin{center}
\includegraphics[width=0.7\textwidth]{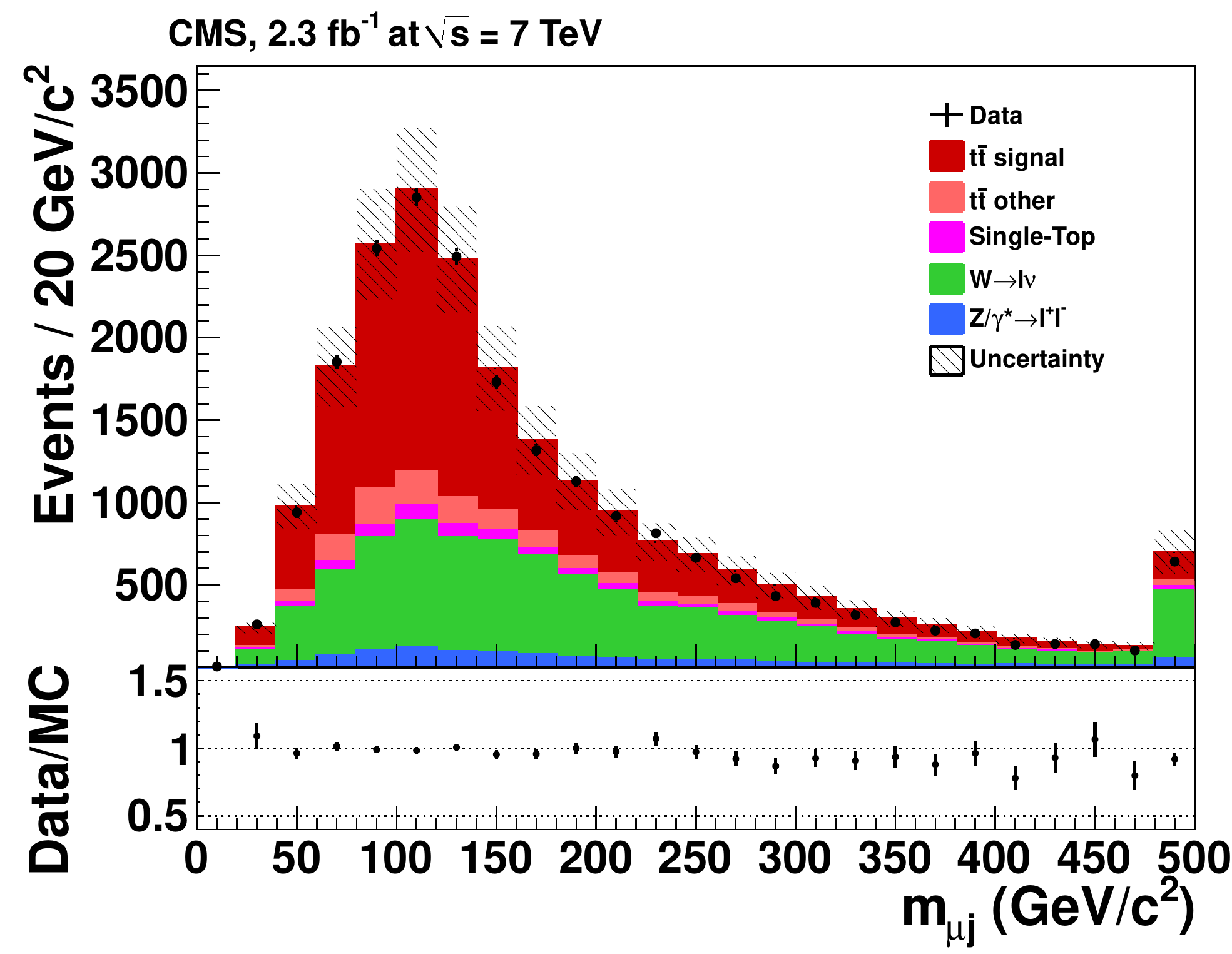}
\caption{
Distribution of the jet-muon mass
in the data (filled circles) and the simulation (solid histograms). The simulated distribution
is normalized to an integrated luminosity of 2.3\fbinv.
The component ``{\em \ttbar\ signal}" stands for the lepton+jet events.
The component ``{\em \ttbar\ other}" contains the events in all other decay channels.
The hatched area corresponds to the combined statistical and systematic uncertainty.
Overflow entries are added to the last bin.
}
\label{fig:mlbdata}
\end{center}
\end{figure}

A generic \cPqb-candidate sample is constructed by taking the
jet assigned to the lepton. This sample is further subdivided
into \cPqb-enriched and \cPqb-depleted subsamples by using the invariant mass of that jet and the muon (called the  jet-muon mass, $m_{\mu j}$). The distribution of this variable is shown in Fig.~\ref{fig:mlbdata}.
For the \cPqb-enriched subsample, the jet-muon mass is required to be in the range $80 < m_{\mu j} <  150 \GeVcc$. For the \cPqb-depleted subsample the jet-muon mass is required to be in the range $150 < m_{\mu j} <  250 \GeVcc$.
Based on the simulation, the purities of the two subsamples are 45\% and 16\%, respectively.

The distribution of the discriminators of the taggers for true b jets, $\hat{\Delta}_\cPqb^\mathrm{enr}$, is obtained by subtracting
the discriminator distribution of the \cPqb-depleted subsample, $\Delta_\cPqb^\mathrm{depl}$
from the discriminator distribution of the \cPqb-enriched subsample, $\Delta_\cPqb^\mathrm{enr}$:
\begin{equation}
\hat{\Delta}_\cPqb^\mathrm{enr} = \Delta_\cPqb^\mathrm{enr} - F \times \Delta_\cPqb^\mathrm{depl}\;.
\label{eq:subtr}
\end{equation}
The factor $F$ represents the ratio of the number of non-\cPqb\ jets in the \cPqb-enriched and \cPqb-depleted subsamples.
It is measured from a background dominated sample composed mainly of light-flavour quark jets.
This sample is obtained by using the jets attributed to the decay of the \PW\ boson and ensuring that they both fail the \bdtag\ requirements of the TCHEM operating point.
Both jets are used to construct a jet-muon mass distribution, and the same subsamples are defined as for the signal sample.
The purity of light-flavour quark jets is  92\% in the region $80 < m_{\mu j} <  150 \GeVcc$ and 95\% in the region $150 < m_{\mu j} <  250 \GeVcc$.
To match the shape of the jet-muon mass distribution of this background sample to that of the signal sample, the
 jets in the background sample are reweighted according to the (\pt, $\eta$) of the signal sample.
After this reweighting, the two samples have similar
jet-muon mass distribution.  The factor $F$ is taken as the ratio of the number of events in the $80  < m_{\mu j} <  150 \GeVcc$ and the  $150 < m_{\mu j} <  250 \GeVcc$ regions in the background sample, and is found to be $1.16 \pm 0.02$.

A small correlation between the jet-muon mass and the discriminators has to be corrected for.
This correlation is attributed to the correlation between the transverse momentum of the jet and the jet-muon mass. This correlation distorts the distribution of the discriminants of the \bdtag algorithms in the \cPqb-depleted subsample with respect to the distribution of the non-\cPqb\ jets in the \cPqb-enhanced subsample.
This effect is corrected by reweighting the jets in the \cPqb-depleted subsample according to the transverse momentum distribution of the jets in the  \cPqb-enhanced subsample.

The systematic uncertainties for the \bdtag efficiency and the scale factors are the absolute differences between the
 nominal simulation sample and the sample with modified parameters.
Additionally, a systematic uncertainty is assigned based on tests of the method in simulation. The tests show no bias in the method with an uncertainty driven by statistical uncertainties on available samples.
For the CSVL operating point, the relative uncertainty is 3.1\%.
The jet energy scale and resolution have a small contribution from the change in the mean masses and widths used for the $\chi^2$.
For the CSVL operating point, the relative uncertainties on the scale factor are  1.4\% and 2.2\%, respectively.
A small uncertainty of 0.5\% is due to the choice of the boundaries of the \cPqb-depleted region.
The  high tail of the jet-muon mass distribution is composed mainly of background events and wrongly combined jets that do not reflect the kinematics of the signal events.
The effect of imposing an upper limit on the region is assessed by varying the boundary between 200 and 300\GeVcc.

%% file: Results.tex
\section{Efficiency measurement results}\label{sec:Results}

\input{EfficienciesTT}

\input{resultsTT}

\subsection{Comparison of results}

The \pt-dependent scale factors
measured in dijet and multijet events have been compared by reweighted them to match the jet
\pt spectrum observed in \ttbar events.
The results are shown in Table~\ref{tab:qcdTTbar}
and are in good agreement with each other.
This justifies the assumption that the scale factors for the muon jets and inclusive jets are compatible.

\begin{table}[bth]
\begin{center}
\topcaption[]{The efficiency scale factors \SFb, and their uncertainties, obtained in multijet and \ttbar events for \cPqb\ jets in the expected \pt range of \ttbar events.
}\label{tab:qcdTTbar}
\begin{tabular}{lcc}
\hline
 \cPqb\ tagger & \SFb\ in multijet events & \SFb\ in \ttbar events \\ \hline
JPL     & $0.98 \pm 0.02$ & $0.97 \pm 0.03$ \\
JBPL    & $0.98 \pm 0.02$ & $0.98 \pm 0.03$ \\
TCHEL   & $0.98 \pm 0.02$ & $0.95 \pm 0.03$ \\
CSVL    & $0.99 \pm 0.02$ & $1.01 \pm 0.03$ \\
\hline
JPM     & $0.92 \pm 0.03$ & $0.95 \pm 0.03$ \\
JBPM    & $0.92 \pm 0.03$ & $0.94 \pm 0.03$ \\
TCHEM   & $0.95 \pm 0.03$ & $0.96 \pm 0.03$ \\
TCHPM   & $0.94 \pm 0.03$ & $0.93 \pm 0.03$ \\
SSVHEM  & $0.95 \pm 0.03$ & $0.96 \pm 0.03$ \\
CSVM    & $0.95 \pm 0.03$ & $0.97 \pm 0.03$ \\
\hline
JPT     & $0.87 \pm 0.04$ & $0.90 \pm 0.03$ \\
JBPT    & $0.87 \pm 0.05$ & $0.89 \pm 0.03$ \\
TCHPT   & $0.91 \pm 0.04$ & $0.93 \pm 0.03$ \\
SSVHPT  & $0.92 \pm 0.03$ & $0.95 \pm 0.03$ \\
CSVT    & $0.91 \pm 0.03$ & $0.96 \pm 0.03$ \\
\hline
\end{tabular}
\end{center}
\end{table}

%% file: EfficienciesTT.tex
\subsection{Results from multijet events}\label{sec:EfficienciesSum}

The methods described in Section~\ref{sec:EfficienciesMuonJets} cover a large range of jet transverse momenta.
The PtRel and the System8 methods provide precise measurements for the lower part of the spectrum. The IP3D and the LT methods have been designed for high jet \pt.
The measured data/MC scale factors are given in
 Table~\ref{tab:effSF80120_PtRelS8} for jets with low \pt\, from 80 to 120\GeVc, and in Table~\ref{tab:effSF80120_IP3DJP} for jets with high \pt, from 160 to 320\GeVc.
In these ranges the methods give compatible results within the quoted uncertainties.
While some of the methods measure the efficiencies and scale factors only for muon jets, and not inclusive \cPqb\ jets, simulation studies have shown that the difference in tagging efficiencies between the two are only a few percent.
We assume that these small differences have no significant effect on the scale factors, which are relative data/MC measurements.

\begin{table}[!htb]
\begin{center}
\topcaption{Data/MC scale factors \SFb\ as measured using the PtRel, System8, and LT
  methods and their combination. Results are given for jet \pt between 80 and
  120\GeVc.
  The first uncertainty on \SFb\  is statistical and the second is
  systematic. For the combination the total uncertainty is quoted.}
  \label{tab:effSF80120_PtRelS8}
\begin{tabular}{@{\extracolsep{\fill}}lcccc}
\hline
 \cPqb\ tagger & \SFb\  (PtRel) & \SFb\  (System8)  & \SFb\  (LT) & \SFb\ (comb.) \\ \hline
JPL    & $0.98 \pm 0.01 \pm 0.03$ & $1.00 \pm 0.02 \pm 0.07$ & $1.00 \pm 0.01 \pm 0.04$ & $0.99 \pm 0.03$  \\
JBPL   & $0.99 \pm 0.01 \pm 0.02$ & $0.98 \pm 0.02 \pm 0.04$ & $1.01 \pm 0.01 \pm 0.04$ & $0.99 \pm 0.02$  \\
TCHEL  & $0.99 \pm 0.01 \pm 0.02$ & $0.97 \pm 0.02 \pm 0.05$ & $1.00 \pm 0.01 \pm 0.01$ & $1.00 \pm 0.01$  \\
CSVL   & $1.00 \pm 0.01 \pm 0.02$ & $1.01 \pm 0.02 \pm 0.06$ & $0.98 \pm 0.01 \pm 0.02$ & $0.99 \pm 0.02$  \\
\hline
JPM    & $0.90 \pm 0.01 \pm 0.03$ & $0.93 \pm 0.03 \pm 0.06$ & $0.99 \pm 0.01 \pm 0.05$ & $0.92 \pm 0.03$  \\
JBPM   & $0.92 \pm 0.01 \pm 0.02$ & $0.96 \pm 0.03 \pm 0.08$ & $0.99 \pm 0.01 \pm 0.05$ & $0.91 \pm 0.03$  \\
TCHEM  & $0.94 \pm 0.01 \pm 0.03$ & $0.99 \pm 0.03 \pm 0.07$ & $0.98 \pm 0.01 \pm 0.03$ & $0.95 \pm 0.02$  \\
TCHPM  & $0.95 \pm 0.01 \pm 0.03$ & $0.94 \pm 0.02 \pm 0.09$ & $0.97 \pm 0.01 \pm 0.02$ & $0.96 \pm 0.02$  \\
SSVHEM & $0.92 \pm 0.01 \pm 0.02$ & $0.92 \pm 0.03 \pm 0.05$ & $0.97 \pm 0.01 \pm 0.02$ & $0.95 \pm 0.02$  \\
CSVM   & $0.93 \pm 0.01 \pm 0.02$ & $0.97 \pm 0.03 \pm 0.06$ & $0.97 \pm 0.01 \pm 0.03$ & $0.95 \pm 0.02$  \\
\hline
JPT    & $0.82 \pm 0.01 \pm 0.05$ & $0.85 \pm 0.03 \pm 0.07$ & $0.96 \pm 0.01 \pm 0.07$ & $0.87 \pm 0.05$  \\
JBPT   & $0.83 \pm 0.01 \pm 0.06$ & $0.89 \pm 0.03 \pm 0.11$ & $0.96 \pm 0.01 \pm 0.08$ & $0.87 \pm 0.06$  \\
TCHPT  & $0.87 \pm 0.01 \pm 0.05$ & $0.91 \pm 0.03 \pm 0.10$ & $0.94 \pm 0.01 \pm 0.04$ & $0.91 \pm 0.04$  \\
SSVHPT & $0.87 \pm 0.01 \pm 0.03$ & $0.84 \pm 0.03 \pm 0.10$ & $0.96 \pm 0.01 \pm 0.03$ & $0.92 \pm 0.03$  \\
CSVT   & $0.86 \pm 0.01 \pm 0.04$ & $0.92 \pm 0.03 \pm 0.07$ & $0.94 \pm 0.01 \pm 0.04$ & $0.90 \pm 0.03$  \\
\hline
\end{tabular}
\end{center}
\end{table}

\begin{table}[bth]
\begin{center}
\topcaption{Data/MC scale factors \SFb\ as measured using the IP3D and LT methods and their combination.
Results are given for jet \pt between 160 and 320\GeVc. The first uncertainty on \SFb\  is statistical and the second is systematic. For the combination the total uncertainty is quoted.}
  \label{tab:effSF80120_IP3DJP}
\begin{tabular}{@{\extracolsep{\fill}}lccc}
\hline
 \cPqb\ tagger & \SFb\  (IP3D) & \SFb\  (LT) & \SFb\ (comb.) \\ \hline
JPL    & $0.99 \pm 0.01 \pm 0.01$ & $1.00 \pm 0.01 \pm 0.06$ & $0.99 \pm 0.02$  \\
JBPL   & $1.00 \pm 0.01 \pm 0.01$ & $1.00 \pm 0.01 \pm 0.05$ & $1.00 \pm 0.02$  \\
TCHEL  & $1.00 \pm 0.01 \pm 0.02$ & $1.00 \pm 0.01 \pm 0.02$ & $1.00 \pm 0.02$  \\
CSVL   & $0.98 \pm 0.02 \pm 0.01$ & $0.96 \pm 0.01 \pm 0.04$ & $0.97 \pm 0.03$  \\
\hline
JPM    & $0.93 \pm 0.02 \pm 0.03$ & $0.99 \pm 0.01 \pm 0.06$ & $0.95 \pm 0.04$  \\
JBPM   & $0.97 \pm 0.02 \pm 0.03$ & $0.99 \pm 0.01 \pm 0.06$ & $0.97 \pm 0.04$  \\
TCHEM  & $0.96 \pm 0.02 \pm 0.03$ & $0.97 \pm 0.01 \pm 0.03$ & $0.96 \pm 0.03$  \\
TCHPM  & $0.97 \pm 0.02 \pm 0.03$ & $0.97 \pm 0.01 \pm 0.02$ & $0.97 \pm 0.02$  \\
SSVHEM & $0.98 \pm 0.02 \pm 0.04$ & $0.98 \pm 0.01 \pm 0.07$ & $0.98 \pm 0.04$  \\
CSVM   & $0.95 \pm 0.02 \pm 0.04$ & $0.97 \pm 0.01 \pm 0.06$ & $0.96 \pm 0.04$  \\
\hline
JPT    & $0.89 \pm 0.02 \pm 0.04$ & $0.95 \pm 0.01 \pm 0.10$ & $0.91 \pm 0.05$  \\
JBPT   & $0.91 \pm 0.02 \pm 0.03$ & $0.96 \pm 0.01 \pm 0.11$ & $0.92 \pm 0.05$  \\
TCHPT  & $0.89 \pm 0.02 \pm 0.03$ & $0.94 \pm 0.01 \pm 0.04$ & $0.92 \pm 0.04$  \\
SSVHPT & $0.92 \pm 0.02 \pm 0.04$ & $0.96 \pm 0.01 \pm 0.05$ & $0.94 \pm 0.04$  \\
CSVT   & $0.90 \pm 0.02 \pm 0.07$ & $0.94 \pm 0.01 \pm 0.09$ & $0.92 \pm 0.07$  \\
\hline
\end{tabular}
\end{center}
\end{table}

The results have been combined to provide the best measurements of the data/MC scale factors for $30 < \pt < 670 \GeVc$.  For each jet
\pt range the most precise results have been used: the PtRel and System8
methods for $\pt < 120 \GeVc$, the IP3D method for $\pt > 120 \GeVc$ and the LT
method for the full momentum range.

The combination is based on a weighted mean of the scale factors in each
jet \pt\ bin~\cite{BLUE}.
However, there are a significant number of jets from QCD dijet and multijet events (with at least one muon associated to a jet) which are shared between the methods.
The shared fraction of jets varies with jet \pt.
Typical values are 10--25\% between the LT and PtRel/IP3D methods, 40--50\% between the PtRel and System8 methods, and 20--50\% between the System8 and LT methods.
This overlap has been taken into account in the combination.

\begin{sidewaysfigure}[p]
\begin{center}
\includegraphics[width=0.49\textwidth]{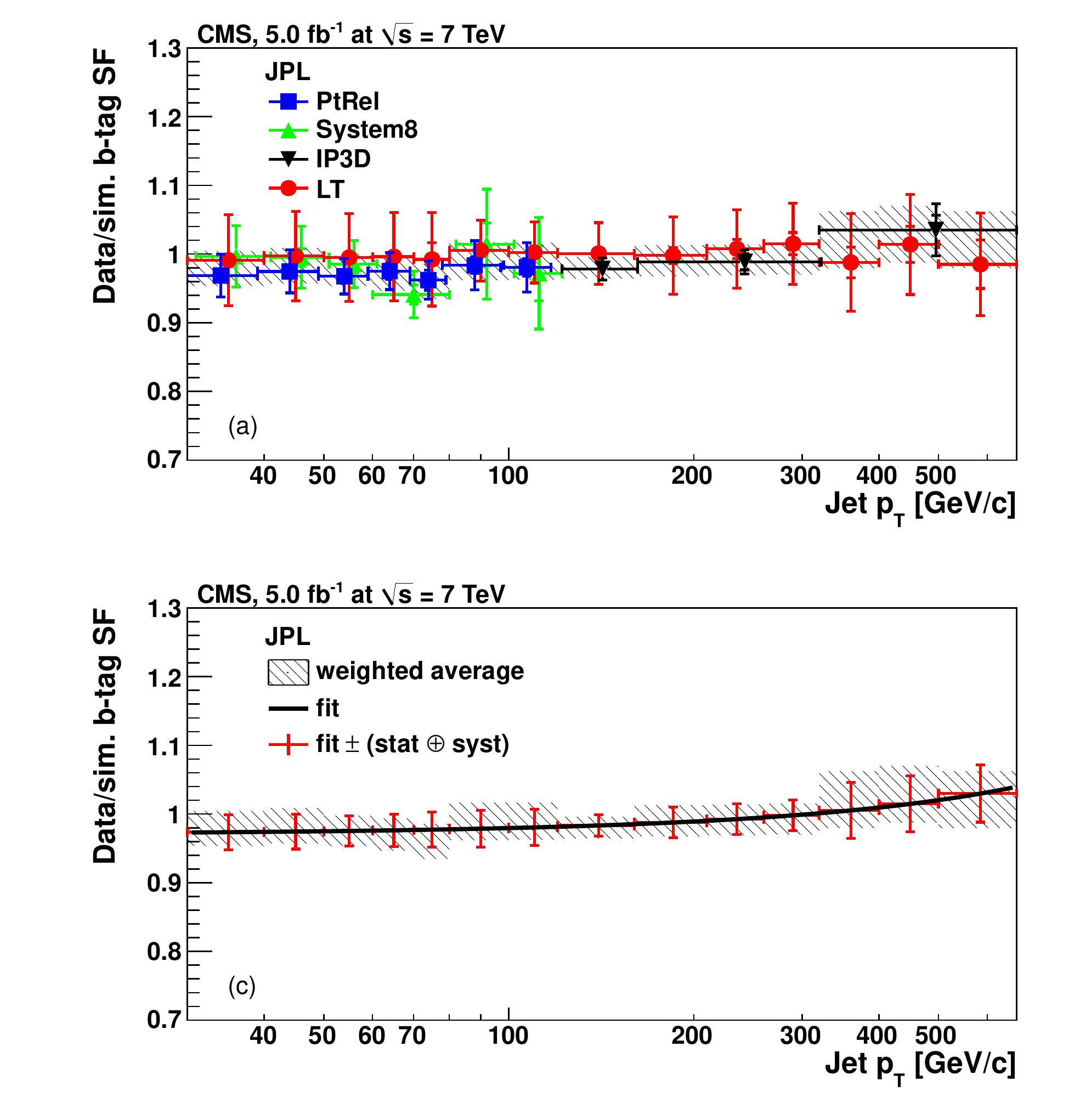} \hfil
\includegraphics[width=0.49\textwidth]{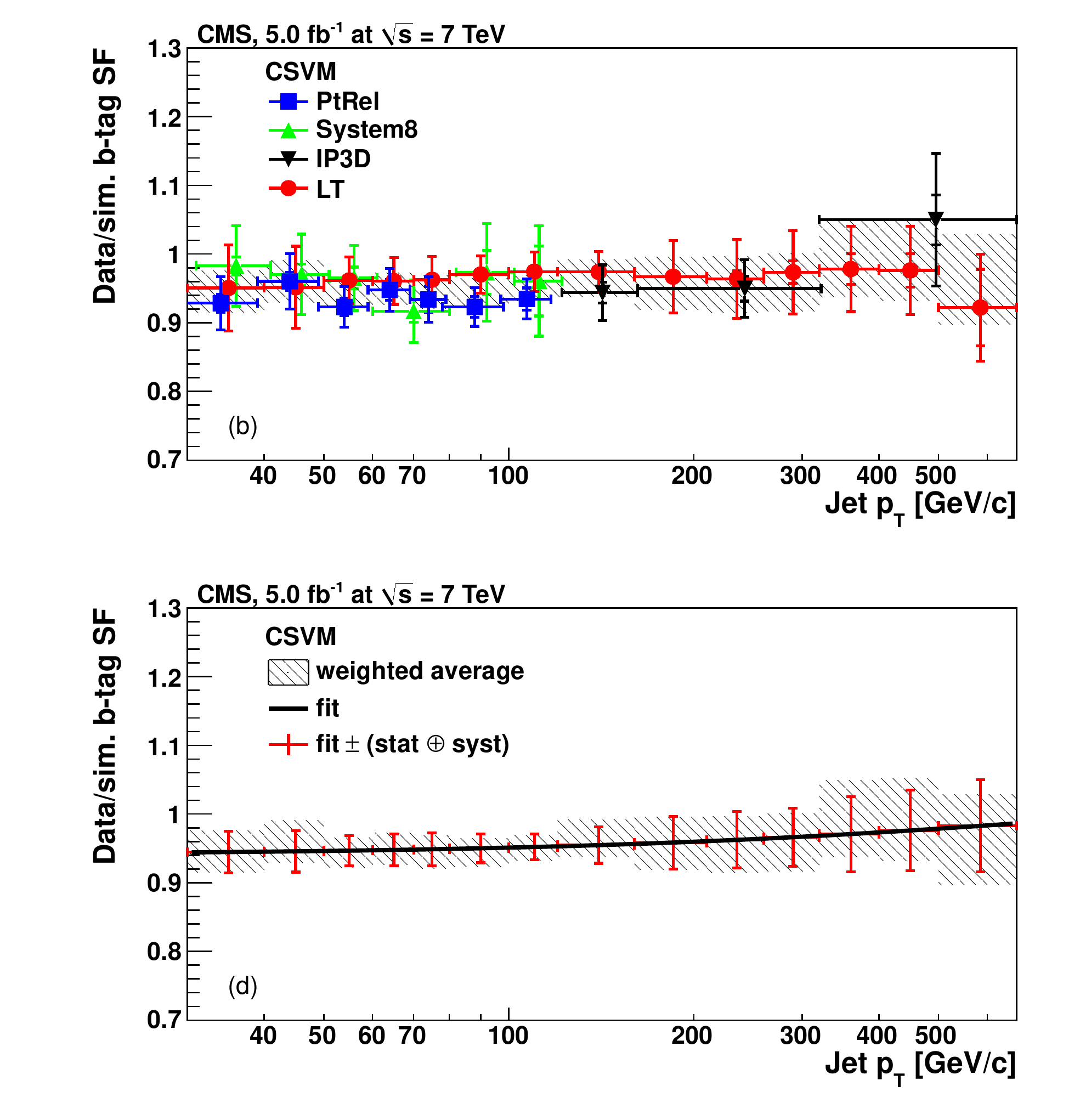}
\caption{ Individual and combined measurements of the ratio of the \bdtag efficiencies of the data to that in simulation for the (a, c) JPL and (b, d) CSVM taggers.
  Figures (a) and (b) show the individual measurements from the muon \ptrel\
  (``PtRel''), System8 (``System8''), muon IP (``IP3D''), and lifetime
  tagger (``LT'') methods.  The inner and outer error bars indicate the statistical and the combined uncertainties, respectively.
  The hatched areas represent the combined measurements.
  In figures (c) and (d) the combined measurements have been parameterized by
  functions of the form $\SFb(\pt) = \alpha
  (1+\beta\pt)/(1+\gamma\pt)$.  The error bars attached to the function
  have the same size as the uncertainties from the combined measurement in
  each bin.  }\label{fig:combinedEfficiencies}
\end{center}
\end{sidewaysfigure}

Several sources of systematic uncertainties are common for all methods: the
effects due to pileup, gluon splitting, and the selection criteria for
muons.  The muon PtRel and IP3D methods have the same sensitivity to the
choice of the away-jet tagger.  The corresponding uncertainties were
assumed to be fully correlated or anticorrelated according to the sign of
the variations observed for the different methods.  All other systematic
effects are specific to individual methods and have been treated as
uncorrelated.  A conservative value for the uncertainty is used if the
$\chi^{2}$ from the fit exceeds the number of degrees of freedom, in which
case the uncertainty is scaled by the square root of the normalised
$\chi^{2}$.  Summaries for the individual and combined scale factor
measurements for the JPL and the CSVM taggers are shown in
Fig.~\ref{fig:combinedEfficiencies}. Also shown are the parameterizations
of the combined scale factor of the form $\SFb(\pt) = \alpha
(1+\beta\pt)/(1+\gamma\pt)$.  Combined values for a low and a high jet \pt\
range are shown in the right hand columns of
Tables~\ref{tab:effSF80120_PtRelS8} and \ref{tab:effSF80120_IP3DJP},
respectively.  
The same studies have been been performed separately for
muon jets with $|\eta|<1.2$ and $1.2<|\eta|<2.4$. Compatible scale factors
values are obtained in both regions.

%% file: resultsTT.tex
\subsection {Results from  \texorpdfstring{\ttbar}{t t-bar} events}
\label{sec:resultsTT}

The statistical properties of the four methods presented in
Sections~\ref{sec:plr} to~\ref{sec:method-b-sample} have been studied using
ensembles of pseudo-experiments based on the expected numbers of signal and
background events.  The distributions of the estimated
values and their uncertainties show that the methods are unbiased.
This is shown by the pull distributions, which have mean values close to zero and standard deviations
close to one.

\begin{figure}[htbp]
\centering
\subfigure{\includegraphics[width=0.6\textwidth]{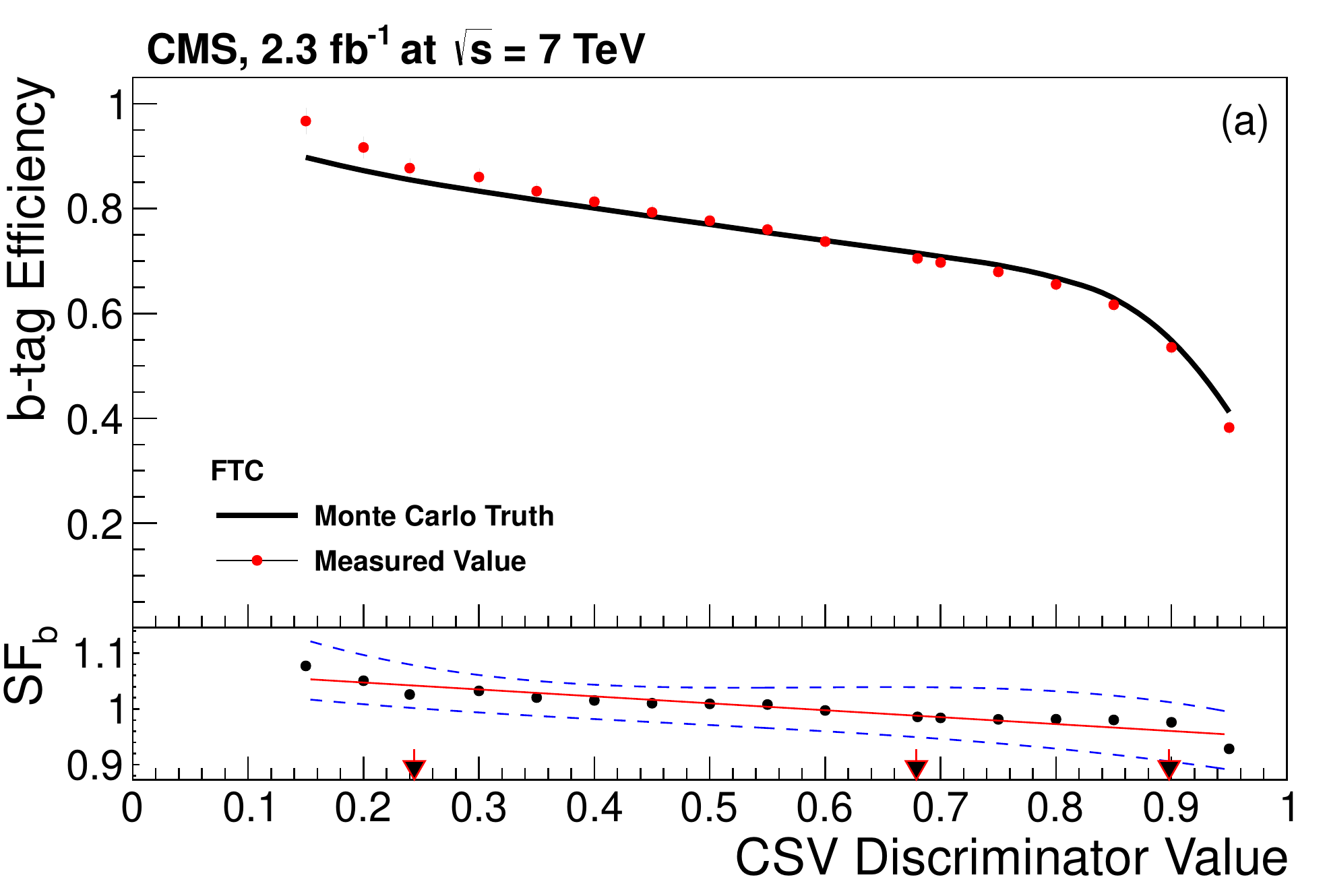}} \\
\subfigure{\includegraphics[width=0.6\textwidth]{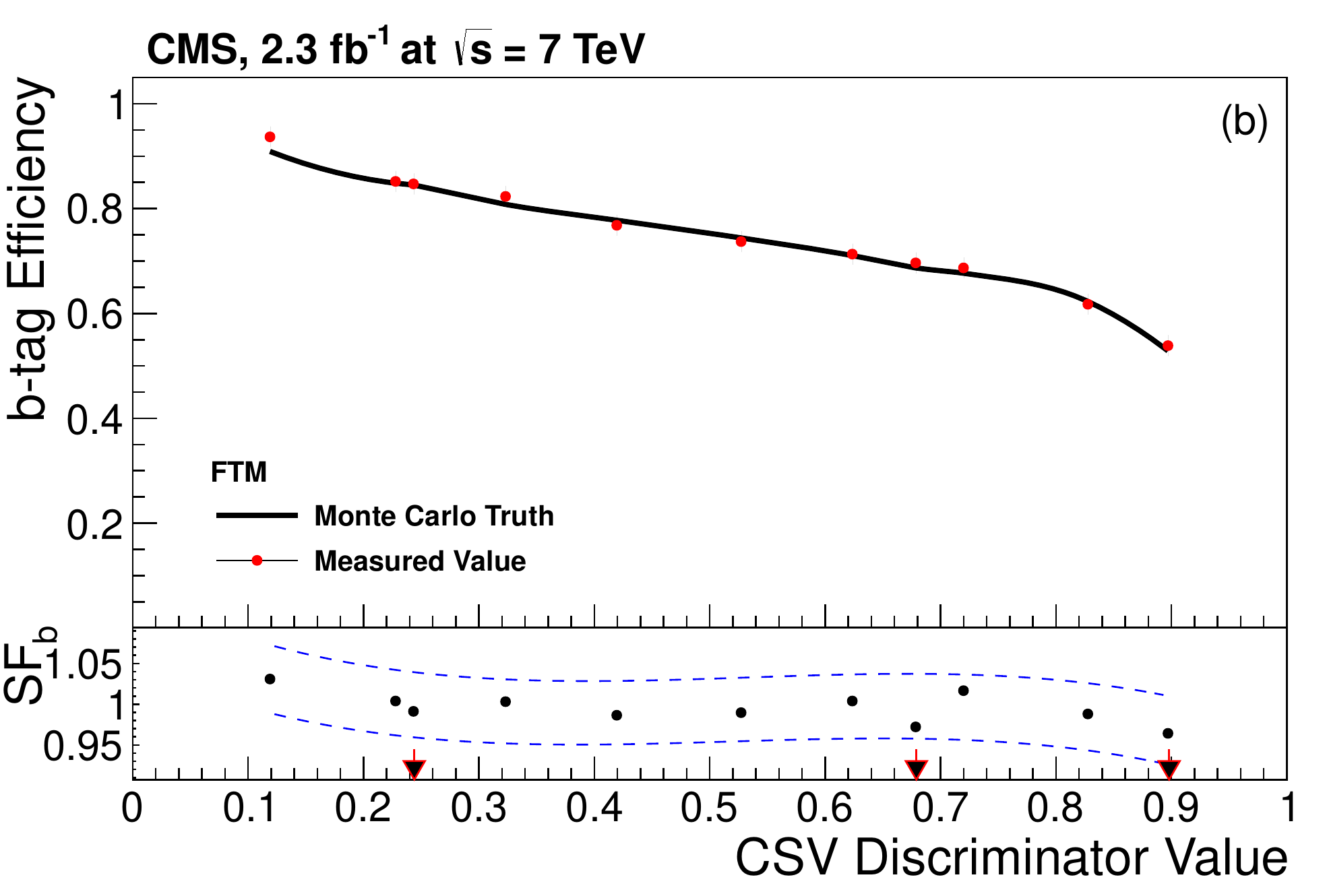}} \\
\subfigure{\includegraphics[width=0.6\textwidth]{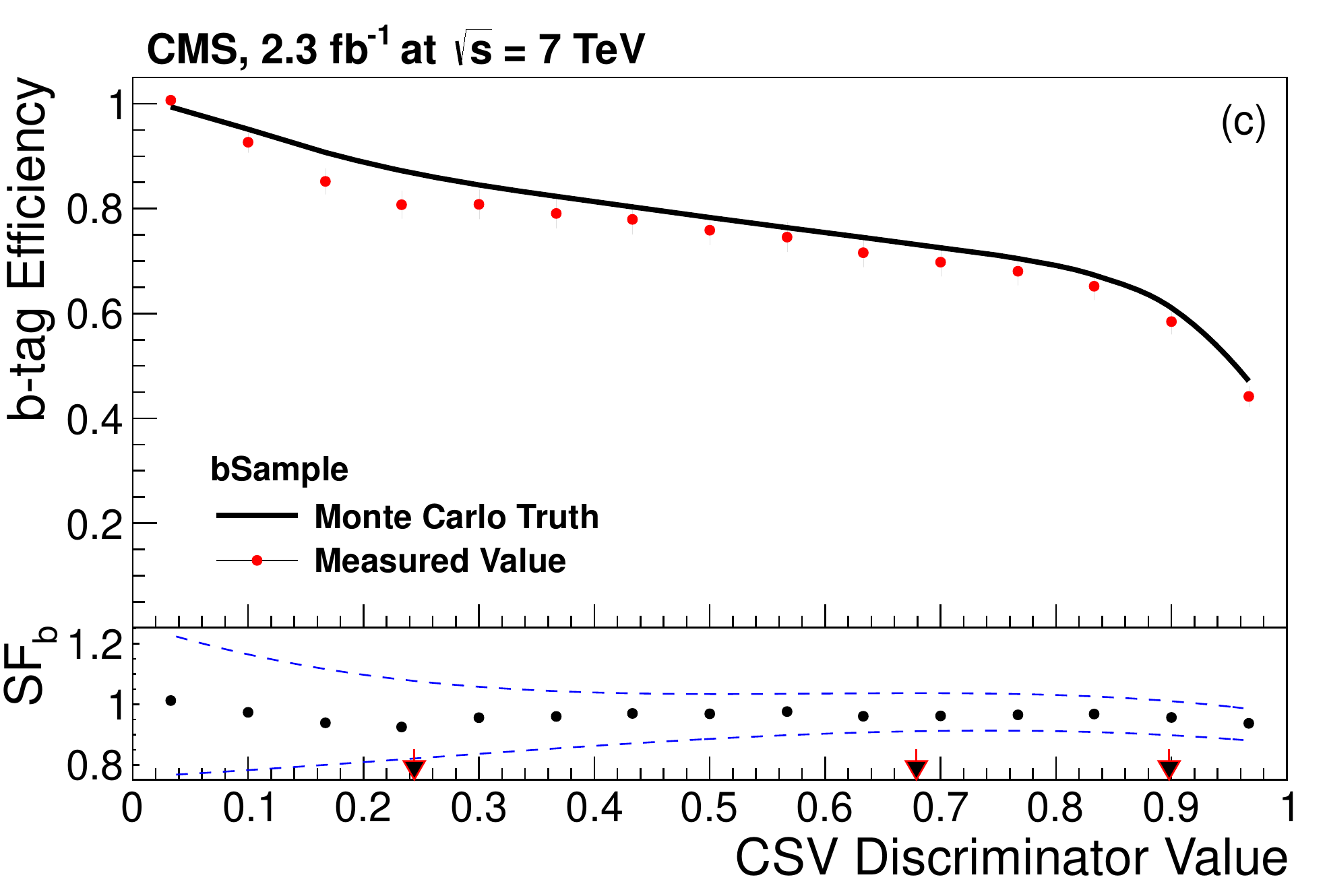}}
\caption{ Measured \cPqb-jet tagging efficiency as a function of the flavour
  discriminator threshold for the CSV algorithm, measured with the (a)
  \FTCm, (b) \FTMm and (c) bSample method.  The absolute \cPqb-jet tagging
  efficiencies measured from data and predicted from simulation are shown in
  the upper histograms of each panel.  The scale factors \SFb\ are shown in
  the lower histogram, where the blue dashed lines represent the
  combined statistical and systematic uncertainty.
  The arrows indicate the standard
  operating points.  For the \FTCm, the red line represents a linear
  function fitted on the distribution of the scale factors.  }
\label{fig:sfCVS}
\end{figure}

The scale factors
$\SFb=\varepsilon_\cPqb^\mathrm{meas}/\varepsilon_\cPqb^\mathrm{MC}$
measured with the different algorithms are shown in Table~\ref{tab:allSFresults} using data with an integrated luminosity of 2.3\fbinv.
The scale factors were stable over the whole data-taking period and can be applied to
the full dataset.  The measured \bdtag efficiencies and scale factors for the CSV
algorithm are shown in Fig.~\ref{fig:sfCVS}.

The PLR and FTC methods are used to calculate a combined scale factor for use in analyses, by taking the weighted mean of the scale factors from each method. The two
methods are chosen because each has the smallest uncertainty among the
analyses in its respective decay channel.  By choosing one analysis in the
dilepton channel and one in the lepton+jets channel, there is no
statistical correlation between the two measurements as the samples are
mutually exclusive.  Based on the statistical and systematic uncertainties
of the PLR and FTC methods, the uncertainty of the resulting scale factor
is ${\pm}0.03$ for all operating points.

\begin{figure}[hbtp]
\begin{center}
\includegraphics[width=0.85\textwidth]{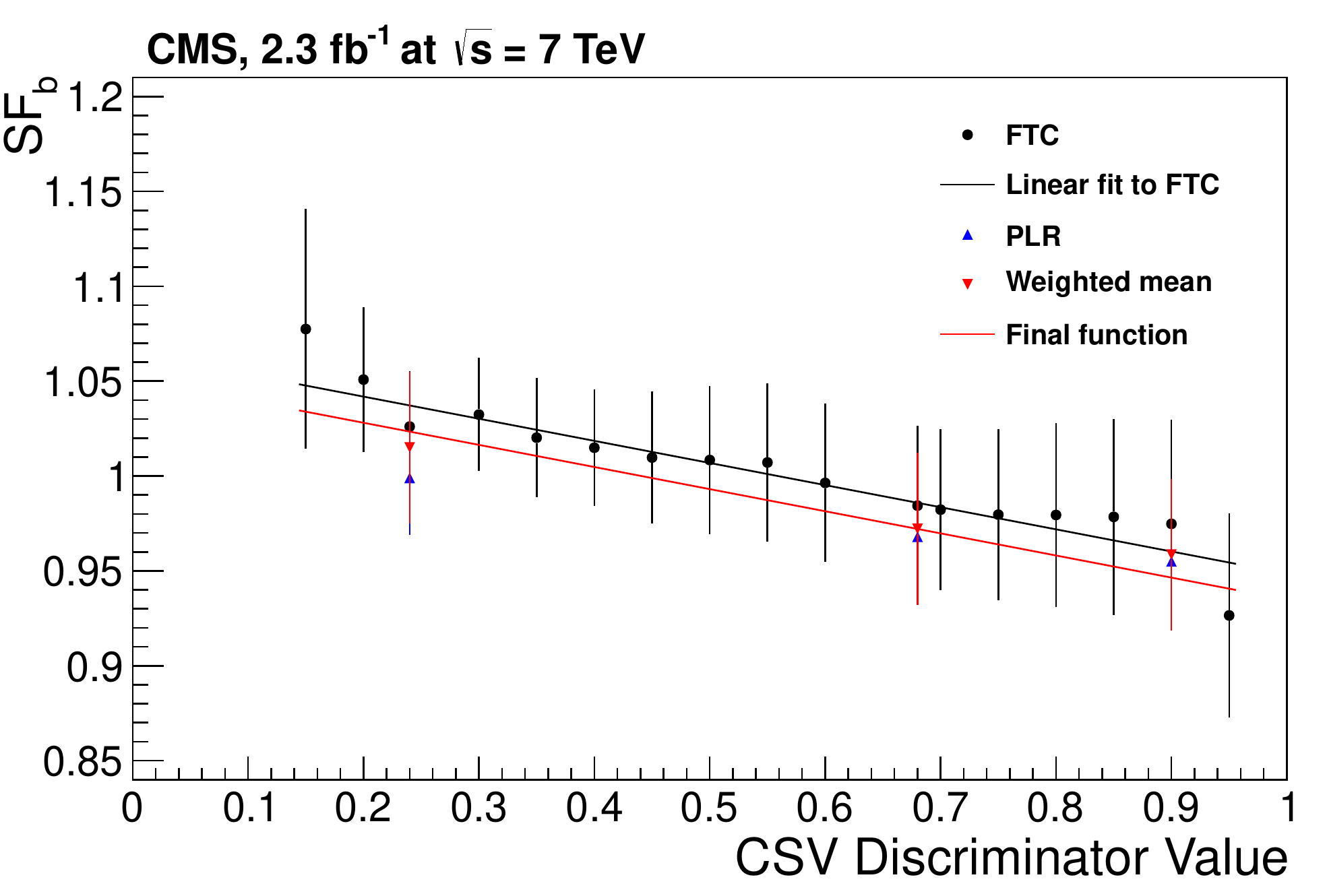}
\caption{
Scale factors measured with the PLR and FTC methods and weighted mean as a
 function of the discriminator threshold for the CSV algorithm.
The black function is derived from a fit to the values measured with the \FTCm. The red function labelled ``Final function" corresponds to the same function offset vertically to match the weighted mean of the medium operating point.
The uncertainties are the combined statistical and systematic uncertainty.
}
\label{fig:sfCVSfinal}
\end{center}
\end{figure}

A continuous function for the scale factors is required in physics analyses that use \bdtag discriminators with multivariate methods.
The function is obtained from a linear fit to the distribution of the scale factors measured with the FTC method. This is offset vertically to match the weighted mean of the medium operating point, as illustrated in Fig.~\ref{fig:sfCVSfinal}.

\begin{table}[tbp]
\begin{center}
\topcaption{The scale factors \SFb\ as measured using the PLR, FTC, FTM and bSample methods, and the weighted mean (WM).
The uncertainties are the combined statistical and systematic uncertainty.}
\label{tab:allSFresults}
\begin{tabular}{lccccc}
\hline
b tagger       &  PLR 	   & FTC	     & FTM 	       & bSample 	  & WM \\
\hline
JPL     &  0.96 $\pm$ 0.03 & 0.99 $\pm$ 0.03 & 0.96 $\pm$ 0.03 & 0.96 $\pm$ 0.05  & 0.97 $\pm$ 0.03 \\
JBPL    &  0.97 $\pm$ 0.03 & 1.00 $\pm$ 0.05 & 0.96 $\pm$ 0.03 & 0.97 $\pm$ 0.05  & 0.98 $\pm$ 0.03 \\
TCHEL	&  0.96 $\pm$ 0.03 & 0.94 $\pm$ 0.04 & 0.95 $\pm$ 0.03 & 0.92 $\pm$ 0.05  & 0.95 $\pm$ 0.03 \\
CSVL    &  1.00 $\pm$ 0.03 & 1.03 $\pm$ 0.03 & 0.99 $\pm$ 0.04 & 0.97 $\pm$ 0.05  & 1.01 $\pm$ 0.03 \\
\hline
JPM     &  0.95 $\pm$ 0.03 & 0.95 $\pm$ 0.04 & 0.93 $\pm$ 0.03 & 0.94 $\pm$ 0.06  & 0.95 $\pm$ 0.03 \\
JBPM    &  0.93 $\pm$ 0.03 & 0.95 $\pm$ 0.04 & 0.91 $\pm$ 0.03 & 0.93 $\pm$ 0.06  & 0.94 $\pm$ 0.03 \\
TCHEM	&  0.96 $\pm$ 0.03 & 0.97 $\pm$ 0.04 & 0.96 $\pm$ 0.03 & 0.93 $\pm$ 0.06  & 0.96 $\pm$ 0.03 \\
TCHPM	&  0.94 $\pm$ 0.03 & 0.93 $\pm$ 0.04 & 0.95 $\pm$ 0.04 & 0.92 $\pm$ 0.06  & 0.93 $\pm$ 0.03 \\
SSVHEM	&  0.95 $\pm$ 0.03 & 0.98 $\pm$ 0.04 & 0.98 $\pm$ 0.04 & 0.95 $\pm$ 0.07  & 0.96 $\pm$ 0.03 \\
CSVM    &  0.97 $\pm$ 0.03 & 0.98 $\pm$ 0.04 & 0.97 $\pm$ 0.04 & 0.95 $\pm$ 0.06  & 0.97 $\pm$ 0.03 \\
\hline
JPT     &  0.90 $\pm$ 0.03 & 0.89 $\pm$ 0.05 & 0.90 $\pm$ 0.03 & 0.95 $\pm$ 0.07  & 0.90 $\pm$ 0.03 \\
JBPT    &  0.90 $\pm$ 0.03 & 0.88 $\pm$ 0.05 & 0.90 $\pm$ 0.03 & 0.90 $\pm$ 0.07  & 0.89 $\pm$ 0.03 \\
TCHPT	&  0.93 $\pm$ 0.03 & 0.92 $\pm$ 0.05 & 0.94 $\pm$ 0.04 & 0.91 $\pm$ 0.07  & 0.93 $\pm$ 0.03 \\
SSVHPT	&  0.95 $\pm$ 0.03 & 0.95 $\pm$ 0.04 & 0.96 $\pm$ 0.04 & 0.89 $\pm$ 0.07  & 0.95 $\pm$ 0.03 \\
CSVT    &  0.95 $\pm$ 0.03 & 0.97 $\pm$ 0.05 & 0.96 $\pm$ 0.03 & 0.95 $\pm$ 0.06  & 0.96 $\pm$ 0.03 \\
\hline
\end{tabular}
\end{center}
\end{table}

%% file: Mistags.tex
\section{Misidentification probability measurement}\label{sec:Mistags}

The measurement of the misidentification probability for
light-parton jets relies on the definition of inverted tagging
algorithms, selecting non-\cPqb\ jets using the same variables and
techniques as the standard versions.  These ``negative taggers" can be
used in the same way as the regular \cPqb-jet tagging algorithms both in data and
in the simulation.  As the negative-tagged jets are enriched in light
flavours, the misidentification probability can be measured from data, with the
simulation used to extract a correction factor.

\begin{figure}[hbtp]
  \begin{center}
\includegraphics[width=.4\textwidth]{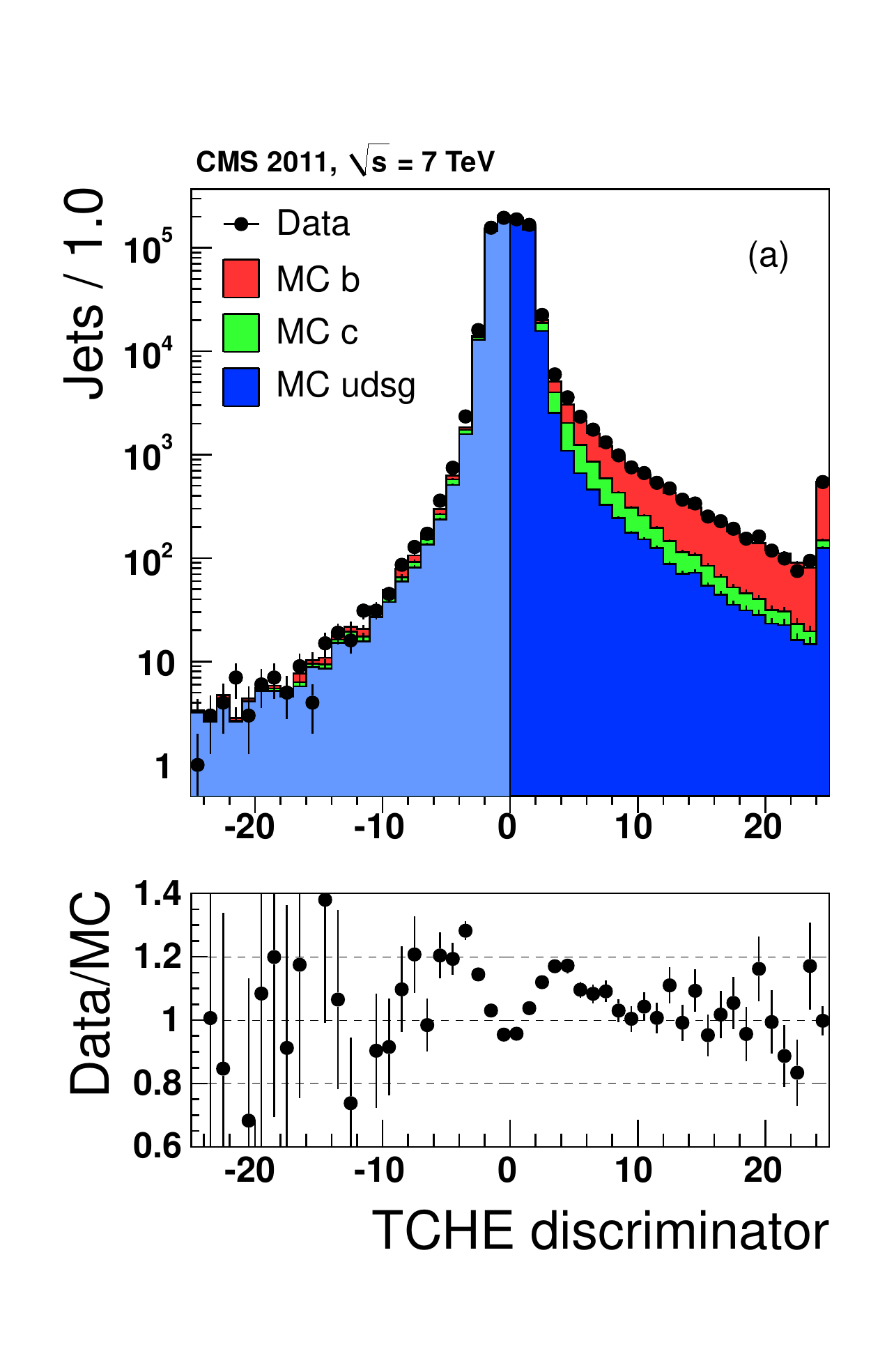} \hfil
\includegraphics[width=.4\textwidth]{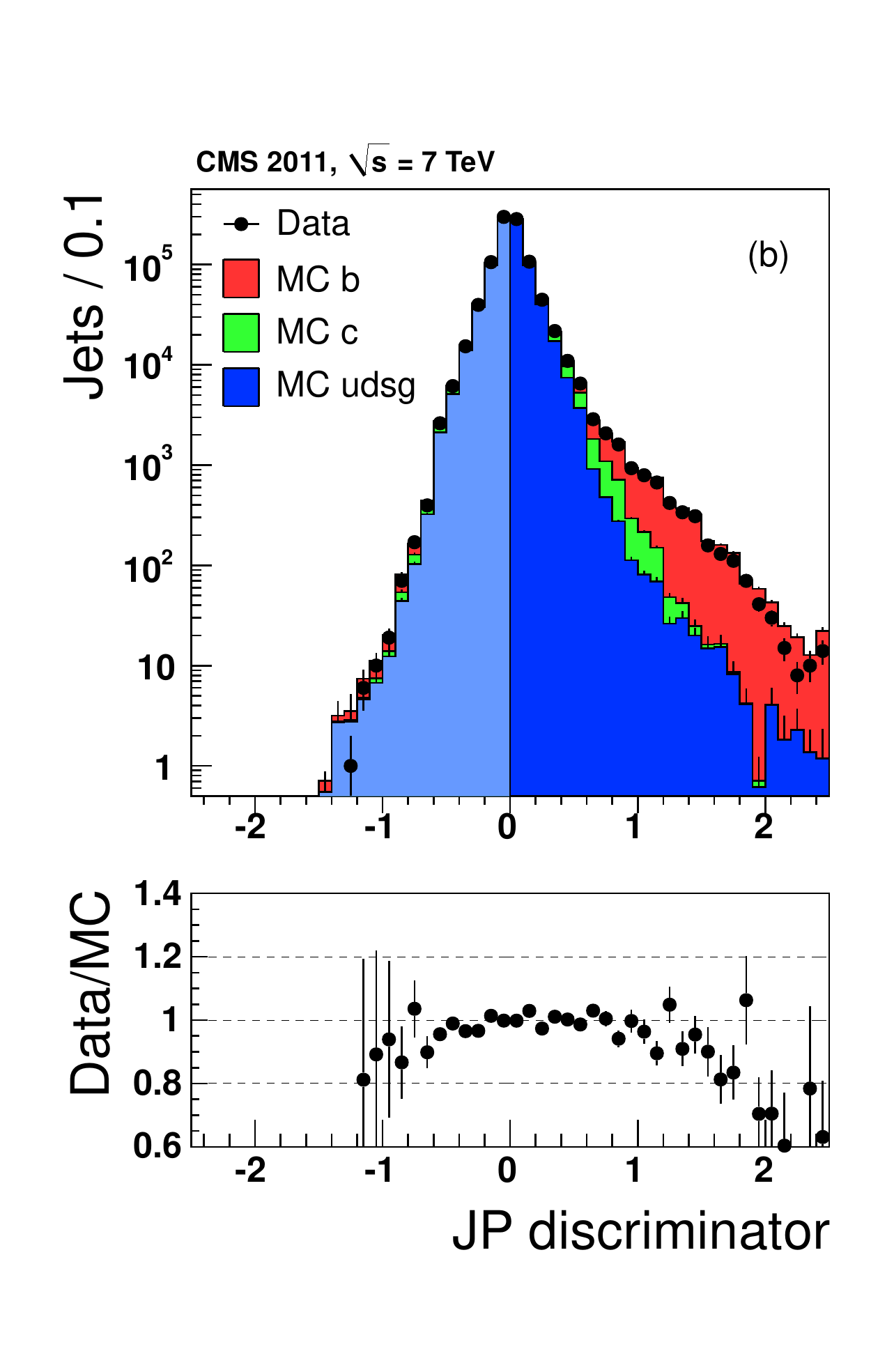} \\
\includegraphics[width=.4\textwidth]{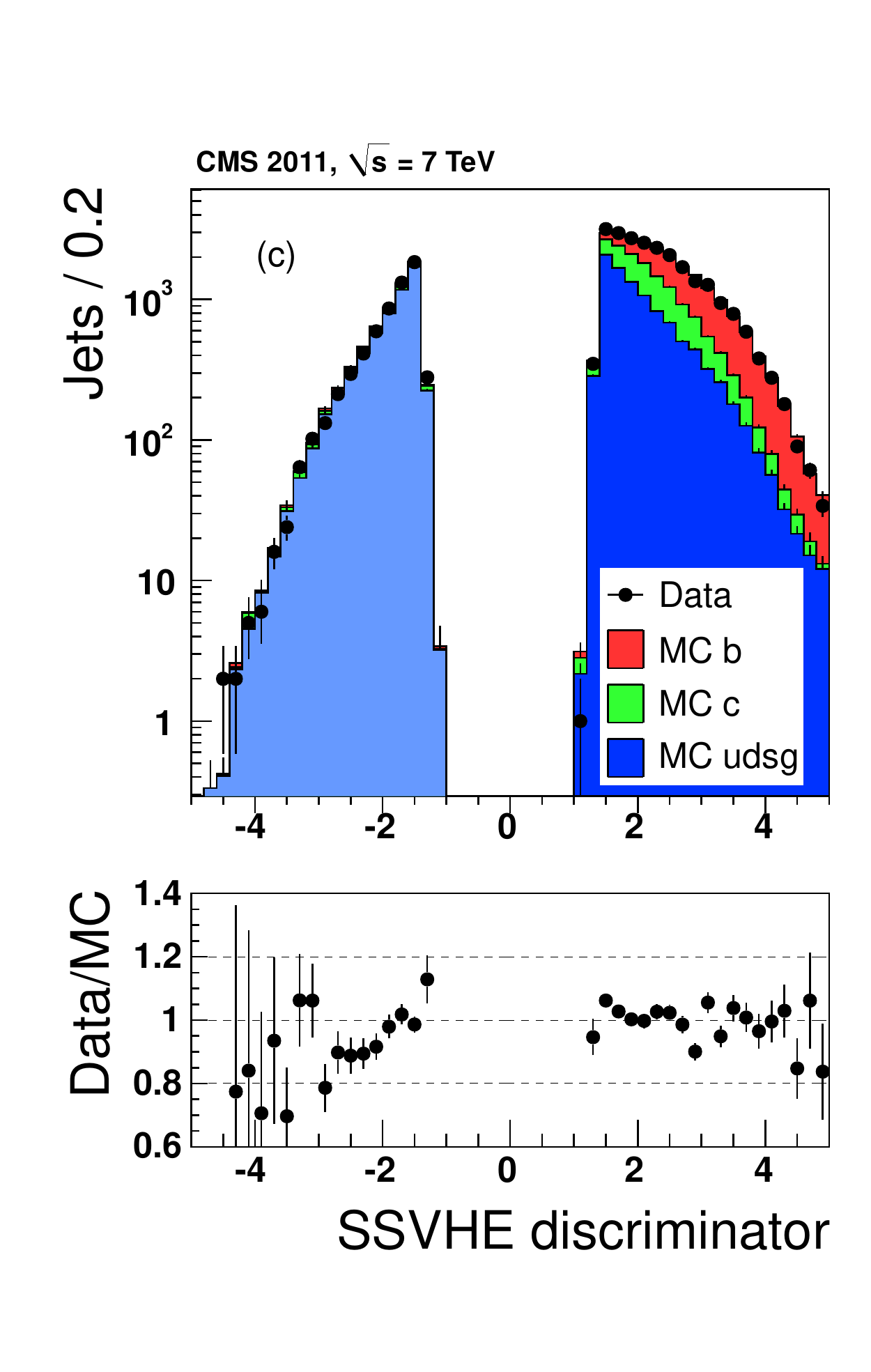} \hfil
\includegraphics[width=.4\textwidth]{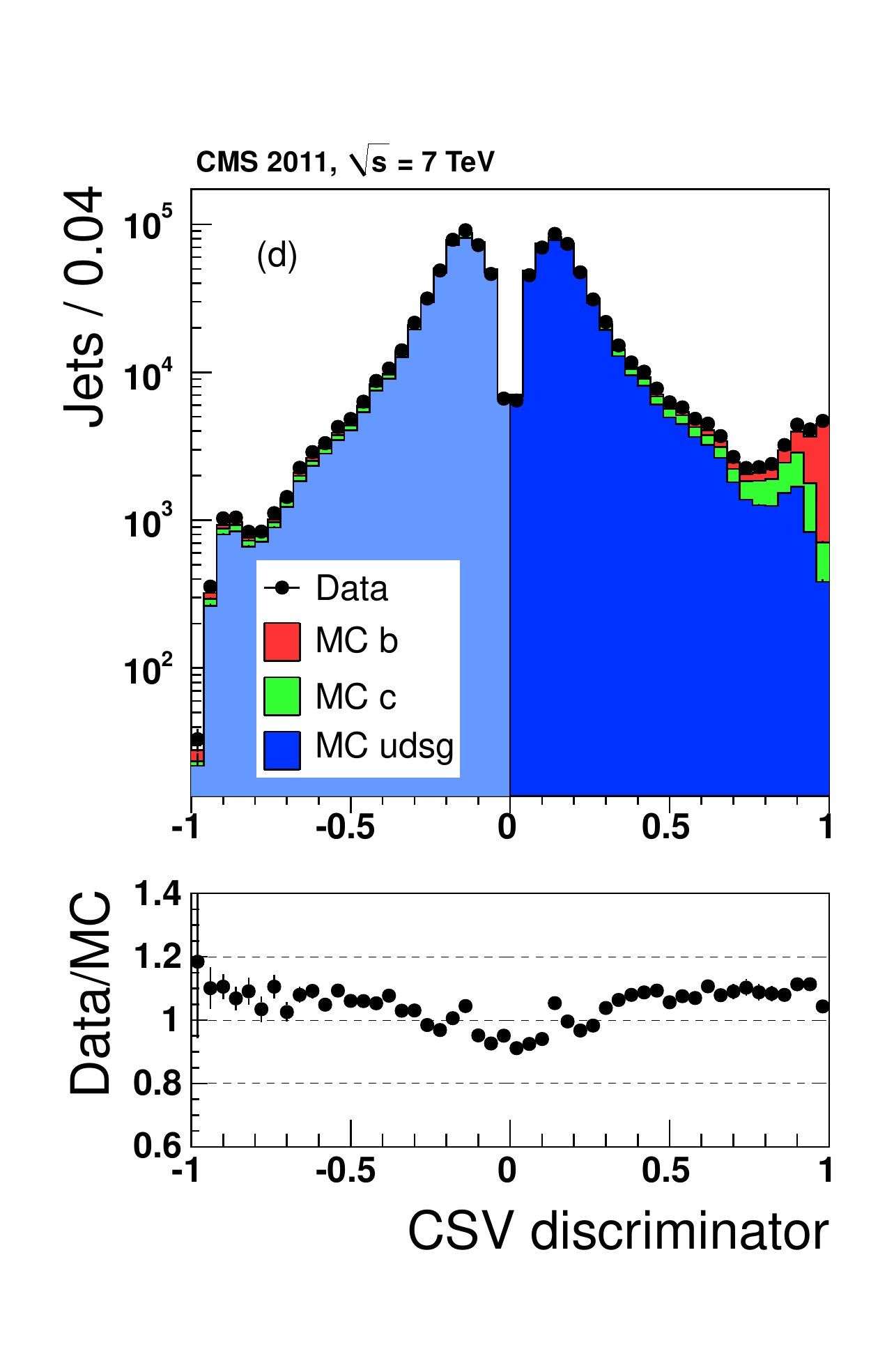}
     \caption{Signed \cPqb-jet tagging discriminators in data (dots) and simulation for
       light-parton jets (blue histogram, with a lighter colour for the
       negative discriminators), \cPqc\ jets (green histogram), and \cPqb\ jets (red histogram)
       for the (a) TCHE, (b) JP, (c) SSVHE, and (d) CSV algorithms.
       A jet-trigger \pt threshold of 30\GeVc is required for both data and
       simulation.  The simulation is normalized to the number of entries in the data.
       Underflow and overflow entries are added to the first and last
       bins, respectively. }
    \label{fig:taggers_jet30}
  \end{center}
\end{figure}

The misidentification probability is evaluated from tracks with a negative impact
parameter or from secondary vertices with a negative decay length (see
Section~\ref{sec:Algorithms}).  When a negative tagger is applied to jets of
any flavour, the corresponding tagging efficiency is denoted ``negative tag
rate".  The negative and positive \cPqb-jet tagging discriminator distributions
in data are compared with the simulation in Fig.~\ref{fig:taggers_jet30}.
The events are selected by requiring jet triggers with a \pt threshold of
30\GeVc, corresponding to an average \pt over all jets in the events of
44\GeVc.
For all \cPqb-jet tagging algorithms, the data and simulation are
found to be in agreement to within about $\pm 20\%$.  Similar results are
found for a sample of events selected by requiring jet triggers with a \pt
threshold of 300\GeVc, in which the average \pt is 213\GeVc.
Depending on the prescales applied, the data correspond to an integrated luminosity of up to 5.0\fbinv.

The misidentification probability is evaluated as:
\begin{equation}
 \varepsilon^\mathrm{misid}_\mathrm{data} = \varepsilon^-_\mathrm{data} \cdot R_\mathrm{light}\;, \label{eq:mistag}
\end{equation}
where $\varepsilon^-_\mathrm{data}$ is the negative tag rate as measured in jet data,
defined as the fraction of jets that are negatively tagged. $R_\mathrm{light} =
\varepsilon^\mathrm{misid}_\mathrm{MC} / \varepsilon^-_\mathrm{MC}$, a correction factor taken
from simulation, is the ratio of the misidentification probability for light-parton jets to
the negative tag rate for jets of all flavours in the simulation.

The rate $\varepsilon^-_\mathrm{data}$ depends
on the numbers of \cPqc\ and \cPqb\ quarks in the negative-tagged jets (which
tend to decrease $R_\mathrm{light}$), on the residual differences between light-flavour quark and gluon jets, the number of tracks from other displaced processes (such as \PKzS\ and
\PgL\  decays, and interactions in the detector material), and mismeasured
tracks (which tend to increase $R_\mathrm{light}$).  Due to
these contributions the simulation predicts ranges of
$R_\mathrm{light}$, for the different algorithms and jet \pt\ values, of
about 1.1 to~1.4, 1 to~2, and 1 to~4, for the loose, medium, and tight
operating points, respectively.

To compare the measured misidentification probability to that predicted by the simulation,
a scale factor $SF_\mathrm{light}$ is defined:
\begin{equation}
SF_\mathrm{light} = \varepsilon^\mathrm{misid}_\mathrm{data} \; / \; \varepsilon^\mathrm{misid}_\mathrm{MC} \; .
\end{equation}

The following systematic effects on the misidentification probability based on negative tags
are considered:
\begin{itemize}

\item \textbf{\cPqb\ and \cPqc\ fractions}:
The fraction of \cPqb-flavour jets has been measured in CMS to agree with the
simulation within a $\pm 20\%$ uncertainty~\cite{BPH-10-009}.
A ${\pm}20\%$ uncertainty is conservatively estimated
for the overall fraction of \cPqb\ and \cPqc\ jets.
The b- and c-flavour fraction is varied in the QCD multijet simulation,
from which a systematic uncertainty on $R_\text{light}$ is inferred.

\item \textbf{Gluon fraction}: This affects both the misidentification probability in
  simulation and the overall negative tag rates.  The average fraction of
  gluon jets depends on the details of the parton density and hadronization
  functions used in the simulation.  An uncertainty of ${\pm}20\%$ is
  extracted from the comparison of simulation with data~\cite{QCD-10-011}.

\item \textbf{Long-lived \boldmath{\PKzS} and \boldmath{\PgL} decays}: The amount
  of reconstructed \PKzS\ and \PgL\  are found to be larger in the data than in the
  simulation~\cite{QCD-10-007}.  To estimate the uncertainty on $R_\text{light}$
  due to the \PKzS\ and \PgL\  contribution, the simulated jets
  are reweighted by factors of $1.3 \pm 0.3$ and $1.5 \pm 0.5$, respectively, in order to match the observed yield of \PKzS\ and  \PgL\  in the data.
  The quoted uncertainties on the factors account for the \pt dependence.
  The yield is varied accordingly and the inferred variation on $R_\text{light}$ is taken as a
  systematic uncertainty.

\item \textbf{Photon conversion and nuclear interactions}:
The rate of secondary interactions in the pixel detector layers has been
measured with ${\pm}5\%$ precision \cite{TRK-10-001,TRK-10-003}.
The corresponding variation implies a systematic uncertainty on $R_\text{light}$.

\item \textbf{Mismeasured tracks}: According to the simulation, jets with a
  reconstructed track not associated with a genuine charged particle also
  present an excess of positive over negative tags.  To correct for
  residual mismeasurement effects, a $\pm 50\%$ variation on this
  contribution is taken into account in the systematic uncertainty on
  $R_\text{light}$.

\item \textbf{Sign flip}: Small differences in the angle between a track and the jet axis can lead to a change of the sign of the impact parameter (``sign flip'') and modify the negative tag rate.
In order to quantify this effect the ratio of the number of negative to positive
  tagged jets is computed in a muon jet sample similar to the one described
  in Section~\ref{sec:EfficienciesMuonJets}, with a larger than 80\% \cPqb\
  purity.  Data and simulation are found to be in good agreement.  From the
  statistical uncertainty on the comparison, the absolute uncertainty on
  this ratio is estimated as 2\%, 1\%, and 0.5\% for loose, medium, and
  tight operating points, respectively.  This sign flip uncertainty can be
  translated into a systematic uncertainty on $R_\text{light}$.

\item \textbf{Pileup:} The misidentification probability depends on the pileup model used
  in the simulation.  The simulated events are reweighted in order to match
  the pileup rate in the data.  Differences between $R_\text{light}$ values
  obtained for different running periods are used to estimate the
  systematic uncertainty, which is about ${\pm}1$\% for all taggers.

\item \textbf{Event sample:} Physics analyses use jets from different event
  topologies.  For a given jet \pt, the misidentification probability is different for the leading jet
or if there are other jets with higher \pt values in
  the same event.  Measured misidentification scale factors for leading and
  subleading jets have a dispersion of about 7\%.  In addition, misidentification scale
  factors vary by 2--7\%, depending on the tagger, for different running
  periods.  These two uncertainties are added in quadrature to account for
  an uncertainty due to sample dependence.  This is the dominant
  contribution to the overall systematic uncertainty on the misidentification probability.
\end{itemize}
The systematic uncertainties are detailed in Table~\ref{tab:syst_mistag}
for the various algorithms and for the example of the medium operating points
in the jet \pt range between 80 and 120\GeVc.
\begin{table}[htbp]
\begin{center}
\topcaption{Relative systematic uncertainties on $SF_\text{light}$ for jet \pt\ in the range 80--120\GeVc.
The columns correspond to the different sources of systematics in the order described in the text.}
\begin{tabular}{lcccccccc}
\hline
\cPqb\ tagger & \cPqb\ and  & gluon & V$^0$ and & mismeas. & sign flip
                                                     & MC stat & pileup and & total \\
              & \cPqc\ jets &  & $2^{nd}$ int. &  &
                                                     &  & evt. sample &  \\
\hline

       JPM & 8.6\% & 0.8\% & 7.9\% & 1.0\% & 6.4\% & 0.9\% & 9.4\% & 16.5\% \\
      JBPM & 6.2\% & 1.2\% & 6.9\% & 0.5\% & 1.6\% & 0.9\% & 9.0\% & 13.2\% \\
     TCHEM & 4.5\% & 0.8\% & 6.2\% & 1.2\% & 5.1\% & 0.7\% & 8.0\% & 12.4\% \\
     TCHPM & 1.6\% & 1.0\% & 3.0\% & 0.6\% & 2.5\% & 0.6\% & 9.2\% & 10.3\% \\
    SSVHEM & 1.0\% & 0.9\% & 3.2\% & 1.9\% & 2.9\% & 0.7\% & 7.3\% & 9.0\% \\
      CSVM & 3.2\% & 1.8\% & 4.4\% & 0.7\% & 4.6\% & 0.7\% & 7.4\% & 10.6\% \\
\hline
\label{tab:syst_mistag}
\end{tabular}
\end{center}
\end{table}

\begin{table}[bth]
\begin{center}
  \topcaption[]{Misidentification probabilities and the corresponding data/MC scale factors $SF_\text{light}$ for different algorithms and operating points for jet \pt\ in the range 80--120\GeVc.
    The statistical uncertainties are quoted for the misidentification probabilities, while both the statistical and the systematic uncertainties are given for the scale factors.}
  \label{tab:SFmistag}
\begin{tabular}{lcccc}
\hline
\cPqb\ tagger  & misidentification probability & $SF_\text{light}$ \\ \hline
       JPL & $0.1000\pm  0.0004$ & $0.99\pm0.01\pm 0.10$\\
      JBPL & $0.1019\pm  0.0004$ & $0.96\pm0.01\pm 0.09$\\
     TCHEL & $0.1989\pm  0.0005$ & $1.10\pm0.01\pm 0.09$\\
      CSVL & $0.1020\pm  0.0004$ & $1.10\pm0.01\pm 0.09$\\
\hline
       JPM & $0.0107\pm  0.0001$ & $1.03\pm0.01\pm 0.17$\\
      JBPM & $0.0110\pm  0.0001$ & $0.95\pm0.01\pm 0.13$\\
     TCHEM & $0.0282\pm  0.0003$ & $1.21\pm0.01\pm 0.15$\\
     TCHPM & $0.0304\pm  0.0003$ & $1.24\pm0.01\pm 0.13$\\
    SSVHEM & $0.0208\pm  0.0002$ & $0.94\pm0.01\pm 0.08$\\
      CSVM & $0.0151\pm  0.0002$ & $1.11\pm0.01\pm 0.12$\\
\hline
       JPT & $0.00116\pm  0.00005$ & $1.03\pm0.04\pm 0.25$\\
      JBPT & $0.00117\pm  0.00004$ & $0.95\pm0.04\pm 0.19$\\
     TCHPT & $0.00284\pm  0.00009$ & $1.26\pm0.04\pm 0.21$\\
    SSVHPT & $0.00207\pm  0.00009$ & $1.02\pm0.04\pm 0.17$\\
      CSVT & $0.00120\pm  0.00005$ & $1.17\pm0.05\pm 0.21$\\
\hline
\end{tabular}
\end{center}
\end{table}

\begin{figure}[hbtp]
  \begin{center}
    \includegraphics[width=.9\textwidth]{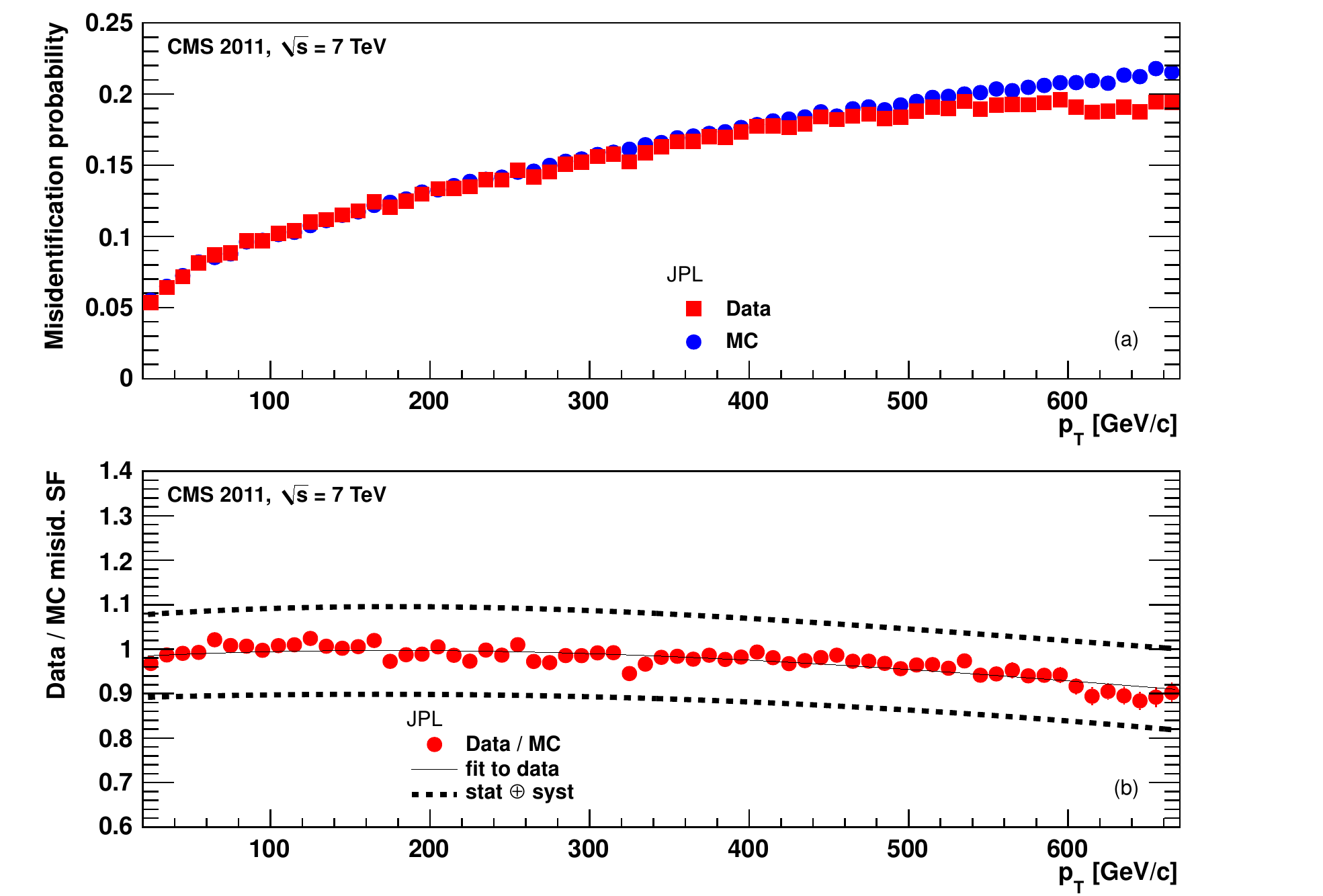}
    \caption{For the JPL tagger:
      (a) misidentification probability in data (red squares) and simulation (blue dots);
      (b) scale factor for the misidentification probability.
      The last \pt bin in each plot includes
      all jets with $\pt > 670$\GeVc.  The solid curve is the result of
      a polynomial fit to the data points.  The dashed curves represent the
      overall statistical and systematic uncertainties on the
      measurements.}
    \label{fig:mistag_jpl}
  \end{center}
\end{figure}

\begin{figure}[hbtp]
  \begin{center}
    \includegraphics[width=.9\textwidth]{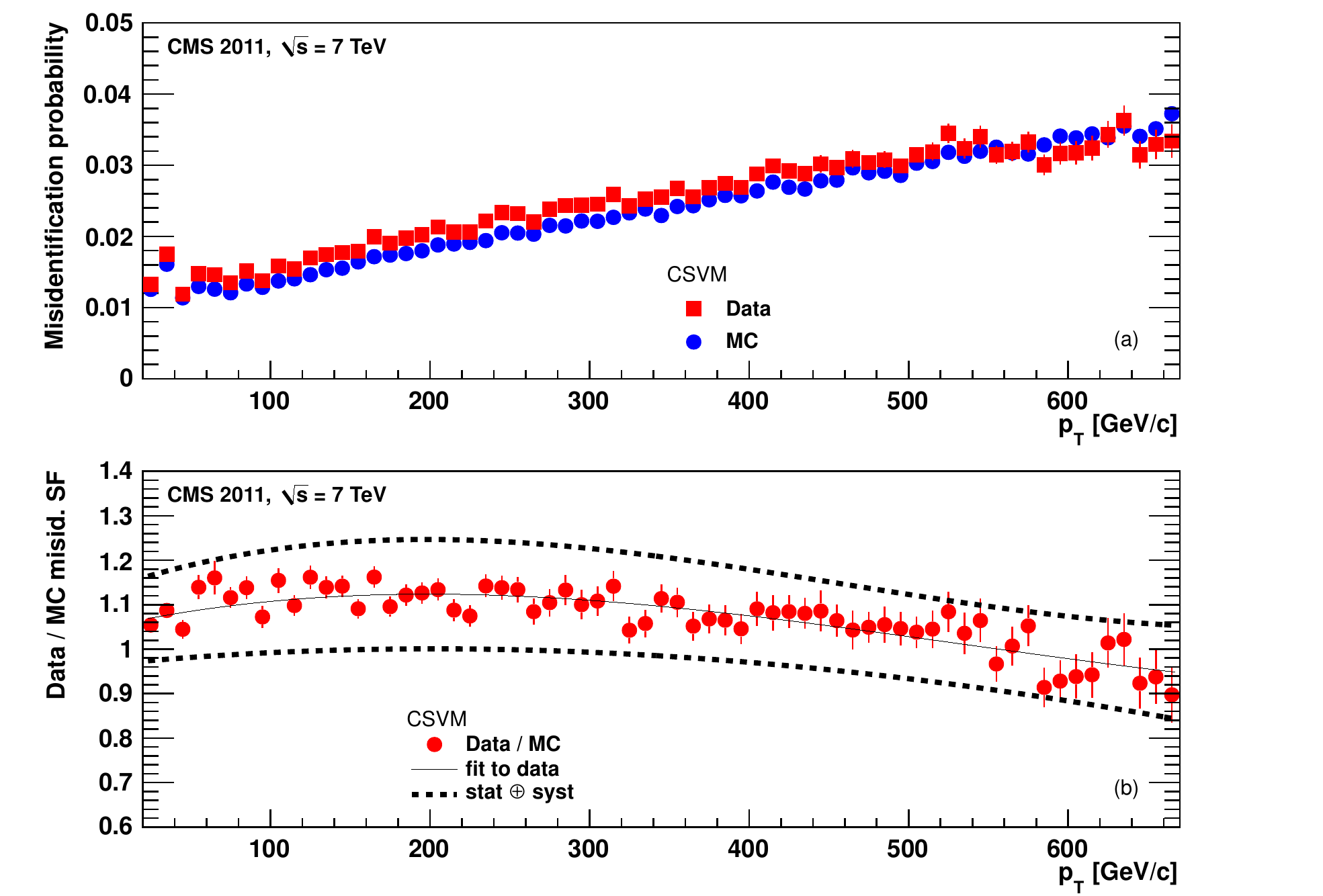}
    \caption{Same as Fig.~\ref{fig:mistag_jpl} but for the CSVM tagger.}
    \label{fig:mistag_csvm}
  \end{center}
\end{figure}

The measured misidentification probabilities and data/MC scale factors are
presented in Figs.~\ref{fig:mistag_jpl} and \ref{fig:mistag_csvm} as a
function of the jet \pt for the JPL and CSVM taggers.  For a jet \pt of
about 80\GeVc the misidentification probabilities are close to 10\% and 1\%
for the loose (JPL) and medium (CSVM) selections, respectively.  Both
algorithms show an increase of the misidentification probability with jet
\pt that can be explained by the higher track densities in collimated jets.
The simulation reproduces this dependence to a large extent.  The observed
scale factors are close to one with a decrease of ${\sim}10\%$ toward the
highest jet \pt.  The misidentification probabilities measured with data
and the data/MC scale factors are given in Table~\ref{tab:SFmistag} for
jets with \pt between 80 and 120\GeVc.  
The scale factors for the
misidentification probability have also been measured as a function of the
jet \pt\ for jets in several pseudorapidity intervals: $|\eta|<0.5$, $0.5
\leq |\eta| <1.0$, $1.0 \leq |\eta|<1.5$ and $1.5 \leq |\eta|<2.4$ for the
loose operating points and $|\eta| <0.8$, $0.8 \leq |\eta|<1.6$ and $1.6
\leq |\eta|<2.4$ for the medium operating points. For each \cPqb-tagging
algorithm, the scale factors are compatible within about 10\%. These
pseudorapidity-dependent scale factors for the misidentification
probabilities are used in physics analyses.


%% file: Conclusions.tex
\clearpage
\section{Conclusions}\label{sec:Conclusions}

The CMS collaboration has developed a variety of algorithms that are used to identify jets that arise from the hadronization of bottom quarks.
Early analyses relied on simple and robust techniques, based on the second or third highest impact parameter significance of the tracks associated to a jet, or the flight distance measured using a reconstructed secondary vertex.
More recent analyses use algorithms with better performance that define a powerful discriminant from the combination of several variables.
The use of these more advanced algorithms is made possible by the high degree of agreement achieved between data and simulation, and by the robustness of the algorithms against variations in the running conditions.

The algorithms provide selections at several operating points.
The efficiencies of the algorithms at these operating points have been measured with a number of methods using multijet and \ttbar events.
A differential measurement of the efficiency as a function of jet \pt, from 30 to 670\GeVc, has been carried out with the multijet sample. This information is used in analyses that require knowledge of the performance of \cPqb-jet tagging over a wide range of transverse momenta. The information is also helpful for analyses such as the measurement of the \ttbar cross section in order to avoid the strong
correlations that can occur if the efficiencies are inferred from the \ttbar event sample itself.

The \ttbar sample provides inclusive results, which are suitable for measurements of top-quark properties and for the analysis of standard model processes with similar jet momentum spectra and multiplicities.
The misidentification probability, that a light-parton jet is mistaken as a \cPqb-quark jet, has been measured by applying inverted tagging algorithms to the multijet events.

The most effective algorithm is the Combined Secondary Vertex tagger.
Using a loose selection, CMS achieves \cPqb-jet tagging efficiencies of about 85\%, for a light-parton misidentification probability of 10\%.
This selection is suited for \ttbar analyses.
In analyses requiring higher \cPqb-jet purity, such as searches for supersymmetric particles, approximately 70\% \cPqb-jet tagging efficiency is achieved, for a light-parton misidentification probability of only 1.5\%.
These values apply for jet transverse momenta typically observed in \ttbar events.

The measured b-jet tagging performance is quantified and implemented in CMS analyses by using scale factor corrections to the MC simulation.
These scale factors have been used extensively to enable studies over a wide range of event topologies that would otherwise not be possible due to limited statistics.
The scale factors for the \cPqb-jet tagging efficiencies are measured with uncertainties of 2--4\%, and 3--8\%, in the jet \pt\ range 30--320\GeVc and 320--670\GeVc respectively.
The maximum deviation of these scale factors from unity is approximately 10\%.
The scale factors for light-parton jet misidentification probabilities are measured to a precision of 8--17\% over the full \pt\ range, and differ from unity by at most 25\%.  
The scale factors for \cPqc-jet tagging efficiency are assumed to be the same as for \cPqb\ jets, with the corresponding uncertainty conservatively doubled.

The  \cPqb-jet identification techniques discussed in this paper have been used in more than 40 analyses published by CMS, including measurements of top-quark properties, the Higgs boson,  and searches for signals of physics beyond the standard model.
The reduction of the uncertainties on the \cPqb-jet tagging scale factors has enabled the CMS experiment to decrease the light-parton background to unprecedented levels while maintaining high \cPqb-jet tagging efficiency for a wide range of processes containing heavy-flavour jets.

%% file: Acknowledgements.tex
\hyphenation{Bundes-ministerium Forschungs-gemeinschaft Forschungs-zentren} We congratulate our colleagues in the CERN accelerator departments for the excellent performance of the LHC machine. We thank the technical and administrative staff at CERN and other CMS institutes. This work was supported by the Austrian Federal Ministry of Science and Research; the Belgium Fonds de la Recherche Scientifique, and Fonds voor Wetenschappelijk Onderzoek; the Brazilian Funding Agencies (CNPq, CAPES, FAPERJ, and FAPESP); the Bulgarian Ministry of Education and Science; CERN; the Chinese Academy of Sciences, Ministry of Science and Technology, and National Natural Science Foundation of China; the Colombian Funding Agency (COLCIENCIAS); the Croatian Ministry of Science, Education and Sport; the Research Promotion Foundation, Cyprus; the Ministry of Education and Research, Recurrent financing contract SF0690030s09 and European Regional Development Fund, Estonia; the Academy of Finland, Finnish Ministry of Education and Culture, and Helsinki Institute of Physics; the Institut National de Physique Nucl\'eaire et de Physique des Particules~/~CNRS, and Commissariat \`a l'\'Energie Atomique et aux \'Energies Alternatives~/~CEA, France; the Bundesministerium f\"ur Bildung und Forschung, Deutsche Forschungsgemeinschaft, and Helmholtz-Gemeinschaft Deutscher Forschungszentren, Germany; the General Secretariat for Research and Technology, Greece; the National Scientific Research Foundation, and National Office for Research and Technology, Hungary; the Department of Atomic Energy and the Department of Science and Technology, India; the Institute for Studies in Theoretical Physics and Mathematics, Iran; the Science Foundation, Ireland; the Istituto Nazionale di Fisica Nucleare, Italy; the Korean Ministry of Education, Science and Technology and the World Class University program of NRF, Korea; the Lithuanian Academy of Sciences; the Mexican Funding Agencies (CINVESTAV, CONACYT, SEP, and UASLP-FAI); the Ministry of Science and Innovation, New Zealand; the Pakistan Atomic Energy Commission; the Ministry of Science and Higher Education and the National Science Centre, Poland; the Funda\c{c}\~ao para a Ci\^encia e a Tecnologia, Portugal; JINR (Armenia, Belarus, Georgia, Ukraine, Uzbekistan); the Ministry of Education and Science of the Russian Federation, the Federal Agency of Atomic Energy of the Russian Federation, Russian Academy of Sciences, and the Russian Foundation for Basic Research; the Ministry of Science and Technological Development of Serbia; the Secretar\'{\i}a de Estado de Investigaci\'on, Desarrollo e Innovaci\'on and Programa Consolider-Ingenio 2010, Spain; the Swiss Funding Agencies (ETH Board, ETH Zurich, PSI, SNF, UniZH, Canton Zurich, and SER); the National Science Council, Taipei; the Scientific and Technical Research Council of Turkey, and Turkish Atomic Energy Authority; the Science and Technology Facilities Council, UK; the US Department of Energy, and the US National Science Foundation.

Individuals have received support from the Marie-Curie programme and the European Research Council (European Union); the Leventis Foundation; the A. P. Sloan Foundation; the Alexander von Humboldt Foundation; the Austrian Science Fund (FWF); the Belgian Federal Science Policy Office; the Fonds pour la Formation \`a la Recherche dans l'Industrie et dans l'Agriculture (FRIA-Belgium); the Agentschap voor Innovatie door Wetenschap en Technologie (IWT-Belgium); the Council of Science and Industrial Research, India; the Compagnia di San Paolo (Torino); and the HOMING PLUS programme of Foundation for Polish Science, cofinanced from European Union, Regional Development Fund.

%% file: Glossary.tex
\newpage
\markboth{Glossary}{Glossary}
\section*{Glossary}

\begin{description}
\item[bSample] Method to measure the \bdtag efficiency in \ttbar events from a \cPqb-enriched sample
\item[CSVL] Combined Secondary Vertex algorithm at the loose operating point
\item[CSVM] Combined Secondary Vertex algorithm at the medium operating point
\item[CSVT] Combined Secondary Vertex algorithm at the tight operating point
\item[FTC] Flavour Tag Consistency method for the measurement of the \bdtag efficiency in \ttbar events
\item[FTM] Flavour Tag Matching method for the measurement of the \bdtag efficiency in \ttbar events
\item[IP3D (method)] Method for the measurement of the \bdtag efficiency in multijet events based on the impact parameters of muons
\item[IP] Impact parameter of a track
\item[JBPL] Jet B Probability algorithm at the loose operating point
\item[JBPM] Jet B Probability algorithm at the medium operating point
\item[JBPT] Jet B Probability algorithm at the tight operating point
\item[JPL] Jet Probability algorithm at the loose operating point
\item[JPM] Jet Probability algorithm at the medium operating point
\item[JPT] Jet Probability algorithm at the tight operating point
\item[LT (method)] Lifetime Tagging method for the measurement of the \bdtag efficiency in multijet events
\item[PLR] Profile Likelihood Ratio method for the measurement of the \bdtag efficiency in \ttbar events
\item[PtRel (method)] Method for the measurement of the \bdtag efficiency in multijet events based on the transverse momenta of muons w.r.t. the jet axis
\item[PV] Primary Vertex (proton-proton interaction point)
\item[SIP] Significance of the impact parameter of a track
\item[SSVHEM] Simple Secondary Vertex High Efficiency algorithm at the medium operating point
\item[SSVHPT] Simple Secondary Vertex High Efficiency algorithm at the tight operating point
\item[SV] Secondary Vertex (decay vertex of a long-lived particle)
\item[TC] Track Counting (TCHE and TCHP) algorithms
\item[TCHEL] Track Counting High Efficiency algorithm at the loose operating point
\item[TCHEM] Track Counting High Efficiency algorithm at the medium operating point
\item[TCHPM] Track Counting High Purity algorithm at the medium operating point
\item[TCHPT] Track Counting High Efficiency algorithm at the tight operating point
\end{description}

%% file: BTV-12-001-authorlist.tex
\textbf{Yerevan Physics Institute,  Yerevan,  Armenia}\\*[0pt]
S.~Chatrchyan, V.~Khachatryan, A.M.~Sirunyan, A.~Tumasyan
\vskip\cmsinstskip
\textbf{Institut f\"{u}r Hochenergiephysik der OeAW,  Wien,  Austria}\\*[0pt]
W.~Adam, E.~Aguilo, T.~Bergauer, M.~Dragicevic, J.~Er\"{o}, C.~Fabjan\cmsAuthorMark{1}, M.~Friedl, R.~Fr\"{u}hwirth\cmsAuthorMark{1}, V.M.~Ghete, J.~Hammer, N.~H\"{o}rmann, J.~Hrubec, M.~Jeitler\cmsAuthorMark{1}, W.~Kiesenhofer, V.~Kn\"{u}nz, M.~Krammer\cmsAuthorMark{1}, I.~Kr\"{a}tschmer, D.~Liko, I.~Mikulec, M.~Pernicka$^{\textrm{\dag}}$, B.~Rahbaran, C.~Rohringer, H.~Rohringer, R.~Sch\"{o}fbeck, J.~Strauss, A.~Taurok, W.~Waltenberger, G.~Walzel, E.~Widl, C.-E.~Wulz\cmsAuthorMark{1}
% ...
\vskip\cmsinstskip
\dag:~Deceased\\
1:~~Also at Vienna University of Technology, Vienna, Austria\\
% ...


%% file: BTV-12-001_temp.bbl
\providecommand{\href}[2]{#2}\begingroup\raggedright\begin{thebibliography}{10}%
\makeatletter
\providecommand{\hrefCMSnoop }[0]{\@secondoftwo}%
\makeatother
\providecommand{\doi}{\texttt{doi:}\begingroup \urlstyle{tt}\Url}

\bibitem{Chatrchyan:2011yy}
\hrefCMSnoop {} {{ CMS} Collaboration, ``{Measurement of the \ttbar\ Production
  Cross Section in pp Collisions at 7~TeV in Lepton + Jets Events Using b-quark
  Jet Identification}'',} \textit{ Phys. Rev. D} \textbf{ 84} (2011) 092004,
  \href{http://dx.doi.org/10.1103/PhysRevD.84.092004}{\doi{10.1103/PhysRevD.84.092004}},
\href{http://www.arXiv.org/abs/1108.3773}{\texttt{ arXiv:1108.3773}}.

\bibitem{Chatrchyan:2011bj}
\hrefCMSnoop {} {{ CMS} Collaboration, ``{Search for supersymmetry in events
  with b jets and missing transverse momentum at the LHC}'',} \textit{ JHEP}
  \textbf{ 07} (2011) 113,
  \href{http://dx.doi.org/10.1007/JHEP07(2011)113}{\doi{10.1007/JHEP07(2011)113}},
\href{http://www.arXiv.org/abs/1106.3272}{\texttt{ arXiv:1106.3272}}.

\bibitem{Chatrchyan:2011vp}
\hrefCMSnoop {} {{ CMS} Collaboration, ``{Measurement of the t-Channel Single
  Top Quark Production Cross Section in $pp$ Collisions at $\sqrt{s} =
  7$~TeV}'',} \textit{ Phys. Rev. Lett.} \textbf{ 107} (2011) 091802,
  \href{http://dx.doi.org/10.1103/PhysRevLett.107.091802}{\doi{10.1103/PhysRevLett.107.091802}},
\href{http://www.arXiv.org/abs/1106.3052}{\texttt{ arXiv:1106.3052}}.

\bibitem{CMS:2008zzk}
\hrefCMSnoop {} {{ CMS} Collaboration, ``The {CMS} experiment at the {CERN}
  {LHC}'',} \textit{ JINST} \textbf{ 3} (2008) S08004,
\href{http://dx.doi.org/10.1088/1748-0221/3/08/S08004}{\doi{10.1088/1748-0221/3/08/S08004}}.

\bibitem{Sjostrand:2006za}
\hrefCMSnoop {} {T.~Sj{\"o}strand, S.~Mrenna, and P.~Z. Skands, ``{PYTHIA} 6.4
  physics and manual'',} \textit{ JHEP} \textbf{ 05} (2006) 026,
  \href{http://dx.doi.org/10.1088/1126-6708/2006/05/026}{\doi{10.1088/1126-6708/2006/05/026}},
\href{http://www.arXiv.org/abs/hep-ph/0603175}{\texttt{ arXiv:hep-ph/0603175}}.

\bibitem{Field:2010}
\hrefCMSnoop {} {R.~Field, ``{Early LHC underlying event data - findings and
  surprises}'',} (2010). \href{http://www.arXiv.org/abs/1010.3558}{\texttt{
  arXiv:1010.3558}}.

\bibitem{Alwall:2011uj}
J.~Alwall\hrefCMSnoop {} { {et~al.}, ``{MadGraph} 5: going beyond'',} \textit{
  JHEP} \textbf{ 06} (2011) 128,
  \href{http://dx.doi.org/10.1007/JHEP06(2011)128}{\doi{10.1007/JHEP06(2011)128}},
\href{http://www.arXiv.org/abs/1106.0522}{\texttt{ arXiv:1106.0522}}.

\bibitem{mlm}
M.~L. Mangano\hrefCMSnoop {} { {et~al.}, ``{Matching matrix elements and shower
  evolution for top- quark production in hadronic collisions}'',} \textit{
  JHEP} \textbf{ 01} (2007) 013,
  \href{http://dx.doi.org/10.1088/1126-6708/2007/01/013}{\doi{10.1088/1126-6708/2007/01/013}},
\href{http://www.arXiv.org/abs/hep-ph/0611129}{\texttt{ arXiv:hep-ph/0611129}}.

\bibitem{Davidson:2010rw}
N.~Davidson\hrefCMSnoop {} { {et~al.}, ``{Universal Interface of TAUOLA
  Technical and Physics Documentation}'',} \textit{ Comput. Phys. Commun.}
  \textbf{ 183} (2012) 821,
  \href{http://dx.doi.org/10.1016/j.cpc.2011.12.009}{\doi{10.1016/j.cpc.2011.12.009}},
\href{http://www.arXiv.org/abs/1002.0543}{\texttt{ arXiv:1002.0543}}.

\bibitem{powheg}
\hrefCMSnoop {} {S.~Frixione, P.~Nason, and C.~Oleari, ``{Matching NLO QCD
  computations with parton shower simulations: the POWHEG method}'',} \textit{
  JHEP} \textbf{ 11} (2007) 070,
  \href{http://dx.doi.org/10.1088/1126-6708/2007/11/070}{\doi{10.1088/1126-6708/2007/11/070}},
  \href{http://www.arXiv.org/abs/0709.2092}{\texttt{ arXiv:0709.2092}}.

\bibitem{Campbell:2010ff}
\hrefCMSnoop {} {J.~M. Campbell and R.~K. Ellis, ``{MCFM for the Tevatron and
  the LHC}'',} \textit{ Nucl. Phys. Proc. Suppl.} \textbf{ 205-206} (2010) 10,
  \href{http://dx.doi.org/10.1016/j.nuclphysbps.2010.08.011}{\doi{10.1016/j.nuclphysbps.2010.08.011}},
\href{http://www.arXiv.org/abs/1007.3492}{\texttt{ arXiv:1007.3492}}.

\bibitem{mstw08}
A.~D. Martin\hrefCMSnoop {} { {et~al.}, ``{Uncertainties on $\alpha_s$ in
  global PDF analyses and implications for predicted hadronic cross
  sections}'',} \textit{ Eur. Phys. J. C} \textbf{ 64} (2009) 653,
  \href{http://dx.doi.org/10.1140/epjc/s10052-009-1164-2}{\doi{10.1140/epjc/s10052-009-1164-2}},
\href{http://www.arXiv.org/abs/0905.3531}{\texttt{ arXiv:0905.3531}}.

\bibitem{Lai:2010nw}
H.-L. Lai\hrefCMSnoop {} { {et~al.}, ``{Uncertainty induced by QCD coupling in
  the CTEQ global analysis of parton distributions}'',} \textit{ Phys. Rev. D}
  \textbf{ 82} (2010) 054021,
  \href{http://dx.doi.org/10.1103/PhysRevD.82.054021}{\doi{10.1103/PhysRevD.82.054021}},
\href{http://www.arXiv.org/abs/1004.4624}{\texttt{ arXiv:1004.4624}}.

\bibitem{Demartin:2010er}
F.~Demartin\hrefCMSnoop {} { {et~al.}, ``{The impact of parton distribution
  function and $\alpha_s$ uncertainties on Higgs boson production in gluon
  fusion at hadron colliders}'',} \textit{ Phys. Rev. D} \textbf{ 82} (2010)
  014002,
  \href{http://dx.doi.org/10.1103/PhysRevD.82.014002}{\doi{10.1103/PhysRevD.82.014002}},
\href{http://www.arXiv.org/abs/1004.0962}{\texttt{ arXiv:1004.0962}}.

\bibitem{Botje:2011sn}
\hrefCMSnoop {} {M.~Botje {et~al.}, ``{The PDF4LHC Working Group Interim
  Recommendations}'',} (2011).
  \href{http://www.arXiv.org/abs/1101.0538}{\texttt{ arXiv:1101.0538}}.

\bibitem{mcfm2}
J.~M. Campbell\hrefCMSnoop {} { {et~al.}, ``{Next-to-Leading-Order Predictions
  for t-Channel Single-Top Production at Hadron Colliders}'',} \textit{ Phys.
  Rev. Lett.} \textbf{ 102} (2009) 182003,
  \href{http://dx.doi.org/10.1103/PhysRevLett.102.182003}{\doi{10.1103/PhysRevLett.102.182003}},
\href{http://www.arXiv.org/abs/0903.0005}{\texttt{ arXiv:0903.0005}}.

\bibitem{mcfm3}
\hrefCMSnoop {} {J.~M. Campbell and F.~Tramontano, ``{Next-to-leading order
  corrections to W t production and decay}'',} \textit{ Nucl. Phys. B} \textbf{
  726} (2005) 109,
  \href{http://dx.doi.org/10.1016/j.nuclphysb.2005.08.015}{\doi{10.1016/j.nuclphysb.2005.08.015}},
\href{http://www.arXiv.org/abs/hep-ph/0506289}{\texttt{ arXiv:hep-ph/0506289}}.

\bibitem{mcfm4}
\hrefCMSnoop {} {J.~M. Campbell, R.~K. Ellis, and F.~Tramontano, ``{Single top
  production and decay at next-to-leading order}'',} \textit{ Phys. Rev. D}
  \textbf{ 70} (2004) 094012,
  \href{http://dx.doi.org/10.1103/PhysRevD.70.094012}{\doi{10.1103/PhysRevD.70.094012}},
\href{http://www.arXiv.org/abs/hep-ph/0408158}{\texttt{ arXiv:hep-ph/0408158}}.

\bibitem{Kidonakis:2010tW}
\hrefCMSnoop {} {N.~Kidonakis, ``{Two-loop soft anomalous dimensions for single
  top quark associated production with a $W^{-}$ or $H^{-}$ }'',} \textit{
  Phys. Rev. D} \textbf{ 82} (2010) 054018,
  \href{http://dx.doi.org/10.1103/PhysRevD.82.054018}{\doi{10.1103/PhysRevD.82.054018}},
  \href{http://www.arXiv.org/abs/1005.4451}{\texttt{ arXiv:1005.4451}}.

\bibitem{NNLLtop}
\hrefCMSnoop {} {N.~Kidonakis, ``Next-to-next-to-leading logarithm resummation
  for $s$-channel single top quark production'',} \textit{ Phys. Rev. D}
  \textbf{ 81} (2010) 054028,
  \href{http://dx.doi.org/10.1103/PhysRevD.81.054028}{\doi{10.1103/PhysRevD.81.054028}},
  \href{http://www.arXiv.org/abs/1001.5034}{\texttt{ arXiv:1001.5034}}.

\bibitem{fewz}
\hrefCMSnoop {} {K.~Melnikov and F.~Petriello, ``{Electroweak gauge boson
  production at hadron colliders through $\mathcal{O}(\alpha_s^2)$}'',}
  \textit{ Phys. Rev. D} \textbf{ 74} (2006) 114017,
  \href{http://dx.doi.org/10.1103/PhysRevD.74.114017}{\doi{10.1103/PhysRevD.74.114017}},
\href{http://www.arXiv.org/abs/hep-ph/0609070}{\texttt{ arXiv:hep-ph/0609070}}.

\bibitem{CMS-PAS-TOP-11-003}
\hrefCMSnoop {} {{ CMS} Collaboration, ``{Measurement of the \ttbar production
  cross section in pp collisions at 7 TeV in lepton+jets events using b-quark
  jet identification}'',} \textit{ Phys. Rev. D} \textbf{ 84} (2011) 092004,
  \href{http://dx.doi.org/10.1103/PhysRevD.84.092004}{\doi{10.1103/PhysRevD.84.092004}},
\href{http://www.arXiv.org/abs/1108.3773}{\texttt{ arXiv:1108.3773}}.

\bibitem{PDG}
\hrefCMSnoop {} {{Particle Data Group}, J.~Beringer {et~al.}, ``Review of
  Particle Physics'',} \textit{ Phys. Rev. D} \textbf{ 86} (2012) 010001,
  \href{http://dx.doi.org/10.1103/PhysRevD.86.010001}{\doi{10.1103/PhysRevD.86.010001}}.

\bibitem{GEANT4}
\hrefCMSnoop {} {{ GEANT4} Collaboration, ``{GEANT4}---a simulation toolkit'',}
  \textit{ Nucl. Instrum. Meth. A} \textbf{ 506} (2003) 250,
\href{http://dx.doi.org/10.1016/S0168-9002(03)01368-8}{\doi{10.1016/S0168-9002(03)01368-8}}.

\bibitem{PFT-09-001}
\href {http://cdsweb.cern.ch/record/1194487} {{ CMS} Collaboration,
  ``Particle-Flow Event Reconstruction in CMS and Performance for Jets, Taus,
  and Missing $E_T$'',} CMS Physics Analysis Summary CMS-PAS-PFT-09-001,
  (2009).

\bibitem{PFT-10-002}
\href {http://cdsweb.cern.ch/record/1279341} {{ CMS} Collaboration,
  ``Commissioning of the Particle-Flow Reconstruction in Minimum-Bias and Jet
  Events from {\Pp\Pp} Collisions at 7 {TeV}'',} CMS Physics Analysis Summary
  CMS-PAS-PFT-10-002, (2010).

\bibitem{antikt}
\hrefCMSnoop {} {M.~Cacciari, G.~P. Salam, and G.~Soyez, ``The anti-$k_t$ jet
  clustering algorithm'',} \textit{ JHEP} \textbf{ 04} (2008) 063,
  \href{http://dx.doi.org/10.1088/1126-6708/2008/04/063}{\doi{10.1088/1126-6708/2008/04/063}},
\href{http://www.arXiv.org/abs/0802.1189}{\texttt{ arXiv:0802.1189}}.

\bibitem{Chatrchyan:2011ds}
\hrefCMSnoop {} {{ CMS} Collaboration, ``Determination of jet energy
  calibration and transverse momentum resolution in {CMS}'',} \textit{ JINST}
  \textbf{ 06} (2011) P11002,
  \href{http://dx.doi.org/10.1088/1748-0221/6/11/P11002}{\doi{10.1088/1748-0221/6/11/P11002}},
\href{http://www.arXiv.org/abs/1107.4277}{\texttt{ arXiv:1107.4277}}.

\bibitem{TRK-10-001}
\hrefCMSnoop {} {{ CMS} Collaboration, ``{CMS Tracking Performance Results from
  Early LHC Operation}'',} \textit{ Eur. Phys. J. C} \textbf{ 70} (2010) 1165,
  \href{http://dx.doi.org/10.1140/epjc/s10052-010-1491-3}{\doi{10.1140/epjc/s10052-010-1491-3}}.

\bibitem{Chatrchyan:2012xi}
\hrefCMSnoop {} {{ CMS} Collaboration, ``{Performance of CMS muon
  reconstruction in pp collision events at sqrt(s) = 7 TeV}'',} \textit{ JINST}
  \textbf{ 7} (2012) P10002,
  \href{http://dx.doi.org/10.1088/1748-0221/7/10/P10002}{\doi{10.1088/1748-0221/7/10/P10002}},
  \href{http://www.arXiv.org/abs/1206.4071}{\texttt{ arXiv:1206.4071}}.

\bibitem{AVFitter}
\hrefCMSnoop {} {W.~Waltenberger, R.~Fr{\"u}hwirth, and P.~Vanlaer, ``Adaptive
  vertex fitting'',} \textit{ J. Phys. G} \textbf{ 34} (2007) N343,
  \href{http://dx.doi.org/10.1088/0954-3899/34/12/N01}{\doi{10.1088/0954-3899/34/12/N01}}.

\bibitem{TRK-11-001}
\hrefCMSnoop {} {{ CMS} Collaboration, ``Description and performance of CMS
  track reconstruction'',} CMS Physics Analysis Summary CMS-PAS-TRK-11-001,
  (2011).

\bibitem{Buskulic:1993}
\hrefCMSnoop {} {{ ALEPH} Collaboration, ``{A precise measurement of $\Gamma_{Z
  \to b \overline{b}}$ / $\Gamma_{Z \to \text{hadrons}}$}'',} \textit{ Phys.
  Lett. B} \textbf{ 313} (1993) 535,
  \href{http://dx.doi.org/10.1016/0370-2693(93)90028-G}{\doi{10.1016/0370-2693(93)90028-G}}.

\bibitem{Borisov:1996}
\hrefCMSnoop {} {G.~Borisov and C.~Mariotti, ``{Fine tuning of track impact
  parameter resolution of the DELPHI detector}'',} \textit{ Nucl. Instrum.
  Meth. A} \textbf{ 372} (1996) 181,
  \href{http://dx.doi.org/10.1016/0168-9002(95)01287-7}{\doi{10.1016/0168-9002(95)01287-7}}.

\bibitem{TkAl_VBlobel}
\hrefCMSnoop {} {V.~Blobel, ``{Software alignment for tracking detectors}'',}
  \textit{ Nucl. Instrum. Meth. A} \textbf{ 566} (2006) 5,
\href{http://dx.doi.org/10.1016/j.nima.2006.05.157}{\doi{10.1016/j.nima.2006.05.157}}.

\bibitem{TkAl_Millepede}
G.~Flucke\hrefCMSnoop {} { {et~al.}, ``{CMS silicon tracker alignment strategy
  with the Millepede II algorithm}'',} \textit{ JINST} \textbf{ 3} (2008)
  P09002,
\href{http://dx.doi.org/10.1088/1748-0221/3/09/P09002}{\doi{10.1088/1748-0221/3/09/P09002}}.

\bibitem{ptrel}
\hrefCMSnoop {} {{ UA1} Collaboration, ``{Study of heavy flavour production in
  events with a muon accompanied by jet(s) at the CERN proton-antiproton
  collider}'',} \textit{ Z. Phys. C} \textbf{ 37} (1988) 489,
  \href{http://dx.doi.org/10.1007/BF01549709}{\doi{10.1007/BF01549709}}.

\bibitem{Benoit}
\href {http://scd-theses.u-strasbg.fr/1103/} {B.~Cl{\'e}ment, ``{Electroweak
  production of the top quark in the Run II of the D0 experiment}''}.
\newblock PhD thesis, IPHC, Universit{\'e} de Strasbourg, France, 2006.

\bibitem{Abazov:2010}
\hrefCMSnoop {} {{ D0} Collaboration, ``{b-Jet Identification in the D0
  Experiment}'',} \textit{ Nucl. Instrum. Meth. A} \textbf{ 620} (2010) 490,
  \href{http://dx.doi.org/10.1016/j.nima.2010.03.118}{\doi{10.1016/j.nima.2010.03.118}},
  \href{http://www.arXiv.org/abs/1002.4224}{\texttt{ arXiv:1002.4224}}.

\bibitem{BPH-10-010}
\hrefCMSnoop {} {{ CMS} Collaboration, ``{Measurement of $B\bar{B}$ angular
  correlations based on secondary vertex reconstruction at $\sqrt{s} =
  7$~TeV}'',} \textit{ JHEP} \textbf{ 03} (2011) 136,
  \href{http://dx.doi.org/10.1007/JHEP03(2011)136}{\doi{10.1007/JHEP03(2011)136}},
  \href{http://www.arXiv.org/abs/1102.3194}{\texttt{ arXiv:1102.3194}}.

\bibitem{CMS-PAS-TOP-11-005}
\hrefCMSnoop {} {{ CMS} Collaboration, ``{Measurement of the
  $\mathrm{\mathrm{t\bar{t}}}$ production cross section in the dilepton channel
  in pp collisions at $\sqrt{s} = 7$~TeV}'',} (2012).
  \href{http://www.arXiv.org/abs/1208.2671}{\texttt{ arXiv:1208.2671}}.
  Submitted to JHEP.

\bibitem{Field:2009zz}
\hrefCMSnoop {} {R.~Field, ``{Studying the underlying event at CDF and the
  LHC}'',} (2009). \href{http://www.arXiv.org/abs/1003.4220}{\texttt{
  arXiv:1003.4220}}. {Proceedings of the First International Workshop on
  Multiple Partonic Interactions at the LHC (MPI08)}.

\bibitem{STATREF}
{F. James}, ``Statistical Methods in Experimental Physics''.
\newblock World Scientific, 2nd edition, 2006.

\bibitem{wilks}
\hrefCMSnoop {} {S.~S. Wilks, ``{The Large-Sample Distribution of the
  Likelihood Ratio for Testing Composite Hypotheses}'',} \textit{ Ann. Math.
  Statist.} \textbf{ 9} (1938) 60,
  \href{http://dx.doi.org/10.1214/aoms/1177732360}{\doi{10.1214/aoms/1177732360}}.

\bibitem{BLUE}
\hrefCMSnoop {} {L.~Lyons, D.~Gibaut, and P.~Clifford, ``How to combine
  correlated estimates of a single physical quantity'',} \textit{ Nucl.
  Instrum. Meth. A} \textbf{ 270} (1988) 110,
  \href{http://dx.doi.org/10.1016/0168-9002(88)90018-6}{\doi{10.1016/0168-9002(88)90018-6}}.

\bibitem{BPH-10-009}
\hrefCMSnoop {} {{ CMS} Collaboration, ``{Inclusive b-jet production in pp
  collisions at $\sqrt{s} = 7$~TeV}'',} \textit{ JHEP} \textbf{ 04} (2012) 084,
  \href{http://dx.doi.org/10.1007/JHEP04(2012)084}{\doi{10.1007/JHEP04(2012)084}},
\href{http://www.arXiv.org/abs/1202.4617}{\texttt{ arXiv:1202.4617}}.

\bibitem{QCD-10-011}
\hrefCMSnoop {} {{ CMS} Collaboration, ``{Measurement of the Inclusive Jet
  Cross Section in pp Collisions at $\sqrt{s} = 7$~TeV}'',} \textit{ Phys. Rev.
  Lett.} \textbf{ 107} (2011) 132001,
  \href{http://dx.doi.org/10.1103/PhysRevLett.107.132001}{\doi{10.1103/PhysRevLett.107.132001}},
  \href{http://www.arXiv.org/abs/1106.0208}{\texttt{ arXiv:1106.0208}}.

\bibitem{QCD-10-007}
\hrefCMSnoop {} {{ CMS} Collaboration, ``{Strange particle production in pp
  collisions at $\sqrt{s} = 0.9$ and $7$~TeV}'',} \textit{ JHEP} \textbf{ 05}
  (2011) 064,
  \href{http://dx.doi.org/10.1007/JHEP05(2011)064}{\doi{10.1007/JHEP05(2011)064}},
  \href{http://www.arXiv.org/abs/1102.4282}{\texttt{ arXiv:1102.4282}}.

\bibitem{TRK-10-003}
\href {http://cdsweb.cern.ch/record/1279138} {{ {CMS}} Collaboration, ``Studies
  of Tracker Material'',} CMS Physics Analysis Summary CMS-PAS-TRK-10-003,
  (2010).

\end{thebibliography}\endgroup
